\documentclass[fleqn,useAMS,usenatbib,usedcolumn]{mnras}

\usepackage{newtxtext,newtxmath}

\usepackage[T1]{fontenc}
\usepackage{ae,aecompl}

\usepackage{graphicx}	
\usepackage{amsmath}	
\usepackage{amssymb}	

\usepackage{pdflscape}
\usepackage{subfig}
\usepackage{afterpage}

\newcommand{\sqdeg}{\,deg$^{2}$}  
\newcommand{\miriad}{\textsc{miriad}}  
\newcommand{\uJy}{\,$\umu$Jy}
\newcommand{\uJybm}{\,$\umu$Jy beam$^{-1}$}

\newcommand{\NBTG}{46}
\newcommand{\NBTGCONF}{22}
\newcommand{\NBTGCAND}{24}
\newcommand{\NBCS}{10}
\newcommand{\NDES}{10}
\newcommand{\NBCSDES}{4}
\newcommand{\NOPTICAL}{16}
\newcommand{\NOPTICALREAL}{8}
\newcommand{\NOPTICALFOOTPRINT}{20}

\newcolumntype{.}[1]{D{.}{.}{#1}}

\title[Spatial Correlation of BT Galaxies and Clusters]{The Spatial Correlation of Bent-Tail Galaxies and Galaxy Clusters}

\author[A. N. O'Brien et al.]{
Andrew N. O'Brien,$^{1,2}$\thanks{E-mail: andrew.obrien@westernsydney.edu.au (AOB)}
Ray P. Norris,$^{1,2}$
Nick F. H. Tothill,$^{1}$
Miroslav D. Filipovi\'c$^{1}$
\\
$^{1}$Western Sydney University, Locked Bag 1797, Penrith NSW 2751, Australia\\
$^{2}$Australia Telescope National Facility, CSIRO Astronomy and Space Science, PO Box 76, Epping, NSW 1710, Australia\\
}

\date{Accepted 2018 September 23. Received 2018 September 23; in original form 2018 January 7}

\pubyear{2018}

\begin{document}
\label{firstpage}
\pagerange{\pageref{firstpage}--\pageref{lastpage}}
\maketitle

\begin{abstract}
We have completed a deep radio continuum survey covering 86 square degrees of the \textit{Spitzer}-South Pole Telescope deep field to test whether bent-tail galaxies are associated with galaxy clusters. We present a new catalogue of \NBTGCONF{} bent-tail galaxies and a further \NBTGCAND{} candidate bent-tail galaxies. Surprisingly, of the \NOPTICALREAL{} bent-tail galaxies with photometric redshifts, only two are associated with known clusters. While the absence of bent-tail sources in known clusters may be explained by effects such as sensitivity, the absence of known clusters associated with most bent-tail galaxies casts doubt upon current models of bent-tail galaxies.
\end{abstract}




\section{Introduction}

Bent-tail (BT) radio sources (which include the classes of head-tail galaxies, wide-angle tails and narrow-angle tails) are a class of radio galaxy in which the jets expelled from the central supermassive black hole have been bent or significantly distorted from their typical linear trajectory. The complex morphologies of BT radio sources can be explained by environmental effects, the most significant of which is the exertion of strong ram pressures on the jets caused by the relative motion of the host galaxy through a dense medium. This motion may be a galaxy moving through a dense medium, or by the medium itself moving across the galaxy. A radio galaxy with a large peculiar velocity moving through the intra-cluster medium (ICM) of its host cluster may produce jet distortions, provided the velocity and density of the ICM are high enough \citep{Miley_1972, Rudnick_1976, Burns_1990}. These distortions may also be caused by violent movement of the ICM due to the merger history of the cluster \citep{Burns_1990, Roettiger_1996, Burns_1996, Filipovic_2010}. This makes BT radio sources a potentially useful tool for probing galaxy interactions on large scales with several studies showing that they tend to reside in galaxy clusters \citep{Blanton_2000, Mao_2009, Mao_2010, Wing_2011, Dehghan_2014}.

Galaxy clusters are the largest gravitationally bound structures in the Universe. They are formed around dark matter concentrations where sheets and filaments of the cosmic web intersect and therefore trace the large-scale structure of the Universe. Little observational evidence exists depicting the formation of galaxy clusters which is critical if we are to understand the evolution of the Universe on large scales.

Galaxy clusters are traditionally detected by optical searches for galaxy over-densities followed by photometric or spectroscopic measurements. Another technique is to detect the thermal emission from the diffuse ICM in the X-ray band. Both of these techniques are limited to detecting clusters within the local Universe ($z \lesssim 0.2$) as cosmological dimming effects make observing more distant clusters difficult. Therefore, studying younger clusters require alternative cluster detection techniques such as the Sunyaev-Zel'dovich (SZ) effect \citep{Sunyaev_1972} which is caused by inverse-Compton scattering of the Cosmic Microwave Background (CMB). As CMB photons pass through a dense plasma (such as an ICM), they collide with the rapidly moving particles in the plasma and receive a boost in energy. This can be observed as a localised increase in higher frequency photons and consequent decrement in lower frequency photons when compared with the surrounding CMB. As this is a direct measurement of the ICM column density, the SZ effect is an extremely useful tool for detecting galaxy clusters. A great benefit of the SZ effect technique is that it is redshift independent since the CMB energy density increases with redshift, cancelling out any cosmological dimming that affects optical or X-ray cluster measurement techniques.

The use of BT radio sources as galaxy cluster indicators shares this ability to detect more distant clusters as even modestly sensitive radio observations are able to detect radio galaxies up to high redshifts. BT radio sources have already been used to find clusters in the local Universe \citep[up to $z \sim 1$, e.g.][]{Blanton_2003}. Evidence of their effectiveness at tracing more distant clusters up to $z \sim 2$ is growing \citep[e.g.][]{Dehghan_2014}, suggesting that BT radio sources could be used to find galaxy clusters during their formative periods. \citet{Norris_2013} has shown that the Evolutionary Map of the Universe (EMU) survey \citep{Norris_2011} may detect hundreds of thousands of BT galaxies, more than the number of currently known clusters, and this is one of the motivations driving the study reported here.

Recent simulations performed by \citet{Mguda_2014} have shown that clusters with masses above $10^{13} h^{-1} \mathrm{M}_{\odot}$ typically host at least one BT radio source at some time. This ability to potentially detect distant clusters is especially attractive as it comes for free in any radio dataset with sufficient resolution to resolve the jets and lobes (i.e. on the order of several arcseconds).

This paper aims to determine the efficacy of detecting distant clusters by correlating BT radio sources identified in the high-resolution ATLAS-SPT radio catalogue with confirmed galaxy clusters from the literature. A preliminary account of this work was reported by \citet{OBrien_2016}. Throughout this paper we assume a flat $\Lambda$CDM cosmology with $H_{0} = 69.3\ \mathrm{km\ s^{-1} Mpc^{-1}}$ and $\Omega_{\mathrm{M}} = 0.286$ unless otherwise stated.

\section{Data}
\subsection{Observation Strategy}
\label{sec:observations}

Our sample of BT radio sources is taken from the Australia Telescope Large Area Survey of the South Pole Telescope \textit{Spitzer} Deep Field (ATLAS-SPT) which consists of 4,787 antenna pointings covering approximately 86\sqdeg{} between 23h $\leq \alpha \leq$ 24h (0h) and $-60\degr \leq \delta \leq -50\degr$ at 1.1--3.1 GHz. The project (ATNF project code: C2788) was awarded a total of 260 hours of observing time on the Australia Telescope Compact Array (ATCA) which were carried out throughout the 2013 April and October semesters. The observations were made with the array in the 6A and 6C configurations which produced images with a synthesised beam full-width half-maximum of approximately 8~arcsec.

\begin{table}
  \caption{ATLAS-SPT survey and imaging summary.}
  \label{tab:survey}
  \begin{tabular}{lr}
    \hline
    Survey & \\
    \hline
    Instrument & ATCA\\
    Project Code & C2788\\
    Frequency & 1.1--3.1 GHz\\
    Channel width & 1 MHz\\
    RA range & 23h $\leq \alpha \leq$ 24h\\
    Declination range & $-60\degr \leq \delta \leq -50\degr$\\
    Total area & $\sim 86$\sqdeg{}\\
    Total observing time & 260 hours\\
    Array configurations & 6A, 6C\\
    Pointings & 4,787\\
    Observing periods & 2013 May 19 -- June 4\\
                      & 2014 Feb 6 -- 11\\
    \hline
    Imaging & \\
    \hline
    Pixel size & 0.86~arcsec\\
    Beam size & $8 \times 8$~arcsec\\
    Mean rms & 108.2 \uJybm\\
    Median rms & 102.8 \uJybm\\
    \hline
  \end{tabular}
\end{table}

\begin{figure}
  \includegraphics[width=\columnwidth]{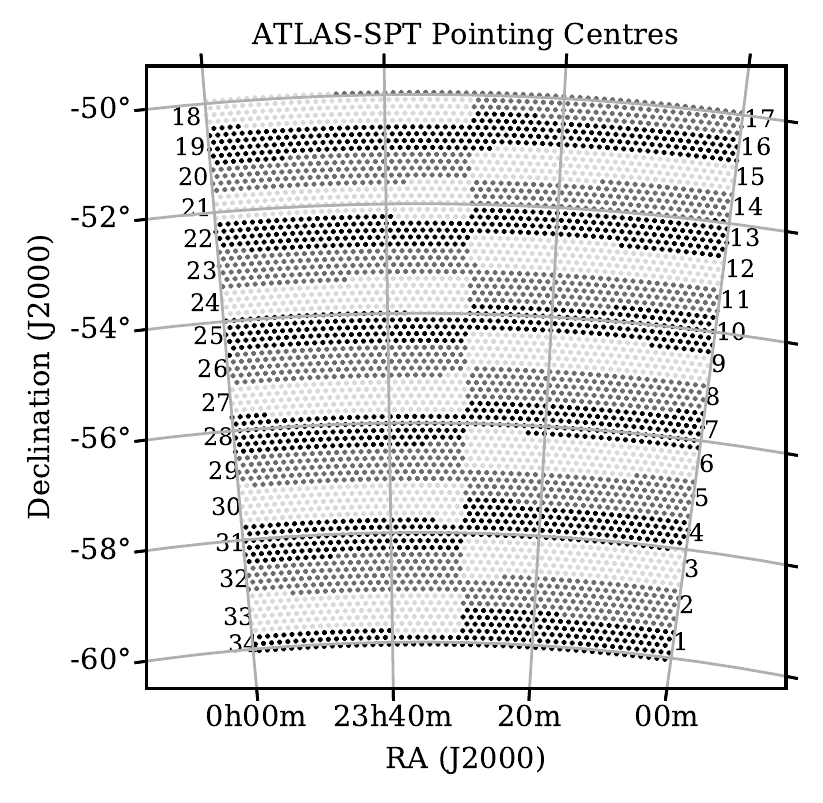}
  \caption{The antenna pointing centres for the ATLAS-SPT survey. Pointings are divided into `blocks' of 143 pointings (except block 34), where each block takes approximately one sidereal hour to observe including phase-calibrator scans. The blocks are shaded to ease distinguishing between blocks and their numbers are indicated on the field boundary.}
  \label{fig:pointings}
\end{figure}

Antenna pointings were arranged in a hexagonal grid where the spacing conforms to Nyquist sampling, i.e. $\theta_{\mathrm{hex}} = \lambda(D\sqrt{3})^{-1}$ where $\theta_{\mathrm{hex}}$ is the hexagonal spacing in radians, $\lambda$ is the observed wavelength and $D$ is the dish diameter. To ensure the pointings were well sampled across the field for the entire band, we used the upper end of the observing band (3.1 GHz) when determining the spacing. This resulted in the lower end of the band (1.1 GHz) being significantly oversampled. The final hexagonal pointing spacing is approximately 8.7~arcmin with horizontal spacings scaled proportional to $\cos{\delta}$ to avoid oversampling at lower declinations. To make the observing scheduling easier, the pointings were then divided into `blocks' that require approximately one sidereal hour to complete, producing 33 blocks of 143 pointings, and one block of 68. Many of the observing slots allocated to the project began before the entire field was above the horizon, so the blocks were arranged so they span half an hour in RA (with the exception of one block) so that observing could begin as soon as possible. The arrangement of the pointings and their respective blocks are illustrated in Figure~\ref{fig:pointings}.

To ensure adequate $uv$ coverage we required that each pointing be observed at least 6 times at equidistant hour angles. After considering various observational overheads such as instrument calibration, we were able to integrate on each pointing for a minimum of 108 seconds (18 seconds per sample of hour angle).

\subsection{Imaging}

We used the \miriad{} software package \citep{Sault_1995} to calibrate and image the data. In this section, we describe the data reduction process, which follows standard calibration procedures outlined in the \miriad{} online manual\footnote{\url{http://www.atnf.csiro.au/computing/software/miriad/userguide/userhtml.html}}.

Bandpass and absolute flux density calibration were conducted using observations of PKS~1934-638. Time-varying gains and polarisation leakage corrections were derived using observations of PKS~2333-528. Radio-frequency interference (RFI) from both internal and terrestrial sources is abundant within the 1.1--3.1~GHz observing window which must be flagged prior to deriving calibration solutions. Internal sources of RFI are well documented for the ATCA and were automatically flagged when loading the raw visibility data into \miriad{}. Other sources of RFI were flagged using statistical thresholding methods implemented in the task \textsc{pgflag} \citep{Offringa_2010}. Once the RFI in the calibrator data had been flagged, the calibration solutions were derived for each observation and applied to the visibilities of each block of observed pointings. We then repeated the automated flagging strategy on the mosaic data, after which approximately 25\% of the data was flagged. Additionally, extended periods of saturating wideband interference rendered approximately 15 hours of observing time unusable. This time was recovered with observations in subsequent semesters.

The visibilities for each observed block of mosaic pointings were then split by pointing and concatenated in time order. Dirty maps of each pointing were created using multi-frequency synthesis gridding and forced to be on the same pixel grid. Each pointing has a different synthesised beam size due to differences in $uv$ coverage and the 10\degr{} variation in Declination across the field. We therefore varied the robust weighting parameter for each pointing to closely match the resolutions between pointings. A spectral dirty beam (i.e. the instrument point-spread function as a function of frequency) was also created to allow wide-bandwidth deconvolution \citep{Sault_1999}.

As each pointing is imaged individually, and the $uv$ coverage per pointing is limited, sources with mJy level flux densities beyond the cutoff of the primary beam model are bright enough to produce linear artefacts which affect the central region of the pointing. We developed a customised outlier subtraction pipeline to identify, model, and subtract these sources from the visibility data during imaging. Since this field has largely not been observed at this frequency, we were unable to use a sky model to accurately subtract these sources. Thus, we imaged affected pointings well beyond the first null of the primary beam to locate the outlier sources. Doing this at the 0.86~arcsec pixel resolution would be extremely computationally expensive, so we did the initial outlier imaging at a coarse resolution of 4.3~arcsec. We then used the \miriad{} source-finding task \textsc{imsad} to extract approximate positions of the outlier sources which we then individually imaged as 200 pixel square `postage stamps' at the full 0.86~arcsec pixel resolution. Each outlier postage stamp was then deconvolved and the clean model was used to extract the source from the pointing visibilities, after which the pointing could be re-imaged without the outlier source artefacts.

Each dirty image was then deconvolved using the spectral dirty beam. We performed three iterations of phase self-calibration on each pointing, where each iteration used a clean model of increasing sensitivity (20, 10, then $5\sigma$). The self-calibration solutions were computed in 8 frequency bins to account for spectral variation. The final clean models were then convolved with a Gaussian fit to the dirty beam and added to the residual images (i.e. an image of the visibilities after subtracting the clean model) to produce a restored image. These restored images were then convolved with a Gaussian such that the final output image of all pointings had a resolution of 8~arcsec.

The pointing imaging was conducted on the \textit{Galaxy} supercomputer at the Pawsey Supercomputing Centre. Since each pointing was imaged independently, this task was ideal for parallelisation. With 472 compute nodes and a fast Lustre filesystem, \textit{Galaxy} is ideally suited for this kind of work, and we were able to image hundreds of pointings simultaneously unsupervised. Automated quality checks and visual inspection were used to find pointings with imaging errors, the most common of which was caused by corrupted self-calibration solutions derived from models containing artefacts. These pointings were re-imaged and re-calibrated manually, confining deconvolution to a user-defined mask around sources.

We then corrected all restored images for primary beam attenuation and linearly mosaicked the pointings. Due to output image size limitations in the linear mosaicking software, we produced 9 sub-mosaics of approximately 20,000 pixels along both coordinate axes, each overlapping with the adjacent sub-mosaics by approximately 1\degr{}. This overlap was chosen such that any given point within the observed field (excluding the outer edges) would be covered by at least one of the sub-mosaics at the lowest possible rms.

The average rms of each sub-mosaic is approximately 108~\uJybm{} with a median rms of 103~\uJybm{}. The sensitivity over the entire field is approximately uniform with only very minor variation. A summary of the survey and image parameters is given in Table~\ref{tab:survey}.

\section{Classification and Cross-Identification}
\subsection{Identification of BT Candidates}
\label{sec:cross-id}
The BT candidates were identified by visual inspection by searching the combined mosaic for FR-I and FR-II radio sources \citep{Fanaroff_1974} that showed significant bending or distortions away from a typical linear trajectory, and radio sources that showed the typical intensity asymmetry of head-tail radio galaxies (i.e. where the source appears to have an intense core with diffuse structure on only one side). If the radio core was clearly visible, the fitted brightness peak of the core component was recorded as the a radio source position. If the core was not easily determined, an estimation of the core position was made based on visual inspection of the source morphology. A total of \NBTG{} BT candidates were identified.

\subsection{Near-infrared Cross-Identification}
\label{sec:cross-id-infrared}

We visually cross-matched our BT candidate sample with catalogued sources from deep ($5\sigma$ sensitivity of 7.0~\uJy) 3.6\micron{} images from the \textit{Spitzer} South Pole Telescope Deep Field \citep[SSDF]{Ashby_2013}. If an AGN core was obvious in the radio image, a corresponding SSDF source was searched for around the radio core position. Otherwise, we selected the nearest SSDF source to an estimated core position based on the source morphology. In cases where the radio core is ambiguous and there are multiple possible SSDF counterparts, we investigated each likely cross-match when determining if the candidate resides in a known cluster (see Section~\ref{sec:matching_clusters}).

To estimate the number of chance cross-identifications, we repeated this cross-matching process using a copy of the SSDF catalogue shifted by $\pm10$~arcmin in both RA and Declination. We counted each occurrence when a shifted SSDF source may have been misidentified as the host galaxy for a BT candidate and find that 34~per~cent of our BT candidates could be falsely matched with an SSDF source. This relatively high fraction is due to a combination of effects. The first is that the SSDF catalogue has a far greater source density and higher resolution compared to our radio BT catalogue. This, combined with the fact that all our BT  candidates are extended, increases the chances of finding false identifications. Secondly, for 56~per~cent of our BT  candidates, the position of the radio core is not always clear from the radio image alone so the core position is identified by the presence of an SSDF source. This means that when matching with the shifted SSDF catalogue there is a much larger area a shifted source may be located within which would result in recording a potential false identification. When matching a BT  candidate which has a clearly distinguishable core in the radio image with its SSDF counterpart, the area the shifted source must be in is much smaller and therefore there is less chance of recording a potential false identification.

An infrared counterpart was recorded for all BT candidates which are the coordinates given in Table~\ref{tab:btg_sample} (columns 2 and 3).

\subsection{Optical Cross-Identification}
\label{sec:xid_optical}

\NOPTICALFOOTPRINT{} of our \NBTG{} BT  candidate sample are covered by the combined footprint of the Blanco Cosmology Survey \citep[BCS]{Desai_2012} and the Dark Energy Survey \citep[DES]{DES_2016} Science Verification (SV) `Gold' data release\footnote{\url{https://des.ncsa.illinois.edu/releases/sva1}} as described in \citet{Rykoff_2016}. The BCS and DES SV catalogues are mostly distinct with a small region of overlap. We cross-matched our BT  candidate sample with the BCS and DES SV catalogues automatically, searching for the nearest match between the SSDF counterparts and optical sources within 1~arcsec. Search radii above this value (up to 4~arcsec) did not increase the number of matches. Only \NOPTICAL{} of our \NOPTICALFOOTPRINT{} BT  candidate sources in the optical survey area have optical counterparts: \NBCS{} matches from BCS and \NDES{} matches from the DES SV catalogue (\NBCSDES{} sources appear in both BCS and DES SV). For both optical catalogues, the cross-matching was repeated 8 times with varying coordinate offsets of $\pm 10$~arcmin in both RA and Declination to estimate the false cross-identification rate. An average of 0.125 offset matches were found within 1~arcsec for both the BCS and DES SV catalogues.

\subsection{Estimating BT Source Redshifts}
\label{sec:redshifts}

Both the BCS and DES SV optical catalogues introduced in Section~\ref{sec:cross-id} contain photometric redshifts which we use to estimate the redshifts for our BT sample. Photometric redshifts provided in the BCS catalogue were estimated using an artificial neural network, \textsc{ANNz} \citep{Collister_2004}, and errors were provided in the catalogue. \citet{Bonnett_2016} provide photometric redshifts for the DES SV catalogue using several different techniques. We use those estimated using the \textsc{ANNz2} \citep{Sadeh_2016} technique to remain as consistent as possible with the BCS estimates. No errors were provided in the catalogue, but a normalised probability distribution function (PDF) for each redshift bin was provided for each source. We estimate the photo-z error by determining the cumulative sum of the PDF then calculating half the difference between the 16th and 86th quartiles.

\begin{figure}
  \includegraphics[width=\columnwidth]{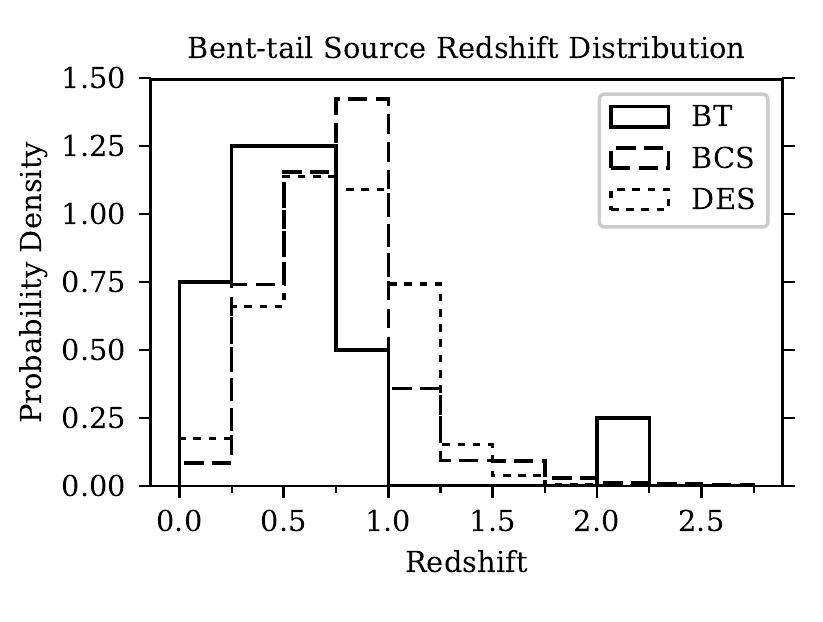}
  \caption{The redshift distribution of the \NOPTICAL{} bent-tail source candidates with optical counterparts in the ATLAS-SPT survey. The bin values are normalised to form a probability distribution such that the integral of each histogram is equal to 1. The redshift distributions of the BCS and DES optical catalogues used for cross-matching as described in Section~\ref{sec:redshifts} are also shown for comparison.}
  \label{fig:btg_z_distribution}
\end{figure}

We use the photometric redshifts for all \NOPTICAL{} optically-matched BT  candidate sources from the optical catalogues described in Section~\ref{sec:xid_optical}. The redshift distribution of the BT source candidates are shown in Figure~\ref{fig:btg_z_distribution} along with the redshift distribution of the BCS and DES optical catalogues. We note that the redshift distribution of our BT  candidate sample appears to be skewed toward low redshifts compared to the optical catalogues. We confirmed that this difference was statistically significant using a Kolmogorov--Smirnov test (BCS-BT p-value: 0.014; DES-BT p-value: 0.013). The \NBCSDES{} sources that have photometric redshift estimates from both BCS and DES SV are in agreement within their errors.

To address the lack of suitable optical data coverage across the ATLAS-SPT field, we limit further discussion of BT candidates and their spatial correlation with known clusters to these \NOPTICAL{} BT candidates with optical cross-identifications.

\subsection{Measuring Bent-tail Flux Densities}
\label{sec:sourcefinding}
We used the source-finding software Python Blob Detection and Source Finder \citep[\textsc{PyBDSF}]{Mohan_2015} to measure the total integrated fluxes of the BT sources. A \mbox{$0.25 \times 0.25\degr$} cutout image was made for each BT source and used as the input to \textsc{PyBDSF}. The background noise was estimated using a sliding box of \mbox{$200 \times 200$~pixels} stepped by 50~pixels. The software then searched for peaks of emission $> 5\sigma$ and then flood-filled down to $3\sigma$ around each peak to create an island. The pixels encompassed by the islands were then fit with multiple Gaussian components.

\begin{figure}
  \includegraphics[width=\columnwidth]{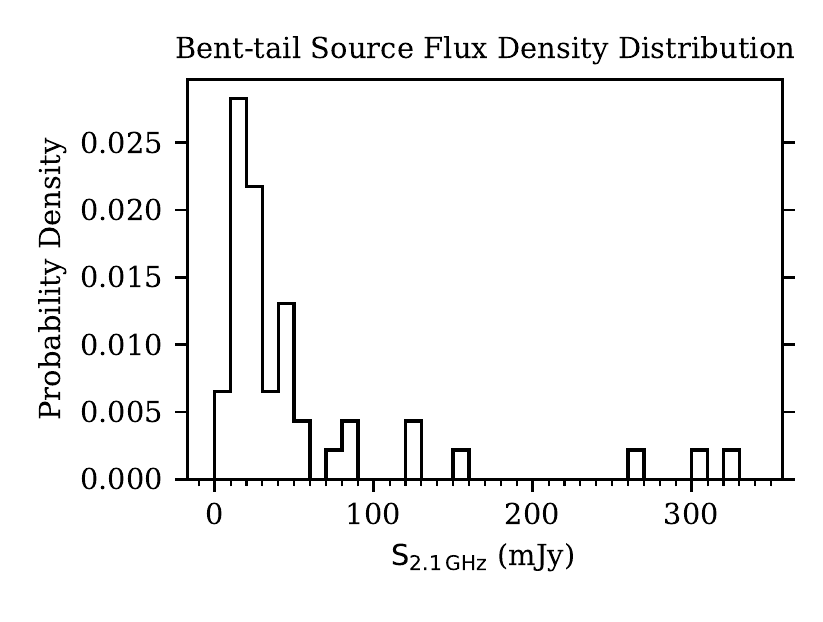}
  \caption{The distribution of the total integrated flux densities for the bent-tail radio sources detected in the ATLAS-SPT survey. The bin widths are 10~mJy and the values are normalised to form a probability distribution such that the integral of the histogram is equal to 1.}
  \label{fig:flux_distribution}
\end{figure}

\textsc{PyBDSF} output a catalogue of Gaussian components and a catalogue of grouped components that the software considers sources. We found that the automated grouping, although configurable with a variety of parameters, did not always group all components from a BT source into a single source. We therefore grouped the components manually and summed the catalogued total flux densities for each component fit to the BT to determine the total integrated flux density. The total integrated flux densities for all BT sources are reported in Table~\ref{tab:btg_sample} (column 9) and the distribution of these flux densities is shown in Figure~\ref{fig:flux_distribution}.

\subsection{Positional Accuracy}
\label{sec:positional-accuracy}

We estimated the positional accuracy of the radio data by inserting 10,000 simulated point-sources of varying flux density distributed across the field, avoiding existing real sources. The source-finding as described in Section~\ref{sec:sourcefinding} was repeated and the true simulated source positions were compared with the extracted source positions. Sources with input flux densities $>1.3$~mJy have positional uncertainties of $\leq 0.5$~arcsec. A more detailed description of this process is given in the ATLAS-SPT survey catalogue paper \citep{OBrien_2018_catalogue}.

\subsection{Matching Bent-tail Sample with Known Clusters}
\label{sec:matching_clusters}

\begin{figure}
  \includegraphics[width=\columnwidth]{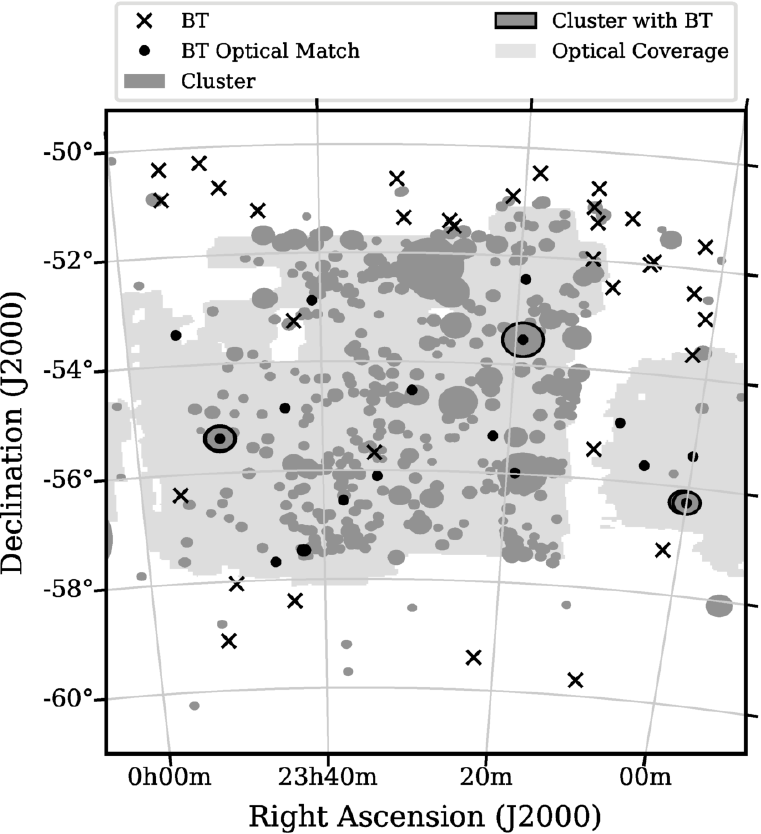}
  \caption{Sky plot of all bent-tail sources (crosses or dots) and clusters (shaded circles) used in this study. Bent-tails with an optical counterpart are marked with a dot. Clusters that were matched with a bent-tail are shown with a solid black outline. The cluster markers are drawn with a 2~Mpc radius. The union of the BCS and DES optical coverage from which photometric redshifts were used is shown in the background as light grey.}
  \label{fig:cluster_matches}
\end{figure}

We matched our BT  candidate sample with cluster catalogues that cover the ATLAS-SPT field from the literature. These clusters have been detected using various methods including optical, IR, X-ray and SZ observations. The cluster catalogues used are listed in Table~\ref{tab:cluster_catalogues}. Using only the \NOPTICAL{} BT  candidate sources with redshifts, we performed a three-dimensional cross-match with the known clusters across the field using a search radius of 2~Mpc within the photometric redshift error. For the cases where there were multiple potential SSDF cross-identifications (as mentioned in Section~\ref{sec:cross-id-infrared}), we cross-matched each potential SSDF source with the optical catalogues. In all of these cases, no alternative cross-identification produced a match within the search area around a cluster. All BTs and clusters used for matching are shown in Figure~\ref{fig:cluster_matches}.

We find matches for 4 BT  candidate sources to 6 known clusters using this technique which are shown in Table~\ref{tab:btg_sample}. Three of these clusters come from X-ray cluster catalogues, two of which provide $\mathrm{M_{500}}$ mass estimates while the other provides a  $\mathrm{M_{200}}$ mass estimate. Two of the clusters are from SZ effect detections (Planck and SPTSZ) which also provide $\mathrm{M_{500}}$ mass estimates. The remaining cluster is from the BCS cluster catalogue which provide optical richness measurements. All cluster matches with mass measurements are of relatively low mass with the most massive being $M_{500} = 3.75 \times 10^{14} \mathrm{M}_{\odot}$.

\begin{figure*}
  \subfloat{\includegraphics[width=0.5\textwidth]{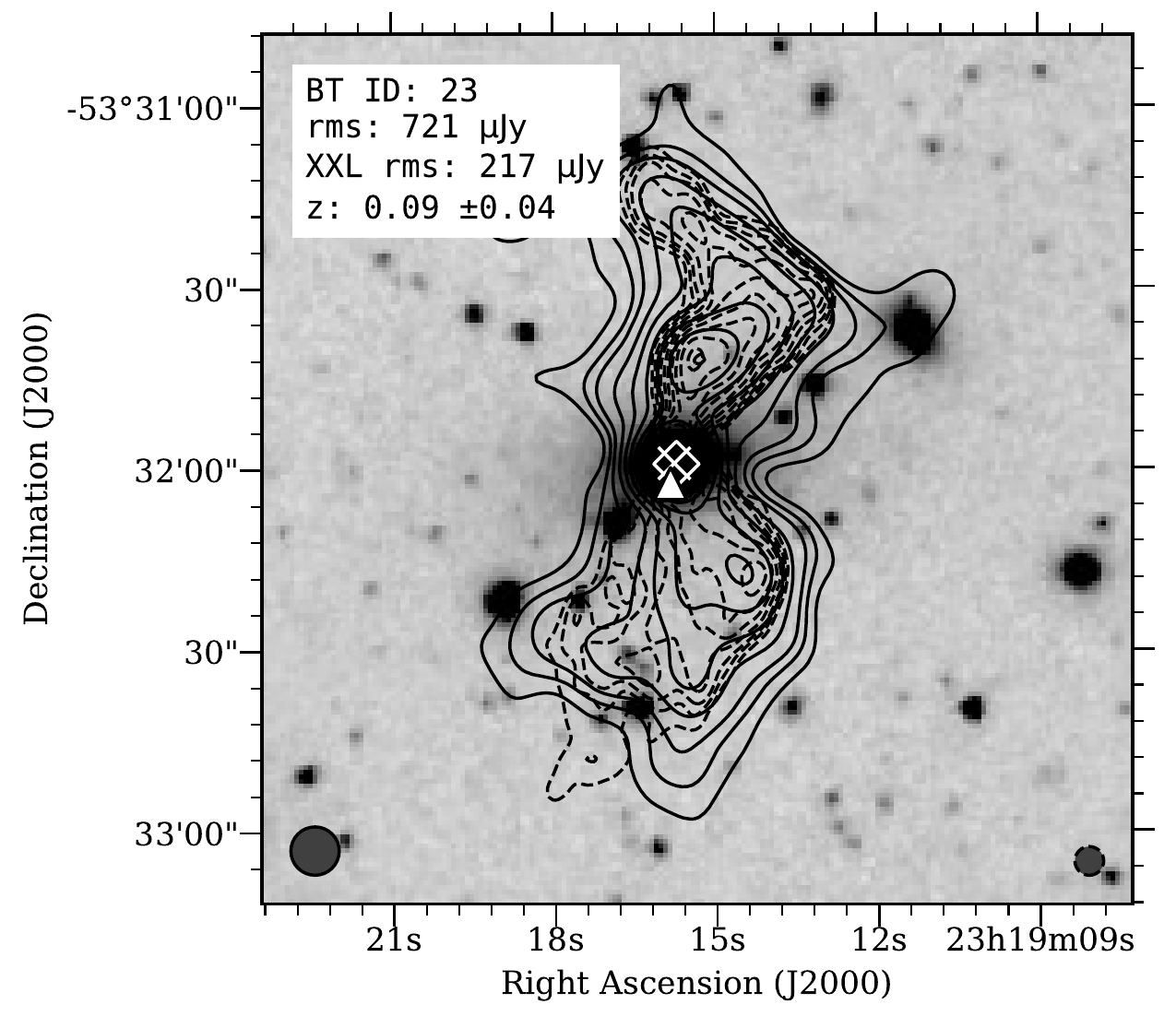}}
  \subfloat{\includegraphics[width=0.5\textwidth]{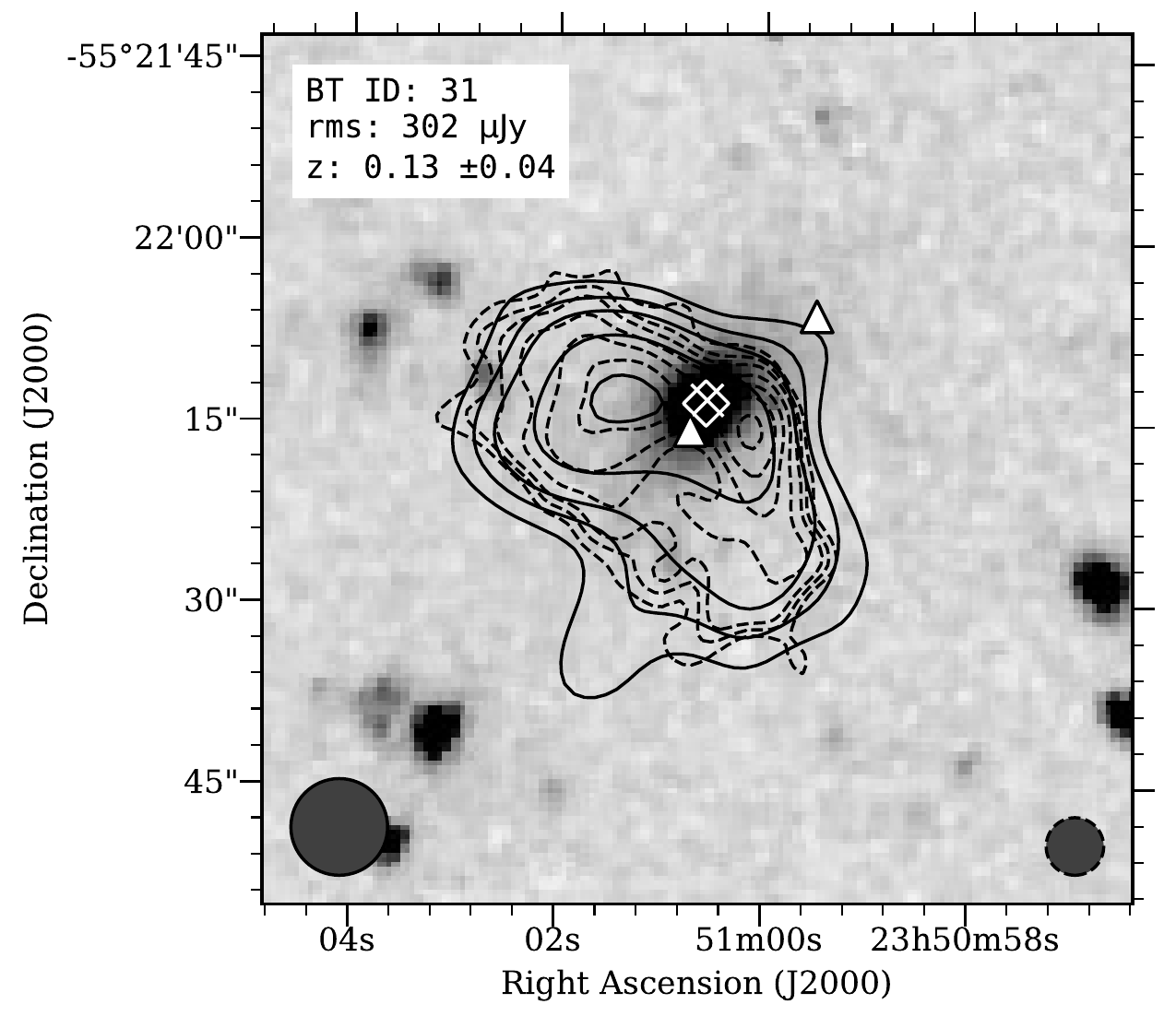}}\\
  \subfloat{\includegraphics[width=0.5\textwidth]{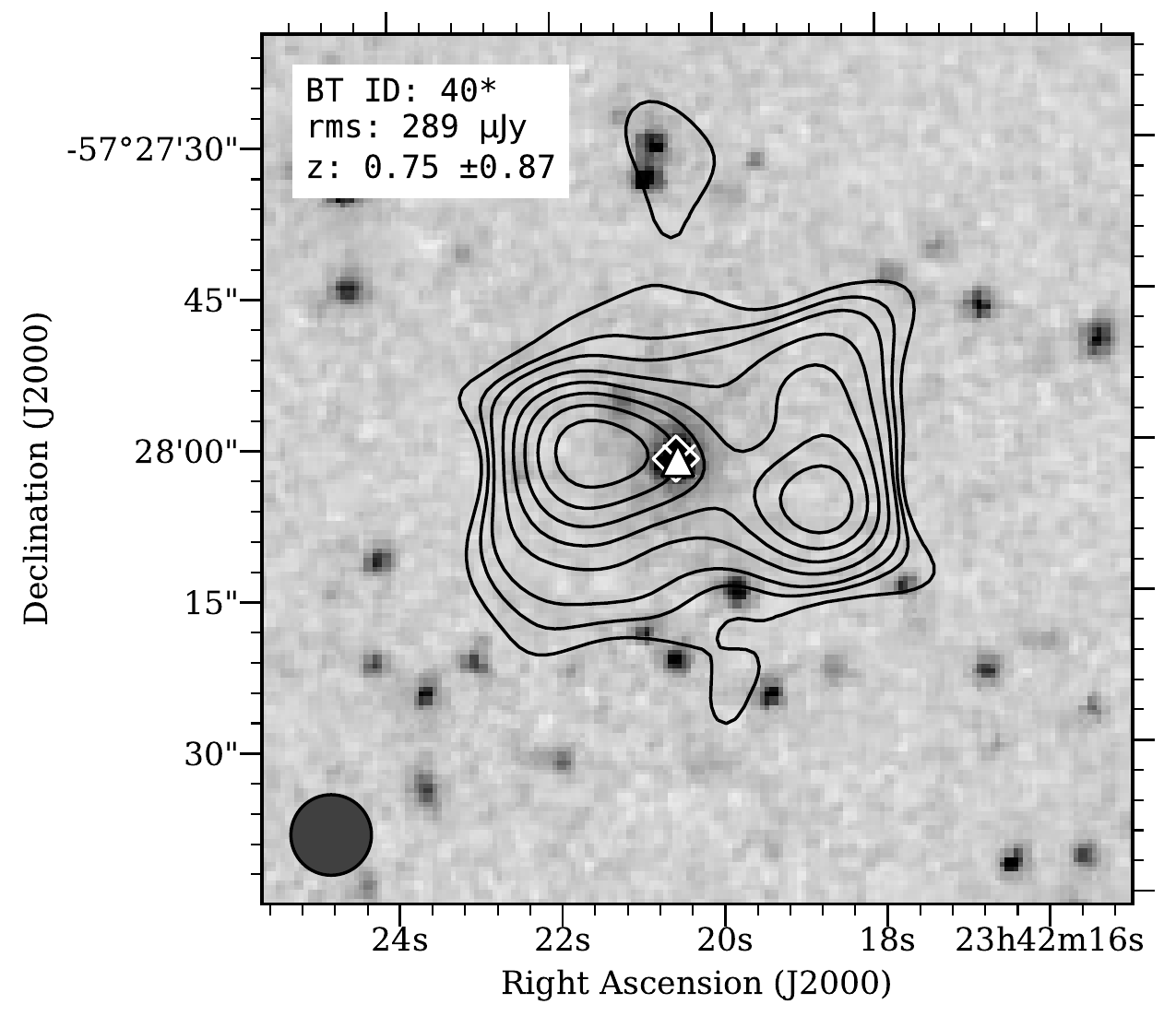}}
  \subfloat{\includegraphics[width=0.5\textwidth]{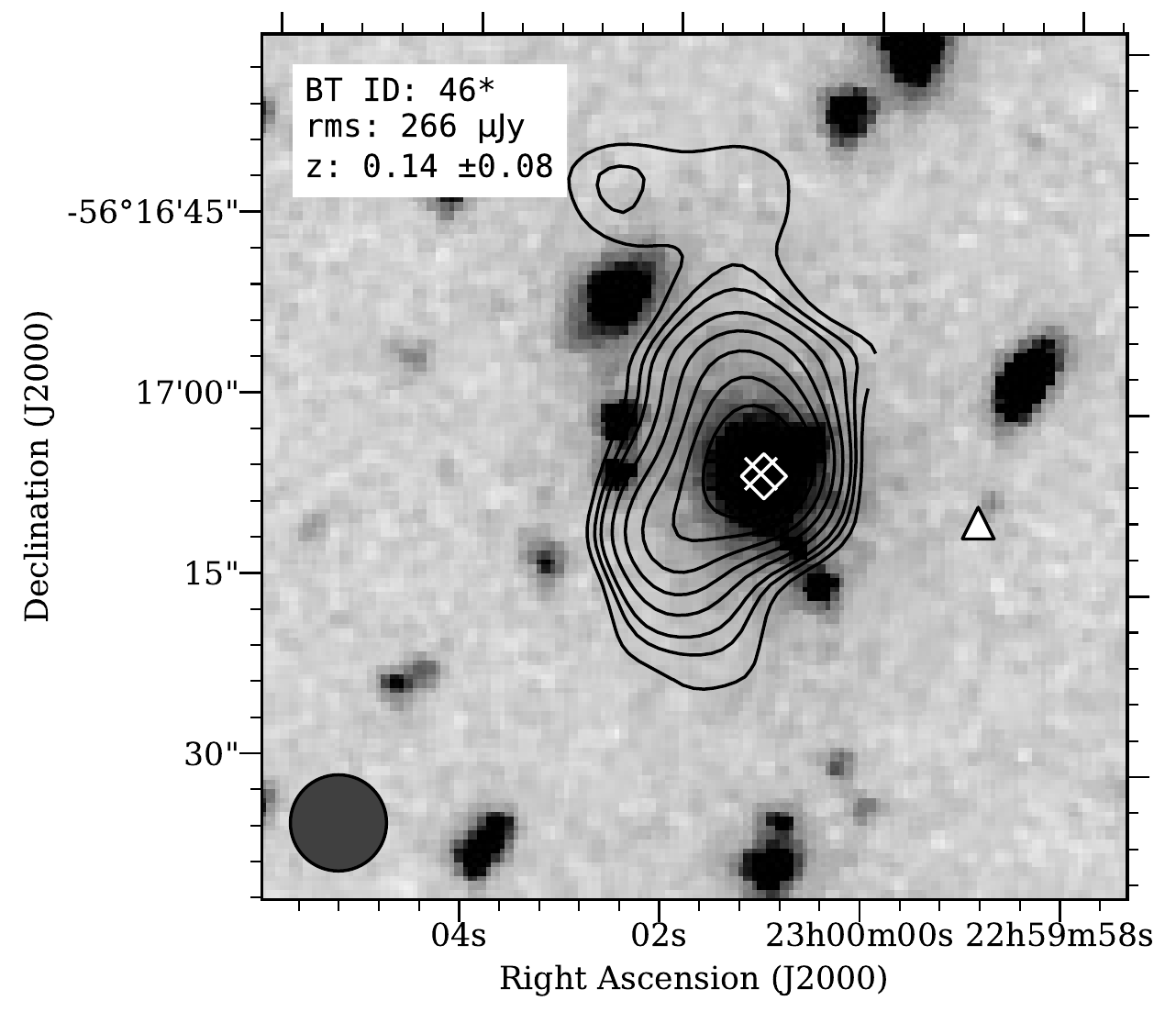}}\\
  \caption{Bent-tail radio candidates identified in this study which reside in known clusters. Radio structure from this work is shown in black contours overlaid on 3.6\micron{} images from the SSDF. When available, higher resolution radio data from the ATCA-XXL survey \citep{butler17} are shown as dashed contours. All contours are drawn at $3, 4, 5\sigma$ and then increase by factors of $\sqrt{2}^{n}$ where $n \in \mathbb{N}$. Radio synthesised beam sizes for the ATLAS-SPT and ATCA-XXL data are shown in the bottom left and right corners, respectively. SSDF sources identified as the likely host galaxy are marked with a diamond. Optical cross-identifications are indicated with an X. Cluster centre positions are marked with a solid white triangle.}
  \label{fig:cluster_btgs}
\end{figure*}

\begin{table*}
  \caption{Cluster catalogues used in this study. The short name is used throughout this paper to refer to the catalogues listed. $N$ is the number of clusters within the ATLAS-SPT field and $N_{z}$ is the number of those clusters with redshifts.}
  \label{tab:cluster_catalogues}
  \begin{tabular}{cp{8cm}lccl}
    \hline
    Short Name & Title & Band & $N$ & $N_{z}$ & Reference\\
    \hline
    XXL 100 GC & The XXL Survey. II. The bright cluster sample: catalogue and luminosity function & X-ray & 52 & 52 & \citet{Pacaud_2016}\\
    MCXC & The MCXC: a meta-catalogue of x-ray detected clusters of galaxies & X-ray & 3 & 3 & \citet{Piffaretti_2011}\\
    Planck & Planck 2015 results. XXVII. The second Planck catalogue of Sunyaev-Zeldovich sources & SZ & 5 & 4 & \citet{Planck_2016}\\
    SPT NIR & Redshifts, Sample Purity, and BCG Positions for the Galaxy Cluster Catalog from the First 720 Square Degrees of the South Pole Telescope Survey & IR & 18 & 10 & \citet{Song_2012}\\
    DES X-ray & Galaxies in X-Ray Selected Clusters and Groups in Dark Energy Survey Data. I. Stellar Mass Growth of Bright Central Galaxies since z~1.2 & X-ray & 28 & 28 & \citet{Zhang_2016}\\
    SCS & Southern Cosmology Survey. II. Massive Optically Selected Clusters from 70 Square Degrees of the Sunyaev-Zel'dovich Effect Common Survey Area & Optical & 44 & 44 & \citet{Menanteau_2010}\\
    BCS & A New Reduction of the Blanco Cosmology Survey: An Optically Selected Galaxy Cluster Catalog and a Public Release of Optical Data Products & Optical & 349 & 349 & \citet{Bleem_2015a}\\
    SPTSZ & Galaxy Clusters Discovered via the Sunyaev-Zel'dovich Effect in the 2500-Square-Degree SPT-SZ Survey & SZ & 27 & 26 & \citet{Bleem_2014}\\
    XMM & The cosmological analysis of X-ray cluster surveys - II. Application of the CR-HR method to the XMM archive & X-ray & 12 & 2 & \citet{Clerc_2012}\\
    Abell & A Catalog of Rich Clusters of Galaxies & Optical & 15 & 0 & \citet{Abell_1989}\\
    2MASS & Groups of Galaxies in the Two Micron All Sky Redshift Survey & Optical, IR & 1 & 1 & \citet{Crook_2007}\\
    FoF & Friends-of-friends galaxy group finder with membership refinement. Application to the local Universe & Optical & 4 & 4 & \citet{Tempel_2016}\\
    Local Groups & The matter distribution in the local Universe as derived from galaxy groups in SDSS DR12 and 2MRS & Optical, IR & 23 & 23 & \citet{Saulder_2016}\\
    \hline
    \hline
    Total & & & 581 & 546 & \\
    \hline
  \end{tabular}
\end{table*}

We find that the majority (12 of \NOPTICAL{} total), of our BT sample have no association with known clusters or groups. Those 4 that are associated with known clusters are found in relatively low-mass clusters (of the order of $10^{14} \mathrm{M}_{\odot}$). We also find that most known high-mass clusters ($> 10^{15} \mathrm{M}_{\odot}$) do not appear to contain a bent-tailed radio galaxy.

We also find that the distances between the BTs found in known clusters and the centre of their cluster hosts are small with most being tens of kpc with the exception of BT 46 which is approximately 0.7~Mpc from its host cluster centre. However, this particular cluster match is from the Planck catalogue which has large positional errors because of the large beam size. A more accurate cluster position is provided in the SPTSZ catalogue which when used to calculate the distance between the cluster and BT gives approximately 66~kpc.

\subsection{Final Classification}

We examined each of our \NBTG{} candidates to identify  those  which are unambiguous BT sources, including using data from the ATCA-XXL survey \citep{butler17}, using the following criteria: (a) the radio morphology is unambiguously a bent-tail or a head-tail galaxy, and cannot easily be explained as a superposition of confusing sources, (b)   there is a clear infrared counterpart for the radio core and no other probable cross-identifications are obvious. \NBTGCONF{} candidates were thus identified as unambiguous BT sources, of which \NOPTICALREAL{} have redshifts. Only one of these is associated with a known cluster.

We include the remaining \NBTGCAND{} sources as BT candidates requiring follow-up observation to confirm their classification (see Section~\ref{sec:future-work}). The whole sample is listed in Table~\ref{tab:btg_sample}, with candidate BTs indicated with an asterisk. A brief justification of each source is given below.

\subsection{Notes on Individual Sources}
\label{sec:bt_source_sample}

BT~1 displays the typical brightness asymmetry of a head-tail galaxy, with the most likely morphology being both radio jets originating from the radio peak and being severely bent toward the south. The closest SSDF source to the radio peak is located 4.4~arcsec to the SE. Follow up observation at higher resolution to resolve the individual jets would confirm our suggested head-tail candidate classification.

BT~2 also displays morphology typical of a head-tail with a wider jet opening angle. The nearest SSDF source to the radio peak is 3.1~arcsec to the SW. We note there is an unrelated radio source to the west with a clear seperate SSDF counterpart. We therefore classify this object as a head-tail.

BT~3 is a clear BT radio source with a wide jet angle. The selected SSDF counterpart is directly aligned with the radio core. We note that the jet to the SW of the core also contains an SSDF source and may contain unrelated, superimposed radio emission, but we do not believe that all of this radio emission to be unrelated. If it were, the source would be a simple radio-double with a core located between the other two radio peaks and there is no evidence of an SSDF counterpart in this region. We also note an SSDF source near the NW radio peak, offset by 4.1~arcsec which may also contribute superimposed radio emission. We suggest this is also unlikely to be dominant because of the positional offset and the sub-arcsecond positional accuracy of the radio data (see Section~\ref{sec:positional-accuracy}.) We therefore classify this object as a bent-tail.

BT~4 is also a clear wide-angle BT radio source. The SSDF counterpart is aligned with the  radio core with no other obvious alternative.

BT~5 is an unambiguous wide-angle BT radio source, for the same reasons as BT~4.

BT~6 is a radio double where the core is undetected. The selected SSDF counterpart is the only clear choice with a trail of radio emission directed toward the counterpart from both lobes. The lobes are mostly aligned, but we note a misalignment in the emission from the eastern lobe which extends toward the north of the core which is possibly caused by ICM movement. Follow up observations at lower frequency may detect more of the lobe emission to help confirm the classification of this BT candidate.

BT~7 is a large, complex radio source with several possible SSDF counterparts. The component to the SE is likely an unrelated source as it is mostly compact and aligned with an SSDF source. We therefore consider the remaining two components form a bent-tail candidate with a wide opening angle and a possible SSDF counterpart between the two radio peaks.

BT~8 is a very complex radio source containing three distinct emission peaks (we note two unrelated radio sources: one to the far west and another to the SW). It is unclear which of the three peaks is the core. We selected the SSDF source nearest to the northern radio peak, but an alternative cross-identification exists within the western-most peak. In either case, an incorrect SSDF cross-identification does not affect our cluster cross-matching in Section~\ref{sec:matching_clusters} as we only consider sources with photometric redshifts, and this source is outside the boundary of available optical data. Follow up observations at higher resolution would help to more accurately determine the location of the radio core for this candidate.

BT~9 is an unambiguous wide-angle BT radio source, for the same reasons as BT~4.

BT~10 is a candidate bent-tail aligned well with an SSDF counterpart. The source is mostly straight with indications of slight bending toward the west for both jets. Low frequency follow up observations may detect more of the lobe emission to confirm the classification.

BT~11 is an atypical bent-tail candidate which shows signs of an abrupt kink in the northern jet. There are several possible SSDF counterparts; we selected the one most central to the radio source. However, it is possible that the host galaxy is one of the SSDF sources at the northern- or southern-most end of the radio emission which would make this source a head-tail. It is also possible that there is superimposed radio emission from other sources making this source appear as an apparent bent-tail. High resolution follow up is required to resolve the jets to accurately classify this source.

BT~12 is an unambiguous wide-angle BT radio source, for the same reasons as BT~4.

BT~13 is another clear bent-tail source. While the radio core is not clearly detected, resolved emission from the western lobe appears to connect the position of an SSDF source which we select as the host galaxy.

BT~14 is a candidate bent-tail with a wide opening angle. The selected SSDF counterpart is offset from the radio peak by 4.8~arcsec. As the cross-identification is not well aligned with the radio core, we list this source as a candidate BT pending follow up observations at higher resolution.

BT~15 is also a candidate bent-tail with a wide opening angle, but the radio core is not easily identified. There are several possible SSDF cross-identifications. Follow up observations at high resolution is required to confirm the classification of this source.

BT~16 is another clear bent-tail with interesting complex morphology. There are several possible SSDF counterparts of which we select the most central one. Higher resolution observations are required to clearly identify the radio core and therefore the true SSDF counterpart. We note that there may be a superimposed radio sources given the large number of SSDF sources within the radio emission contours. However, given the complex morphology of the radio emission, any of the possible counterparts result in this object being identified as a bent-tail, or head-tail.

BT~17 is a bent-tail candidate with a wide opening angle and three distinct radio peaks. The central component contains a possible SSDF counterpart offset from the peak by 8.7~arcsec. The southern component is isolated from the other two, but the lack of an aligned SSDF source indicates that it may not be an unrelated source. We note that the faint radio emission around this source is most likely artefacts from incomplete deconvolution as they match the shape of the dirty beam. Due to the unclear SSDF counterpart, we cannot be certain with the classification of this source as a bent-tail and therefore include it as a candidate for follow up observation at high resolution.

BT~18 is an unambiguous wide-angle BT radio source, for the same reasons as BT~4.

BT~19: see BT~4, but note the complex, extended structure and crowded SSDF field, increasing the possibility of superimposed emission.

BT~20 is a clear bent-tail source, but there are two clear possible SSDF counterparts: the selected SSDF source, and another the north of the selected source. In either case, the radio morphology clearly depicts jet bending, therefore we classify this source as a bent-tail.

BT~21 is a candidate head-tail galaxy. There are two adjacent possible SSDF counterparts; we select the one that aligns directly with the radio peak. Higher resolution follow up observations to resolve the individual jets is required to confirm this classification.

BT~22 is a bent-tail source with a wide opening angle and a clear SSDF counterpart located at the radio peak. We note the interesting curvature of the western jet, possibly indicating strong cluster winds. There is some low-level radio emission to the south of the western jet which we believe to be an image artefact upon closer inspection of the radio image.

BT~23 is an extremely complex radio source exhibiting signs of a residing in a turbulent environment. While the average direction of the jets appears straight, the higher-resolution data from the ATCA-XXL survey reveal dramatic kinks in both jets. We therefore classify this source as a bent-tail.

BT~24 is a candidate bent-tail with a faint radio core aligned with an SSDF counterpart. The NE lobe shows signs of bending toward the NW, but the presence of another SSDF source suggests this may be another superimposed radio source. Higher resolution observations are required to disentangle these sources to confirm the classification.

BT~25 is a head-tail source with a clear SSDF counterpart aligned with the radio peak. We therefore classify this object as a head-tail.

BT~26 is a radio double source with significant bending in the SE lobe. The radio core is not detected and there are no clear SSDF counterparts visible within the radio contours. We therefore classify this source as a bent-tail candidate. Follow up high resolution observations at low frequency may resolve more of the lobes to confirm this classification.

BT~27 is a complex radio source showing signs of jet bending. Like BT~23, the average jet direction appears straight, but the high resolution ATCA-XXL data shows a wide-angle kink in the southern jet. We note that the ATCA-XXL data, and the presence of an aligned SSDF source, appears to suggest that the emission to the SE is an unrelated point source.

BT~28 appears to be a head-tail candidate. There appears to be only one clear SSDF counterpart which is located in the northern side of the radio emission. However, a nearby saturating source in the SSDF image lowers the sensitivity in this area making it difficult to identify alternate cross-identifications. The selected counterpart fits our classification of a head-tail candidate, but higher resolution follow up is required to confirm.

BT~29 is a clear bent-tail with a very wide opening angle. The lack of alternate SSDF counterparts suggests that are no superimposed sources.

BT~30 appears to be a clear bent-tail but the eastern lobe sits just above the radio sensitivity limit. More sensitive follow up observations to detect more of the extended structure of this source are required to confirm this classification.

BT~31 is an asymmetrical radio source with a clear SSDF counterpart, suggestive of a bent-tail with a wide opening angle. The higher resolution ATCA-XXL image resolves the jets and depicts the bending more clearly. We therefore classify this source as a bent-tail.

BT~32 appears to be three-component bent-tail with an SSDF counterpart aligned with the central component. However, the presence of an SSDF source in the western lobe casts doubt on whether this component is part of the same radio source. High resolution follow up at low frequency could resolve more of the lobes and provide stronger evidence that the three components form a single bent-tail. We therefore classify this source as a bent-tail candidate.

BT~33 is an asymmetrical radio source with two possible SSDF cross-identifications. We have selected the source nearest to the radio peak with a separation of 6.5~arcsec. If this is a true cross-identification, the source is likely a head-tail with the host galaxy on the NW edge of the radio emission, with the jets severely bent toward the SE and curving toward the NE.

BT~34 is a clear bent-tail with a wide opening angle. There is a clear radio core aligned with an SSDF source with no other viable SSDF cross-identification is present within the radio contours. The higher resolution ATCA-XXL data reveals some interesting structure in the SE lobe as it appears to bend back in the same direction as the opposite lobe. This suggests that relative ICM movement may be pushing these lobes to the west. We therefore classify this source as a bent-tail.

BT~35 is an extended, single component radio source with two distinct peaks. There are two possible classification for this source: it may be a head-tail with both jets being severely bent toward the north; or it may be a bent-tail with a wide opening angle, with one jet descending to the SW and the other extending to the NE and bending toward the north. Both classifications may be justified with a SSDF counterpart. We select the most central SSDF source and mark this source as a bent-tail candidate requiring follow up observation to confirm.

BT~36 is a faint bent-tail with a wide opening angle. There is an SSDF counterpart aligned with the radio peak with no other obvious alternative cross-identification.

BT~37 is a complex but clear bent-tail with a wide opening angle. The western lobe appears more collimated and longer than its counterpart on the eastern side which appears more extended and frustrated. This may be explained by a relative ICM movement pushing the jets to the SW. There are two possible adjacent SSDF cross-identifications; we have selected the one aligned with the radio peak.

BT~38 displays an asymmetrical radio structure, suggestive of a head-tail. There is a clear SSDF counterpart aligned with the radio peak with a majority of the radio emission extending out to the south and bending toward the west. Close inspection of the faint emission to the NE and SW suggests it is a residual of incomplete deconvolution. Higher resolution follow up observations are required to confirm this classification.

BT~39 is a head-tail candidate with a clear SSDF counterpart within the radio peak. We suggest that both jets are tightly bent toward the SW before bending to the south. We therefore classify this object as a head-tail.

BT~40 is a mostly straight radio double with a clear SSDF counterpart located between the radio peaks. We classify this source as a bent-tail candidate as the western lobe shows signs of bending toward the north. High resolution follow up observation at lower frequency may resolve more of the lobe structure to confirm this classification.

BT~41 is a clear bent-tail with interesting structure. While no radio core is detected, a single SSDF source is located between the two radio components which we select as the probable host galaxy. The northern lobe shows signs of a bending trajectory originating from the core, extending to the NE and bending around toward the NW. Similarly, the southern lobe appears to originate from the core, extend to the south, and bend toward the SE. We suggest this structure is caused by complex environmental effects.

BT~42 appears to be a bent-tail with a clear SSDF counterpart between the two radio peaks. The SW lobe appears to bend toward the NW. However, the presence of an SSDF source within this lobe indicates this may be superimposed radio emission from a separate source. We therefore classify this source as a bent-tail candidate as higher resolution follow up observations are required to disentangle the radio emission.

BT~43 is a clear bent-tail source with a wide opening angle. An SSDF counterpart is aligned with the radio peak which we select as the probably host galaxy. There are other SSDF sources within the radio contours but they are offset from the jet emission. We therefore classify this source as a bent-tail.

BT~44 is an asymmetrical radio source with an SSDF counterpart located 4.3~arcsec from the radio peak. We suggest two possible classifications for this source: a head-tail where the jets are bent tightly back from the radio peak toward the SW before bending toward the west; or a bent-tail with a wide opening angle with one lobe extending to the NE and the other to the west. The offset SSDF counterpart from the radio peak casts some doubt on our head-tail classification, and the lack of a clear radio core detection makes us unable to confirm this source is a bent-tail. In either case, the radio morphology is most likely affected by environmental effects and thus we include this source as a bent-tail candidate.

BT~45 is an extended radio source with complex, asymmetrical structure. There appears to be no obvious SSDF counterparts within the radio peaks to the east and west, and thus we select the SSDF source near the centre as the putative host galaxy. The complex structure in this source and the crowded SSDF field makes it difficult to classify this source. We therefore include this source as a candidate for follow up observation.

BT~46 is an extended, single component radio source where the peak is offset toward the west. We suggest this is most likely a bent-tail with a wide opening angle, with jets bent toward the north-NE and SE. There is a clear SSDF counterpart aligned with the radio peak. Low frequency observations at high resolution may reveal more of the lobe structure for this source to confirm this classification.

\section{Discussion}

Most previous work on BT sources (see Section 1) assumes that  BT sources are associated with massive clusters, and this is confirmed by simulations such as those of \citet{Mguda_2014}. However, the number of BT sources we have detected within clusters (4 of \NOPTICAL{} candidates, and 2 out of \NOPTICALREAL{} unambiguous sources) is contrary to these expectations, and specifically contrary to simulations by \citet{Mguda_2014} which predict that clusters with masses in excess of $10^{15} \mathrm{M}_{\odot}$ should contain approximately 7 BT galaxies (subject to AGN duty-cycle and projection effects), and that few BT galaxies should be found outside massive clusters.

However, in those cases where a BT source is associated with a cluster, the BT cluster-centric distances are in agreement with the simulations which predict that BT galaxies may be found up to $\sim 400$~kpc from the centre of clusters with \mbox{$13.5 \le \log \mathrm{M}_{\odot} \le 14.0$}.

It is clear that our results do not agree with simulations predicting the number of BT sources within clusters. While our sample size of \NOPTICAL{} BT sources is small and greater optical data coverage across the whole ATLAS-SPT field may significantly impact our results, a similar result has recently been obtained by \citet{Garon_2018} on a completely different data set. Two main questions arising from these results are:

\begin{enumerate}
\item Why don't we see BT sources in known clusters?
\item Why don't most BT sources reside in clusters?
\end{enumerate}

We discuss possible explanations for these questions in the subsections below.

\subsection{Missing Bent-tail Sources}

BT sources are expected to reside in galaxy clusters. Our results suggest that the completeness of BT sources as a tracer of clusters may be very low. Below we discuss putative effects that may result in BT sources being excluded from our sample.

Projection effects may have an impact on the classification of BT candidates. While a symmetrical source will remain symmetrical when reprojected, a bent source when reprojected may appear straight. As our sample of BT candidates include any radio source that appears bent, we expect only a small fraction of BT jets appear straight due to their orientation on the sky. Only BTs which have jets aligned within a few degrees of the line-of-sight would appear straight and therefore be excluded from our sample. Therefore, we do not think this will have a significant effect on our sample. We also note that projection effects significantly impact measurements such as jet opening angles (see figure 15 in \cite{Pratley_2013} for an excellent example) which have previously been used to classify BT AGN as either wide- or narrow-angle tailed (WAT and NAT respectively). We avoid use of the WAT/NAT nomenclature for this reason.

Resolution effects may also be a significant contributor to the apparent lack of BT sources in clusters. The \citet{Mguda_2014} simulations look for BT sources with jet sizes greater than 50~kpc. The resolution of our BT sample is approximately 8~arcsec, which is sufficient to detect 50~kpc jets up to $z \sim 0.5$. This jet size is a minimum limit so our data are sensitive to jet sizes larger than this over a wider range of redshifts. We noted in Section~\ref{sec:redshifts} that the redshift distribution of our BT sample is biased to lower redshifts compared to the optical catalogues. This could be due to the resolution of the radio data limiting the number of BT detections at higher redshifts.

The number of observed BT sources at a given time is heavily dependent on the duty cycle of the AGN producing the bent jets. \citet{Mguda_2014} predict that clusters above $10^{14.5} \mathrm{M}_{\odot}$ will host at least one BT at some point in their lifetime. It is possible that many of the clusters in this study \textit{will} host or \textit{have} hosted a BT source at some point that is no longer visible in our dataset.

All of these effects (projection, resolution and duty cycle) are difficult to model. We cannot eliminate the possibility that clusters investigated in this study contain BT sources that are too faint or unresolved to be detected in our survey.

\subsection{Missing Clusters}
If BT radio galaxies are tracers of clusters, then one would expect to find most BT galaxies to be associated with clusters. Our results indicate this is not the case and we discuss putative explanations for this below.

A strong signal within the mm-wavelength bands observed by the South Pole Telescope can contaminate the SZ signal used to detect galaxy clusters leading to the possibility that the SPTSZ cluster catalogue may not be able to detect clusters near strong radio sources. The SPTSZ cluster finder places a 4~arcmin mask around point sources with $S_{\mathrm{150 GHz}} > 5 \sigma$, discards any clusters found within 8~arcmin around these masks and records them in a separate catalogue. We cross-matched the SPT point source catalogue \citep{Mocanu_2013} with our BT sample and find that only 3 of our \NBTG{} sources are potentially affected by masked point sources over the defined threshold. Therefore, we do not think this have a significant effect on our sample.

\begin{figure}
  \includegraphics[width=\columnwidth]{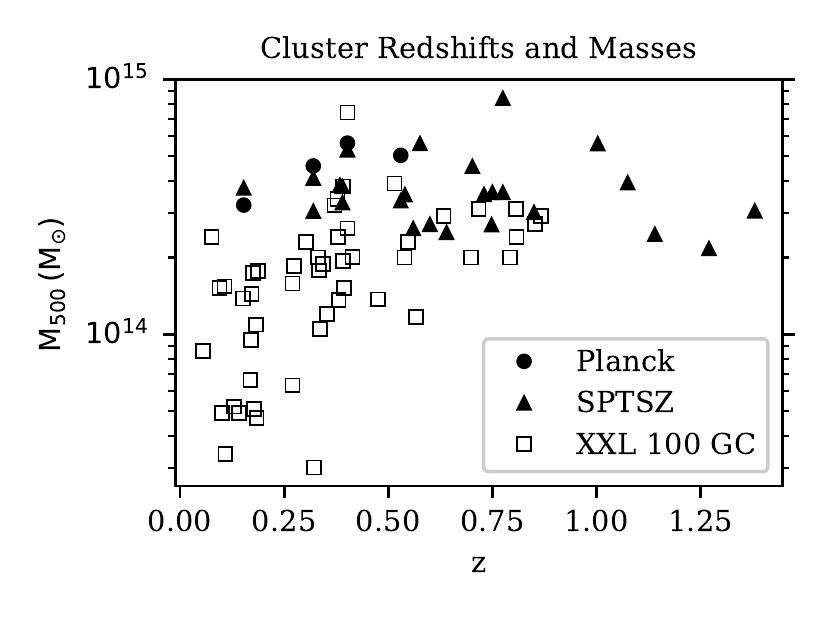}
  \caption{Comparison of redshifts and masses of clusters used in this study. Only the catalogues which provided $M_{500}$ masses are shown.}
  \label{fig:cluster_redshift_mass}
\end{figure}

An alternative explanation is that with the exception of the SPTSZ cluster catalogue, all cluster catalogues used in this study are limited to low redshifts because of sensitivity. The Planck catalogue consists of clusters detected with the mostly redshift independent SZ effect, but the large beam dilutes the SZ signal which constrains cluster detections to low redshift. This means that the only available high-redshift cluster catalogue for this area is SPTSZ which itself is mass limited and expected to be nearly entirely complete for clusters with $M_{500} > 7 \times 10^{14} h^{-1} \mathrm{M}_{\odot}$ at $z > 0.25$. This relatively high mass limit means that the available catalogues of clusters for the ATLAS-SPT field are not sensitive to low-mass clusters at $z > 0.25$. This can clearly be seen in the redshift-mass distribution of the clusters used in this study in Figure~\ref{fig:cluster_redshift_mass}.

\citet[Figure 5]{Mguda_2014} shows that $\sim 40$~per~cent of BT galaxies are in clusters with $M > 10^{14.5} h^{-1} \mathrm{M}_{\odot}$. Applying this fraction to our \NOPTICAL{} BTs with redshifts, we expect that $\sim 6$ BT galaxies will be associated with SPTSZ clusters. Instead we find only one (BT ID 46). We therefore conclude that BT galaxies are found in lower mass host clusters than predicted by the \citet{Mguda_2014} model. This is supported by evidence presented in \cite{Garon_2018} that suggests BT galaxies are poor generic cluster tracers, as they find more bent sources in the far outer regions of clusters than near the cluster centre. However, \cite{Garon_2018} further note that bent radio sources are statistically associated with local overdensities of optical galaxies. This suggests that the medium around even small galaxy overdensities is sufficient to produce radio jet bending.

It is also possible that we do not find many clusters around our BT objects because they have been misclassified i.e. our candidate BT galaxies are not really BT galaxies. Our method for finding BT candidates as explained in Section~\ref{sec:bt_source_sample} is subjective and some objects show only minor signs of jet bending or HT asymmetry.
However,  our smaller sample of unambiguous BT sources also shows this  effect. Even if some of these were misclassified, the mismatch between the predicted number (6) and the actual number (2) of associations in SPTSZ is too large to be explained by some of our classifications being incorrect.

We further note that several putative BT objects identified by \citet{Dehghan_2014} similarly showed only minor signs of jet bending, but when imaged using higher-resolution (2~arcsec) data obtained with the Very Large Array (VLA), most were confirmed to be BTs. 

\section{Conclusions}
\label{sec:conclusions}
\label{sec:future-work}

Bent-tail galaxies are expected to be associated with high-mass clusters. We present a new catalogue of \NBTG{} BT candidate sources, \NBTGCONF{} of which we classify as unambiguous BT sources, detected in the ATLAS-SPT field and compare the \NOPTICAL{} BT galaxies with photometric redshifts with the 546 clusters detected using other techniques. We find that only 4 of our \NOPTICAL{} candidate BT sources, and 2 of our \NOPTICALREAL{} unambiguous sources, are associated with known clusters. This raises two questions: (i) why don't we see BT sources in known clusters? (ii) why don't most BT sources reside in clusters? The first of these may be caused by selection effects which means our sample of BT galaxies may be incomplete. The second question is more problematic, and our data suggest that BT galaxies may be found in lower-mass clusters (or lower-density environments) than expected from models.

As mentioned throughout Section~\ref{sec:bt_source_sample}, further observations are required to confirm the classifications of the BT candidates presented in this work. In many cases, the radio data lacks sufficient resolution to exclude the possibility of superimposed radio sources causing an apparent BT morphology. Higher resolution radio data is required to reduce the confusion of the radio emission. This data would also aid in confirming the candidate head-tail classifications if it were able to resolve the individual jets. In other cases, the radio data depicts only minor signs of bending. Lower frequency observations with sufficient resolution may resolve more of the lobe emission for these sources which could then confirm these classifications. We intend to follow up the sources presented in this work using observations from the Australian Square Kilometre Array Pathfinder (ASKAP) at 850 and 1650~MHz (yielding a resolution of $\sim20$ and $\sim10$~arcsec, respectively). These data combined with increased optical data coverage from the full DES survey will provide more redshifts for sources across the entire ATLAS-SPT field, increasing our sample size of \NOPTICAL{} BT sources and candidates. Additional redshift estimates may also be provided with machine-learning techniques on radio data as described in \citet{Luken_2018}.

Additionally, future cluster surveys from the SPT and eROSITA will expand the number of known clusters immensely, especially into the currently unexplored low-mass high-redshift phase space. Matching these clusters with larger samples of BT galaxies from EMU will provide stronger statistical evidence on the efficacy of BTs as tracers of galaxy clusters.

\afterpage{
\clearpage
\begin{landscape}
\begin{table}
  \caption{Bent-tail sources identified in the ATLAS-SPT survey. The ID of candidate BTs requiring follow-up observation are marked with an asterisk (*). The RA and Dec coordinates given in columns 2 and 3 are for the 3.6\micron{} counterparts from SSDF. The coordinates given in columns 4 and 5 are for the optical counterparts from either the BCS or the DES catalogues. Columns 6, 7 and 8 are the photometric redshifts for the BT, the $1\sigma$ error for the redshift, and the source of the redshift value, respectively. The total integrated flux densities of the BT sources and their $1\sigma$ errors are given in columns 9 and 10, respectively. If a BT is matched with a cluster, the cluster catalogue and its name are given in columns 11 and 12, respectively. The redshift of the matched cluster is given in column 13 and the mass measured within an aperture enclosing a region 500 times the critical density of the Universe (i.e. $M_{500}$) is given in column 14. Masses marked with an asterisk (*) are $M_{200}$ measurements which by definition must be greater than $M_{500}$. The distance between the BT and the cluster centre is given in column 15.}
  \label{tab:btg_sample}
  \scalebox{0.85}{
  \begin{tabular}{lccccccc.{3.2}.{1.2}|ccl.{2}.{3.3}}
\hline
ID  & RA                                            & Dec                                           & Optical RA                                    & Optical Dec                                   & $z$  & $z_{\sigma}$ & $z_{\mathrm{source}}$ & \multicolumn{1}{c}{$S$}   & \multicolumn{1}{c}{$S_\sigma$} & Cluster Catalogue                      & Cluster ID                 & \multicolumn{1}{c}{$z_{\mathrm{cl}}$} & \multicolumn{1}{c}{$\mathrm{M_{500,cl}}$}                 & \multicolumn{1}{c}{Distance$_{\mathrm{BT,cl}}$} \\
    & \multicolumn{2}{c}{J2000 (deg)}                                                               & \multicolumn{2}{c}{J2000 (deg)}                                                               &      &              &                       & \multicolumn{1}{c}{(mJy)} & \multicolumn{1}{c}{(mJy)}      & See Table \ref{tab:cluster_catalogues} &                            &                                       & \multicolumn{1}{c}{($\times 10^{14} \mathrm{M_{\odot}}$)} & \multicolumn{1}{c}{(kpc)}                       \\
(1) & (2)                                           & (3)                                           & (4)                                           & (5)                                           & (6)  & (7)          & (8)                   & \multicolumn{1}{c}{(9)}   & \multicolumn{1}{c}{(10)}       & (11)                                   & (12)                       & (13)                                  & (14)                                                      & \multicolumn{1}{c}{(15)}                        \\
\hline
1*  & $23^\mathrm{h}18^\mathrm{m}41.596^\mathrm{s}$ & $-50^\circ27{}^\prime16.420{}^{\prime\prime}$ &                                               &                                               &      &              &                       & 48.36                     & 0.54                           &                                        &                            &                                       &                                                           &                                                 \\
2   & $23^\mathrm{h}51^\mathrm{m}21.874^\mathrm{s}$ & $-50^\circ17{}^\prime45.647{}^{\prime\prime}$ &                                               &                                               &      &              &                       & 58.90                     & 0.72                           &                                        &                            &                                       &                                                           &                                                 \\
3   & $23^\mathrm{h}55^\mathrm{m}15.138^\mathrm{s}$ & $-50^\circ23{}^\prime04.326{}^{\prime\prime}$ &                                               &                                               &      &              &                       & 48.50                     & 0.74                           &                                        &                            &                                       &                                                           &                                                 \\
4   & $23^\mathrm{h}32^\mathrm{m}27.128^\mathrm{s}$ & $-50^\circ38{}^\prime06.655{}^{\prime\prime}$ &                                               &                                               &      &              &                       & 17.07                     & 0.32                           &                                        &                            &                                       &                                                           &                                                 \\
5   & $23^\mathrm{h}21^\mathrm{m}10.182^\mathrm{s}$ & $-50^\circ54{}^\prime06.635{}^{\prime\prime}$ &                                               &                                               &      &              &                       & 29.69                     & 0.96                           &                                        &                            &                                       &                                                           &                                                 \\
6*  & $23^\mathrm{h}12^\mathrm{m}56.403^\mathrm{s}$ & $-50^\circ40{}^\prime16.756{}^{\prime\prime}$ &                                               &                                               &      &              &                       & 155.45                    & 0.83                           &                                        &                            &                                       &                                                           &                                                 \\
7*  & $23^\mathrm{h}13^\mathrm{m}14.692^\mathrm{s}$ & $-51^\circ01{}^\prime11.964{}^{\prime\prime}$ &                                               &                                               &      &              &                       & 327.41                    & 2.50                           &                                        &                            &                                       &                                                           &                                                 \\
8*  & $23^\mathrm{h}55^\mathrm{m}13.799^\mathrm{s}$ & $-50^\circ56{}^\prime28.403{}^{\prime\prime}$ &                                               &                                               &      &              &                       & 23.61                     & 0.74                           &                                        &                            &                                       &                                                           &                                                 \\
9   & $23^\mathrm{h}45^\mathrm{m}55.770^\mathrm{s}$ & $-51^\circ12{}^\prime14.929{}^{\prime\prime}$ &                                               &                                               &      &              &                       & 42.89                     & 1.46                           &                                        &                            &                                       &                                                           &                                                 \\
10* & $23^\mathrm{h}31^\mathrm{m}42.520^\mathrm{s}$ & $-51^\circ20{}^\prime56.897{}^{\prime\prime}$ &                                               &                                               &      &              &                       & 8.56                      & 0.48                           &                                        &                            &                                       &                                                           &                                                 \\
11* & $23^\mathrm{h}27^\mathrm{m}13.602^\mathrm{s}$ & $-51^\circ23{}^\prime34.260{}^{\prime\prime}$ &                                               &                                               &      &              &                       & 49.20                     & 0.55                           &                                        &                            &                                       &                                                           &                                                 \\
12  & $23^\mathrm{h}26^\mathrm{m}46.136^\mathrm{s}$ & $-51^\circ29{}^\prime38.684{}^{\prime\prime}$ &                                               &                                               &      &              &                       & 25.21                     & 0.78                           &                                        &                            &                                       &                                                           &                                                 \\
13  & $23^\mathrm{h}12^\mathrm{m}45.966^\mathrm{s}$ & $-51^\circ18{}^\prime12.938{}^{\prime\prime}$ &                                               &                                               &      &              &                       & 19.57                     & 0.62                           &                                        &                            &                                       &                                                           &                                                 \\
14* & $23^\mathrm{h}09^\mathrm{m}25.649^\mathrm{s}$ & $-51^\circ10{}^\prime55.477{}^{\prime\prime}$ &                                               &                                               &      &              &                       & 19.24                     & 0.49                           &                                        &                            &                                       &                                                           &                                                 \\
15* & $23^\mathrm{h}01^\mathrm{m}58.751^\mathrm{s}$ & $-51^\circ34{}^\prime41.545{}^{\prime\prime}$ &                                               &                                               &      &              &                       & 10.19                     & 0.73                           &                                        &                            &                                       &                                                           &                                                 \\
16  & $23^\mathrm{h}07^\mathrm{m}09.434^\mathrm{s}$ & $-52^\circ00{}^\prime06.566{}^{\prime\prime}$ &                                               &                                               &      &              &                       & 124.23                    & 1.26                           &                                        &                            &                                       &                                                           &                                                 \\
17* & $23^\mathrm{h}12^\mathrm{m}53.981^\mathrm{s}$ & $-51^\circ58{}^\prime40.645{}^{\prime\prime}$ &                                               &                                               &      &              &                       & 52.62                     & 0.73                           &                                        &                            &                                       &                                                           &                                                 \\
18  & $23^\mathrm{h}19^\mathrm{m}23.479^\mathrm{s}$ & $-52^\circ25{}^\prime17.479{}^{\prime\prime}$ & $23^\mathrm{h}19^\mathrm{m}23.554^\mathrm{s}$ & $-52^\circ25{}^\prime17.292{}^{\prime\prime}$ & 0.42 & 0.10         & BCS                   & 17.88                     & 0.83                           &                                        &                            &                                       &                                                           &                                                 \\
19  & $23^\mathrm{h}10^\mathrm{m}39.750^\mathrm{s}$ & $-52^\circ28{}^\prime24.884{}^{\prime\prime}$ &                                               &                                               &      &              &                       & 33.71                     & 0.61                           &                                        &                            &                                       &                                                           &                                                 \\
20  & $23^\mathrm{h}40^\mathrm{m}51.313^\mathrm{s}$ & $-52^\circ52{}^\prime26.497{}^{\prime\prime}$ & $23^\mathrm{h}40^\mathrm{m}51.247^\mathrm{s}$ & $-52^\circ52{}^\prime26.040{}^{\prime\prime}$ & 0.70 & 0.04         & BCS                   & 12.74                     & 0.95                           &                                        &                            &                                       &                                                           &                                                 \\
21* & $23^\mathrm{h}00^\mathrm{m}56.971^\mathrm{s}$ & $-52^\circ53{}^\prime58.621{}^{\prime\prime}$ &                                               &                                               &      &              &                       & 121.80                    & 1.73                           &                                        &                            &                                       &                                                           &                                                 \\
22  & $23^\mathrm{h}42^\mathrm{m}44.818^\mathrm{s}$ & $-53^\circ15{}^\prime04.651{}^{\prime\prime}$ &                                               &                                               &      &              &                       & 31.39                     & 1.04                           &                                        &                            &                                       &                                                           &                                                 \\
23  & $23^\mathrm{h}19^\mathrm{m}15.730^\mathrm{s}$ & $-53^\circ31{}^\prime59.164{}^{\prime\prime}$ & $23^\mathrm{h}19^\mathrm{m}15.768^\mathrm{s}$ & $-53^\circ31{}^\prime59.052{}^{\prime\prime}$ & 0.09 & 0.04         & BCS                   & 267.82                    & 2.20                           & XXL GC                                 & XLSSC 544                  & 0.095                                 & 1.52                                                      & 7.280                                           \\
24* & $23^\mathrm{h}54^\mathrm{m}50.484^\mathrm{s}$ & $-53^\circ26{}^\prime00.042{}^{\prime\prime}$ & $23^\mathrm{h}54^\mathrm{m}50.464^\mathrm{s}$ & $-53^\circ25{}^\prime59.740{}^{\prime\prime}$ & 0.27 & 0.06         & DES                   & 27.26                     & 0.64                           &                                        &                            &                                       &                                                           &                                                 \\
25  & $23^\mathrm{h}01^\mathrm{m}44.217^\mathrm{s}$ & $-53^\circ34{}^\prime48.864{}^{\prime\prime}$ &                                               &                                               &      &              &                       & 80.98                     & 0.78                           &                                        &                            &                                       &                                                           &                                                 \\
26* & $23^\mathrm{h}30^\mathrm{m}31.488^\mathrm{s}$ & $-54^\circ31{}^\prime28.621{}^{\prime\prime}$ & $23^\mathrm{h}30^\mathrm{m}31.460^\mathrm{s}$ & $-54^\circ31{}^\prime28.888{}^{\prime\prime}$ & 0.68 & 0.09         & DES                   & 20.42                     & 0.87                           &                                        &                            &                                       &                                                           &                                                 \\
27* & $23^\mathrm{h}43^\mathrm{m}55.261^\mathrm{s}$ & $-54^\circ50{}^\prime56.782{}^{\prime\prime}$ & $23^\mathrm{h}43^\mathrm{m}55.253^\mathrm{s}$ & $-54^\circ50{}^\prime56.760{}^{\prime\prime}$ & 0.85 & 0.07         & BCS                   & 16.22                     & 0.57                           &                                        &                            &                                       &                                                           &                                                 \\
28* & $23^\mathrm{h}08^\mathrm{m}20.266^\mathrm{s}$ & $-54^\circ56{}^\prime19.572{}^{\prime\prime}$ & $23^\mathrm{h}08^\mathrm{m}20.232^\mathrm{s}$ & $-54^\circ56{}^\prime18.481{}^{\prime\prime}$ & 0.66 & 0.07         & DES                   & 32.36                     & 0.56                           &                                        &                            &                                       &                                                           &                                                 \\
29  & $23^\mathrm{h}21^\mathrm{m}44.959^\mathrm{s}$ & $-55^\circ19{}^\prime20.737{}^{\prime\prime}$ & $23^\mathrm{h}21^\mathrm{m}44.978^\mathrm{s}$ & $-55^\circ19{}^\prime20.388{}^{\prime\prime}$ & 2.22 & 0.40         & BCS                   & 9.89                      & 0.64                           &                                        &                            &                                       &                                                           &                                                 \\
30* & $23^\mathrm{h}10^\mathrm{m}49.553^\mathrm{s}$ & $-55^\circ27{}^\prime25.290{}^{\prime\prime}$ &                                               &                                               &      &              &                       & 23.28                     & 0.84                           &                                        &                            &                                       &                                                           &                                                 \\
31  & $23^\mathrm{h}51^\mathrm{m}00.568^\mathrm{s}$ & $-55^\circ22{}^\prime13.372{}^{\prime\prime}$ & $23^\mathrm{h}51^\mathrm{m}00.557^\mathrm{s}$ & $-55^\circ22{}^\prime13.116{}^{\prime\prime}$ & 0.13 & 0.04         & BCS                   & 29.68                     & 0.41                           & XXL GC                                 & XLSSC 511                  & 0.130                                 & 0.52                                                      & 8.517                                           \\
    &                                               &                                               &                                               &                                               &      &              &                       &                           &                                & DES X-ray                              & XMMXCS J235059.5--552206.1 & 0.140                                 & < 1.58^{*}                                                & 37.017                                          \\
32* & $23^\mathrm{h}34^\mathrm{m}25.640^\mathrm{s}$ & $-55^\circ40{}^\prime39.907{}^{\prime\prime}$ &                                               &                                               &      &              &                       & 14.88                     & 0.85                           &                                        &                            &                                       &                                                           &                                                 \\
33* & $23^\mathrm{h}00^\mathrm{m}04.836^\mathrm{s}$ & $-55^\circ25{}^\prime36.512{}^{\prime\prime}$ & $23^\mathrm{h}00^\mathrm{m}04.831^\mathrm{s}$ & $-55^\circ25{}^\prime36.415{}^{\prime\prime}$ & 0.29 & 0.06         & DES                   & 13.58                     & 0.71                           &                                        &                            &                                       &                                                           &                                                 \\
34  & $23^\mathrm{h}19^\mathrm{m}07.033^\mathrm{s}$ & $-55^\circ58{}^\prime56.791{}^{\prime\prime}$ & $23^\mathrm{h}19^\mathrm{m}07.032^\mathrm{s}$ & $-55^\circ58{}^\prime56.784{}^{\prime\prime}$ & 0.63 & 0.04         & BCS                   & 18.24                     & 0.85                           &                                        &                            &                                       &                                                           &                                                 \\
35* & $23^\mathrm{h}05^\mathrm{m}10.370^\mathrm{s}$ & $-55^\circ40{}^\prime46.862{}^{\prime\prime}$ & $23^\mathrm{h}05^\mathrm{m}10.366^\mathrm{s}$ & $-55^\circ40{}^\prime47.039{}^{\prime\prime}$ & 0.71 & 0.09         & DES                   & 22.23                     & 0.48                           &                                        &                            &                                       &                                                           &                                                 \\
36  & $23^\mathrm{h}55^\mathrm{m}43.238^\mathrm{s}$ & $-56^\circ22{}^\prime21.731{}^{\prime\prime}$ &                                               &                                               &      &              &                       & 5.34                      & 0.47                           &                                        &                            &                                       &                                                           &                                                 \\
37  & $23^\mathrm{h}37^\mathrm{m}47.680^\mathrm{s}$ & $-56^\circ33{}^\prime18.425{}^{\prime\prime}$ & $23^\mathrm{h}37^\mathrm{m}47.666^\mathrm{s}$ & $-56^\circ33{}^\prime18.360{}^{\prime\prime}$ & 0.35 & 0.04         & BCS                   & 18.36                     & 0.70                           &                                        &                            &                                       &                                                           &                                                 \\
38* & $23^\mathrm{h}01^\mathrm{m}49.421^\mathrm{s}$ & $-57^\circ11{}^\prime16.048{}^{\prime\prime}$ &                                               &                                               &      &              &                       & 26.76                     & 0.52                           &                                        &                            &                                       &                                                           &                                                 \\
39  & $23^\mathrm{h}45^\mathrm{m}33.467^\mathrm{s}$ & $-57^\circ40{}^\prime12.097{}^{\prime\prime}$ & $23^\mathrm{h}45^\mathrm{m}33.426^\mathrm{s}$ & $-57^\circ40{}^\prime12.410{}^{\prime\prime}$ & 0.34 & 0.05         & DES                   & 10.61                     & 0.33                           &                                        &                            &                                       &                                                           &                                                 \\
40* & $23^\mathrm{h}42^\mathrm{m}20.520^\mathrm{s}$ & $-57^\circ28{}^\prime01.391{}^{\prime\prime}$ & $23^\mathrm{h}42^\mathrm{m}20.479^\mathrm{s}$ & $-57^\circ28{}^\prime01.632{}^{\prime\prime}$ & 0.75 & 0.87         & BCS                   & 42.69                     & 0.96                           & BCS                                    & LCS-CL J234220--5728.0     & 0.49                                  &                                                           & 2.087                                           \\
41  & $23^\mathrm{h}50^\mathrm{m}10.199^\mathrm{s}$ & $-58^\circ02{}^\prime49.621{}^{\prime\prime}$ &                                               &                                               &      &              &                       & 87.16                     & 0.82                           &                                        &                            &                                       &                                                           &                                                 \\
42* & $23^\mathrm{h}43^\mathrm{m}33.704^\mathrm{s}$ & $-58^\circ23{}^\prime25.512{}^{\prime\prime}$ &                                               &                                               &      &              &                       & 300.97                    & 2.62                           &                                        &                            &                                       &                                                           &                                                 \\
43  & $23^\mathrm{h}51^\mathrm{m}33.654^\mathrm{s}$ & $-59^\circ05{}^\prime15.259{}^{\prime\prime}$ &                                               &                                               &      &              &                       & 10.43                     & 0.52                           &                                        &                            &                                       &                                                           &                                                 \\
44* & $23^\mathrm{h}22^\mathrm{m}21.118^\mathrm{s}$ & $-59^\circ24{}^\prime07.970{}^{\prime\prime}$ &                                               &                                               &      &              &                       & 79.77                     & 0.95                           &                                        &                            &                                       &                                                           &                                                 \\
45* & $23^\mathrm{h}09^\mathrm{m}53.865^\mathrm{s}$ & $-59^\circ42{}^\prime34.718{}^{\prime\prime}$ &                                               &                                               &      &              &                       & 42.57                     & 0.43                           &                                        &                            &                                       &                                                           &                                                 \\
46* & $23^\mathrm{h}00^\mathrm{m}01.072^\mathrm{s}$ & $-56^\circ17{}^\prime05.845{}^{\prime\prime}$ & $23^\mathrm{h}00^\mathrm{m}01.103^\mathrm{s}$ & $-56^\circ17{}^\prime05.640{}^{\prime\prime}$ & 0.14 & 0.08         & DES                   & 22.07                     & 0.78                           & Planck                                 & PSZ2 G329.53--54.73        & 0.153                                 & 3.21                                                      & 712.796                                         \\
    &                                               &                                               &                                               &                                               &      &              &                       &                           &                                & SPTSZ                                  & SPT-CL J2259--5617         & 0.153                                 & 3.75                                                      & 66.211                                          \\
\hline
\end{tabular}
  }
\end{table}
\end{landscape}
}

\section*{Acknowledgements}

The Australia Telescope Compact Array is part of the Australia Telescope National Facility which is funded by the Australian Government for operation as a National Facility managed by CSIRO.

We acknowledge the Pawsey Supercomputing Centre which is supported by the Western Australian and Australian Governments.

This project used public archival data from the Dark Energy Survey (DES).
Funding for the DES Projects has been provided by 
the U.S. Department of Energy, 
the U.S. National Science Foundation, 
the Ministry of Science and Education of Spain, 
the Science and Technology Facilities Council of the United Kingdom, 
the Higher Education Funding Council for England, 
the National Center for Supercomputing Applications at the University of Illinois at Urbana-Champaign, 
the Kavli Institute of Cosmological Physics at the University of Chicago, 
the Center for Cosmology and Astro-Particle Physics at the Ohio State University, 
the Mitchell Institute for Fundamental Physics and Astronomy at Texas A\&M University, 
Financiadora de Estudos e Projetos, Funda{\c c}{\~a}o Carlos Chagas Filho de Amparo {\`a} Pesquisa do Estado do Rio de Janeiro, 
Conselho Nacional de Desenvolvimento Cient{\'i}fico e Tecnol{\'o}gico and the Minist{\'e}rio da Ci{\^e}ncia, Tecnologia e Inovac{\~a}o, 
the Deutsche Forschungsgemeinschaft, 
and the Collaborating Institutions in the Dark Energy Survey. 
The Collaborating Institutions are 
Argonne National Laboratory, 
the University of California at Santa Cruz, 
the University of Cambridge, 
Centro de Investigaciones En{\'e}rgeticas, Medioambientales y Tecnol{\'o}gicas-Madrid, 
the University of Chicago, 
University College London, 
the DES-Brazil Consortium, 
the University of Edinburgh, 
the Eidgen{\"o}ssische Technische Hoch\-schule (ETH) Z{\"u}rich, 
Fermi National Accelerator Laboratory, 
the University of Illinois at Urbana-Champaign, 
the Institut de Ci{\`e}ncies de l'Espai (IEEC/CSIC), 
the Institut de F{\'i}sica d'Altes Energies, 
Lawrence Berkeley National Laboratory, 
the Ludwig-Maximilians Universit{\"a}t M{\"u}nchen and the associated Excellence Cluster Universe, 
the University of Michigan, 
{the} National Optical Astronomy Observatory, 
the University of Nottingham, 
the Ohio State University, 
the University of Pennsylvania, 
the University of Portsmouth, 
SLAC National Accelerator Laboratory, 
Stanford University, 
the University of Sussex, 
and Texas A\&M University.

We thank the  anonymous referee whose comments helped improve this manuscript. 




\bibliographystyle{mnras}
\bibliography{atlasspt_btgs_mendeley}


\clearpage
\appendix

\section{Cutout Images of Bent-Tail Sources}

\begin{figure*}
\subfloat{\includegraphics[width=0.3\textwidth]{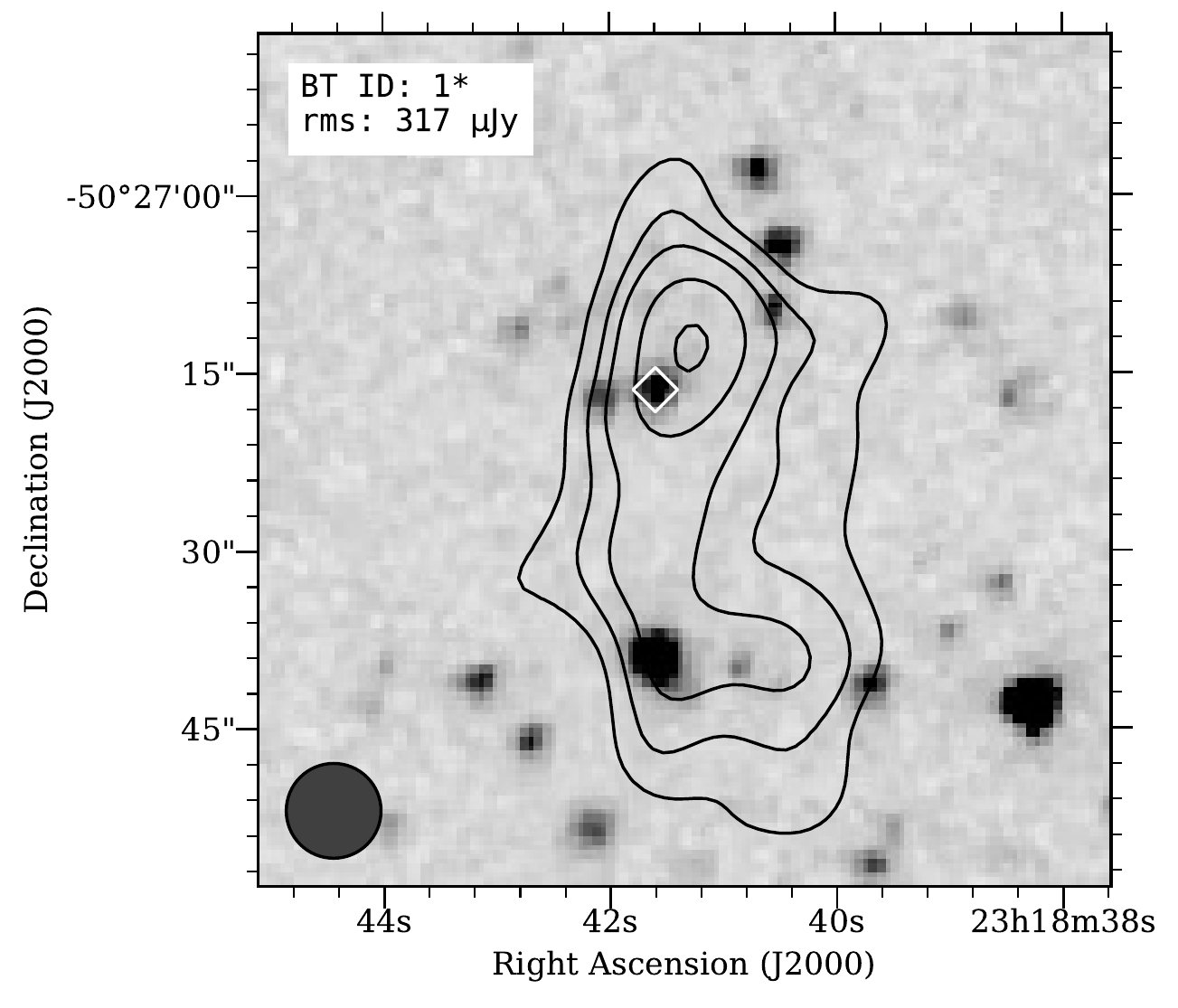}}
\subfloat{\includegraphics[width=0.3\textwidth]{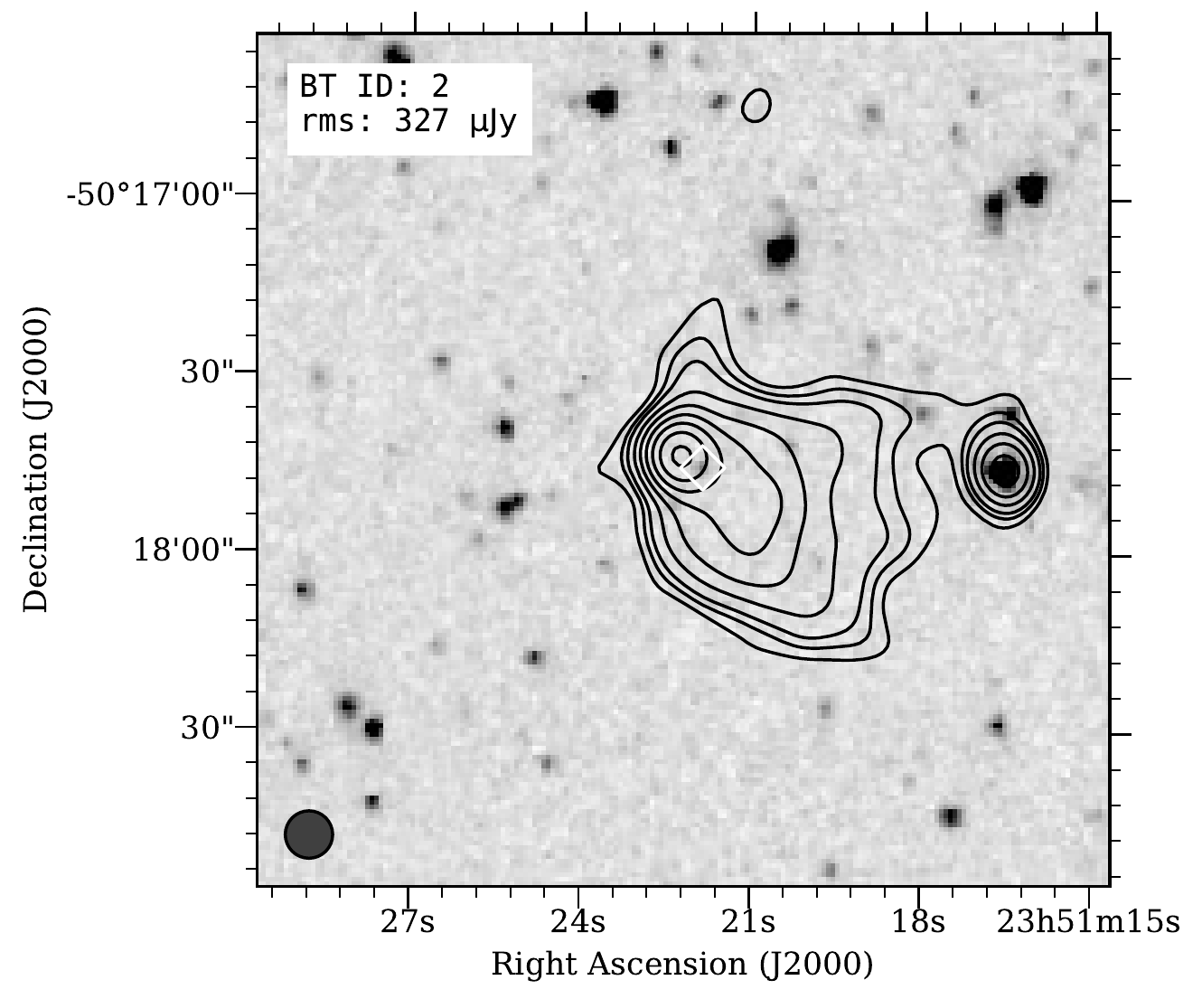}}
\subfloat{\includegraphics[width=0.3\textwidth]{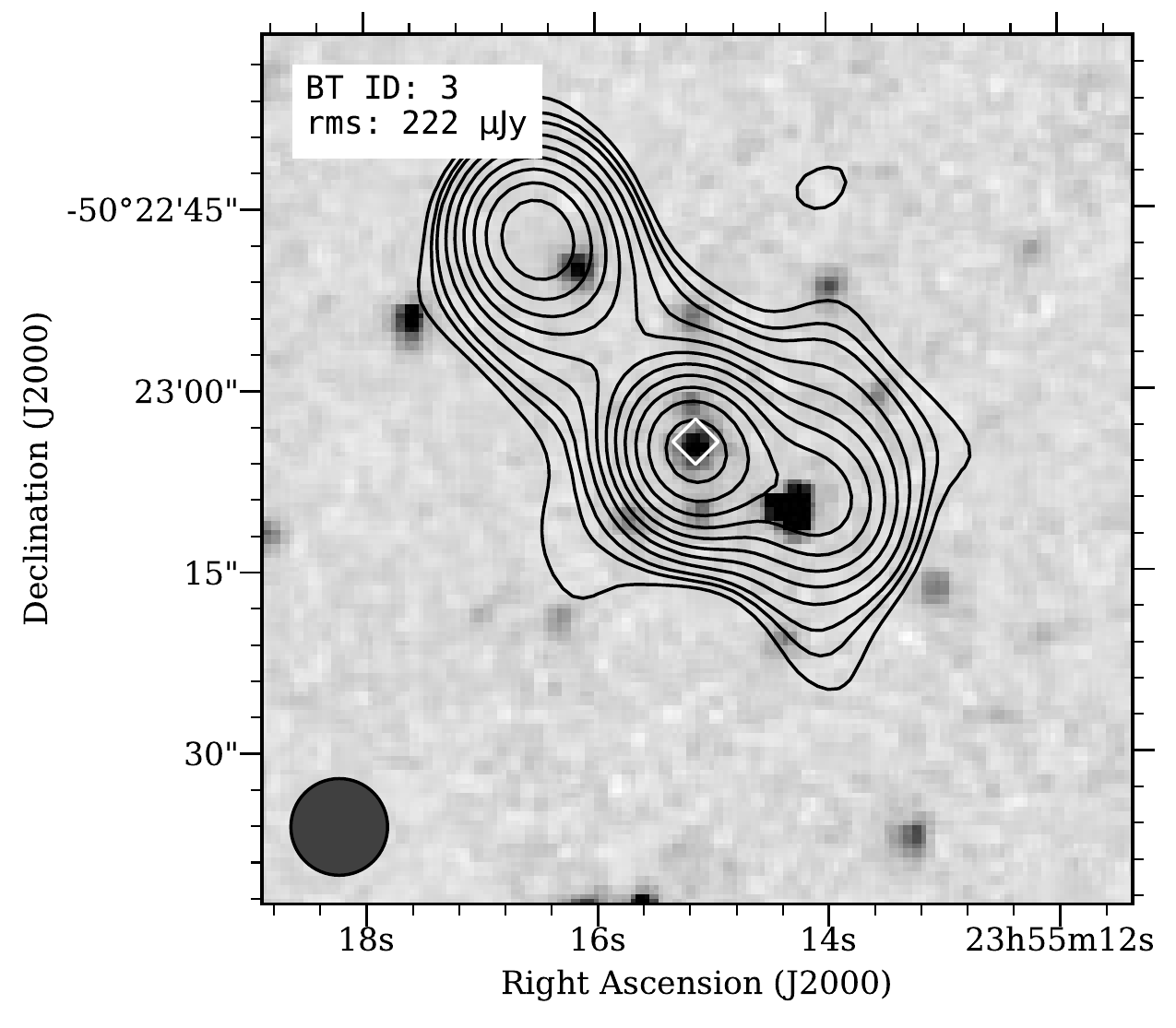}}\\
\subfloat{\includegraphics[width=0.3\textwidth]{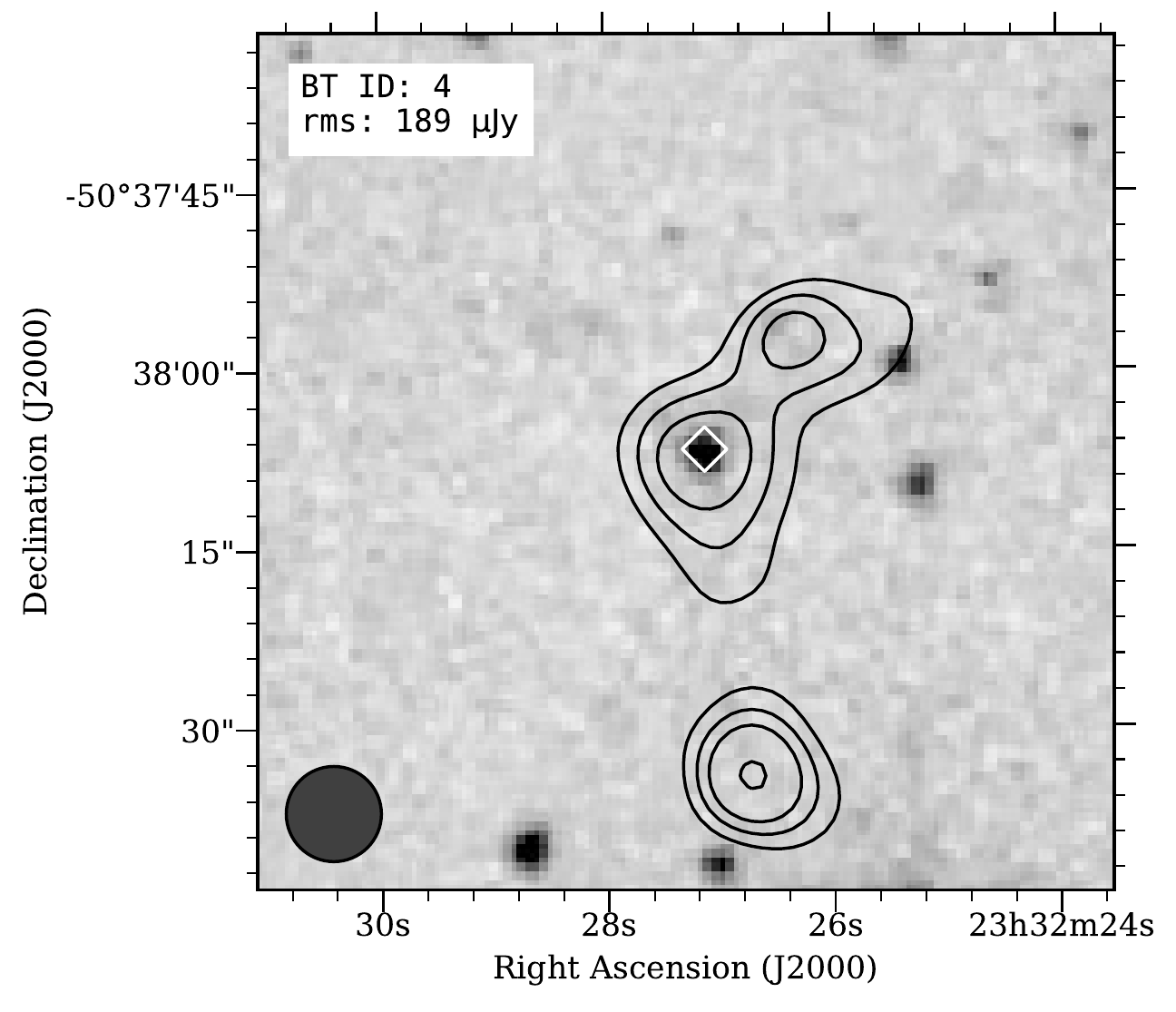}}
\subfloat{\includegraphics[width=0.3\textwidth]{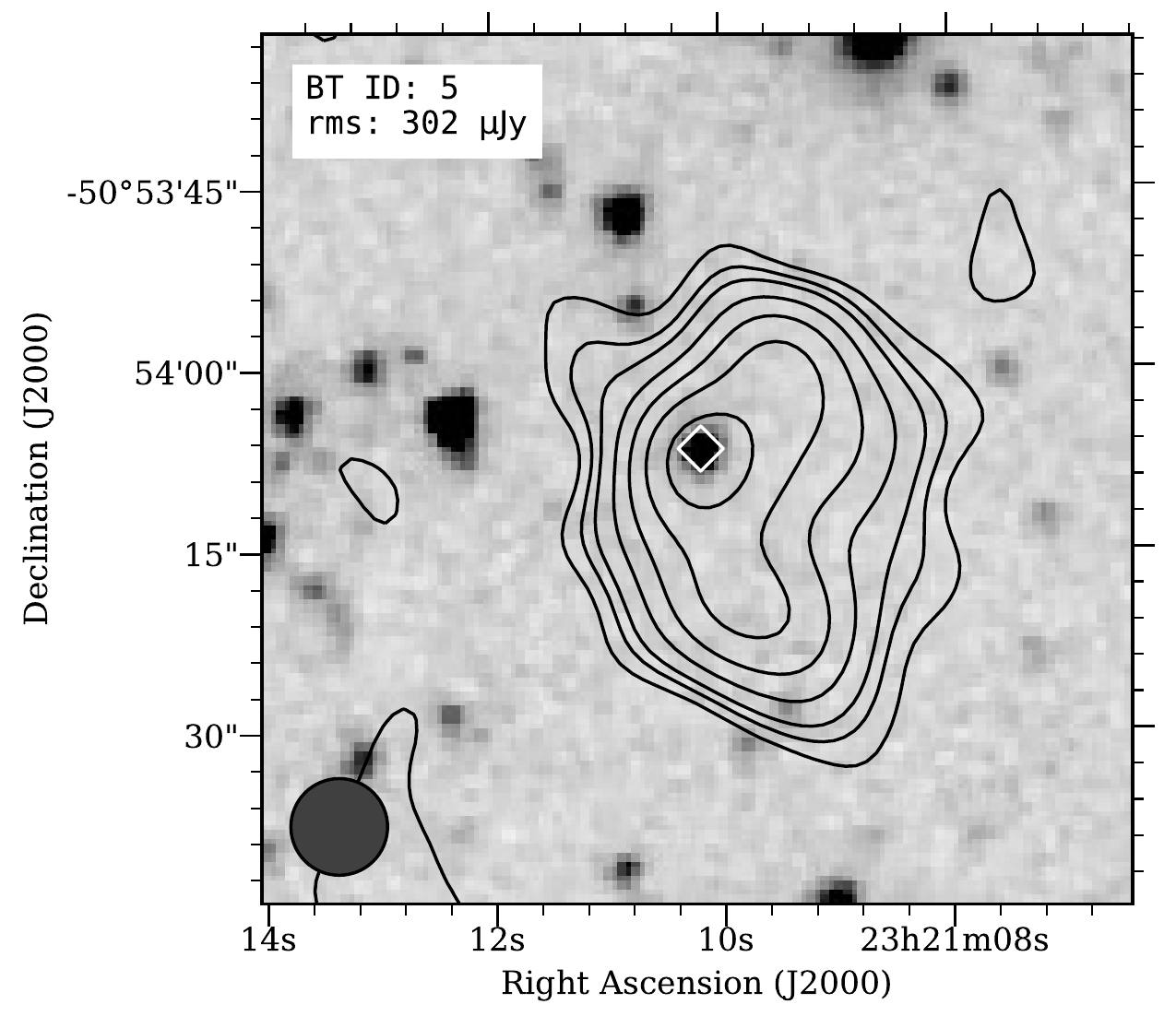}}
\subfloat{\includegraphics[width=0.3\textwidth]{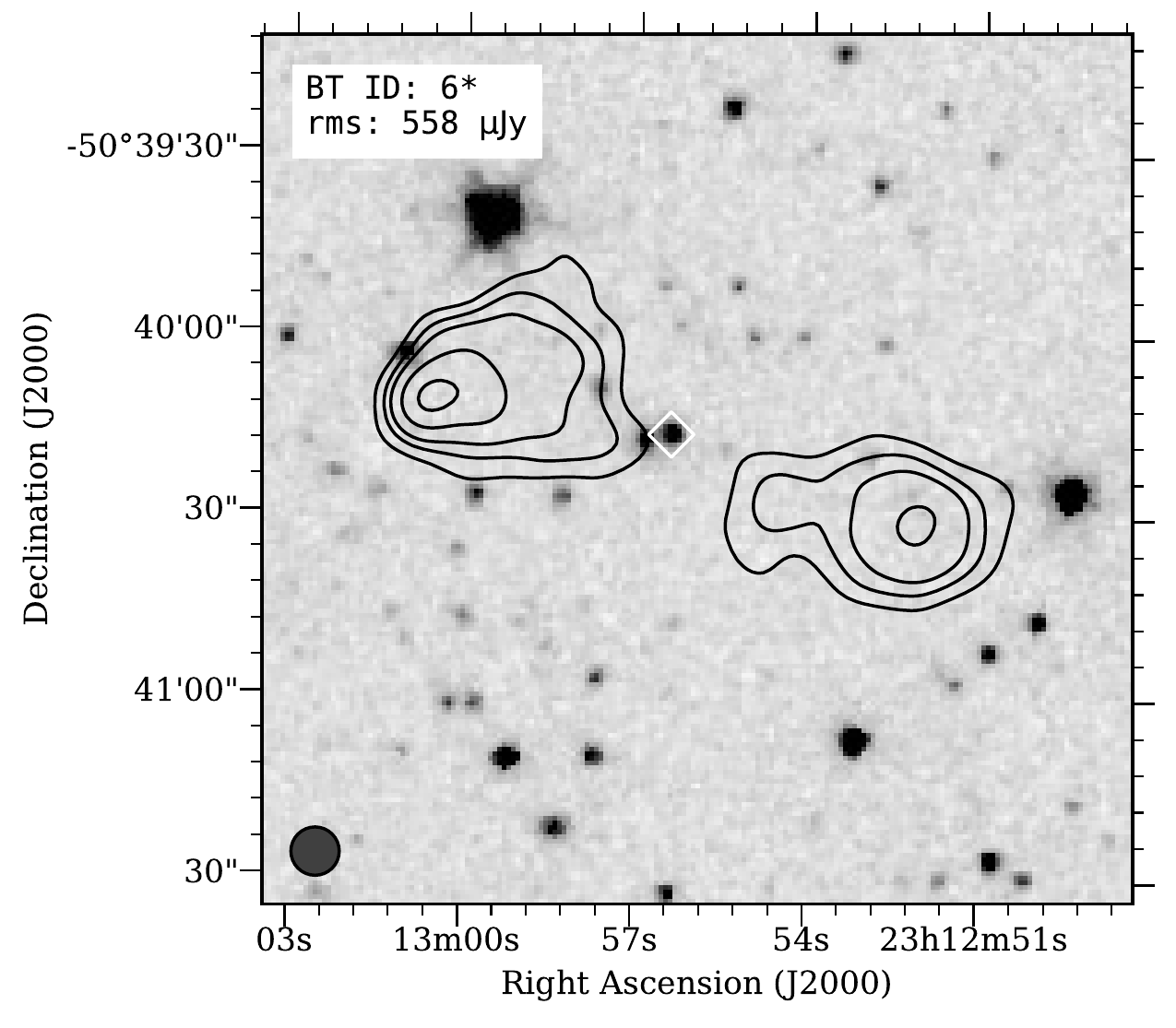}}\\
\subfloat{\includegraphics[width=0.3\textwidth]{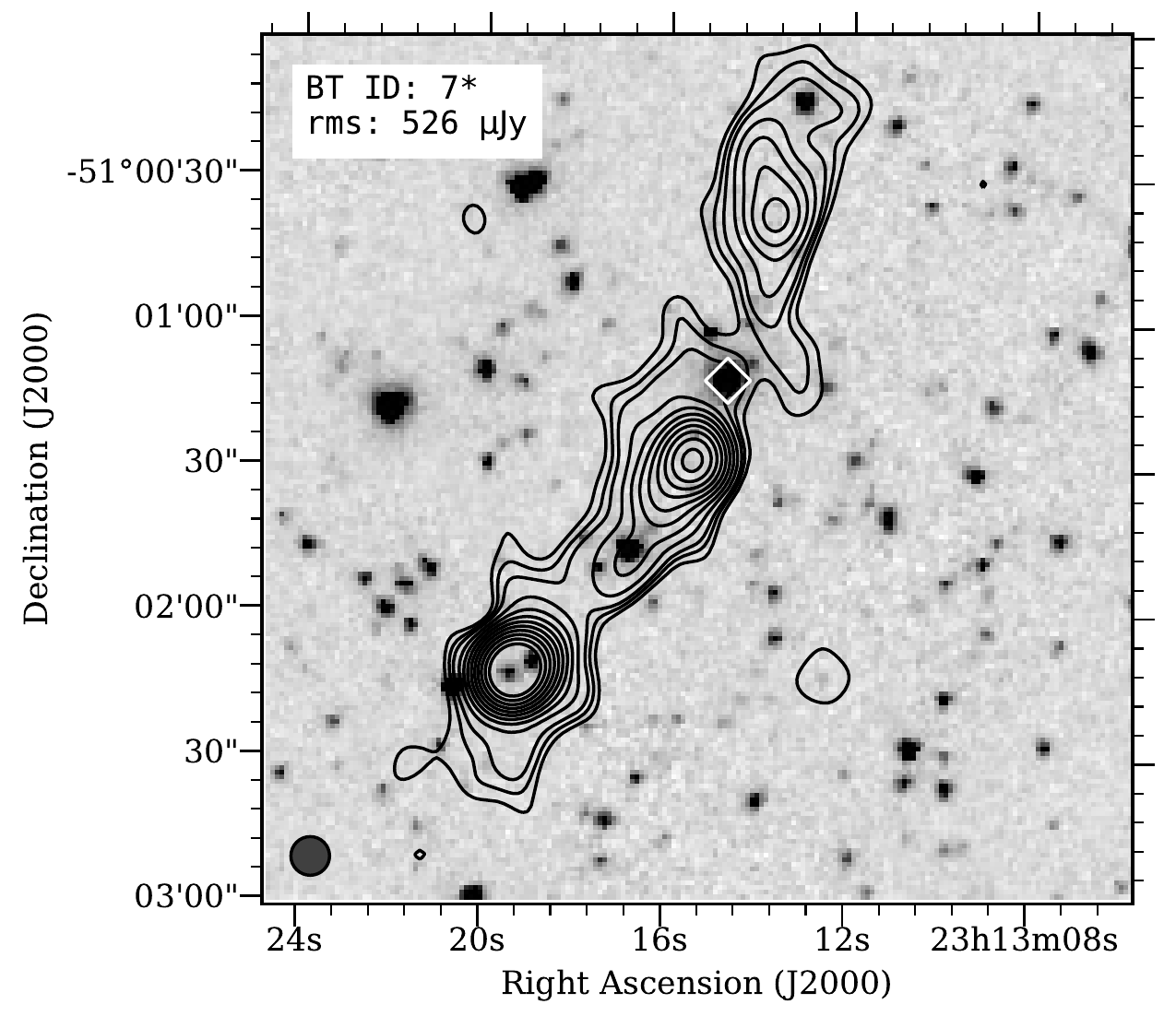}}
\subfloat{\includegraphics[width=0.3\textwidth]{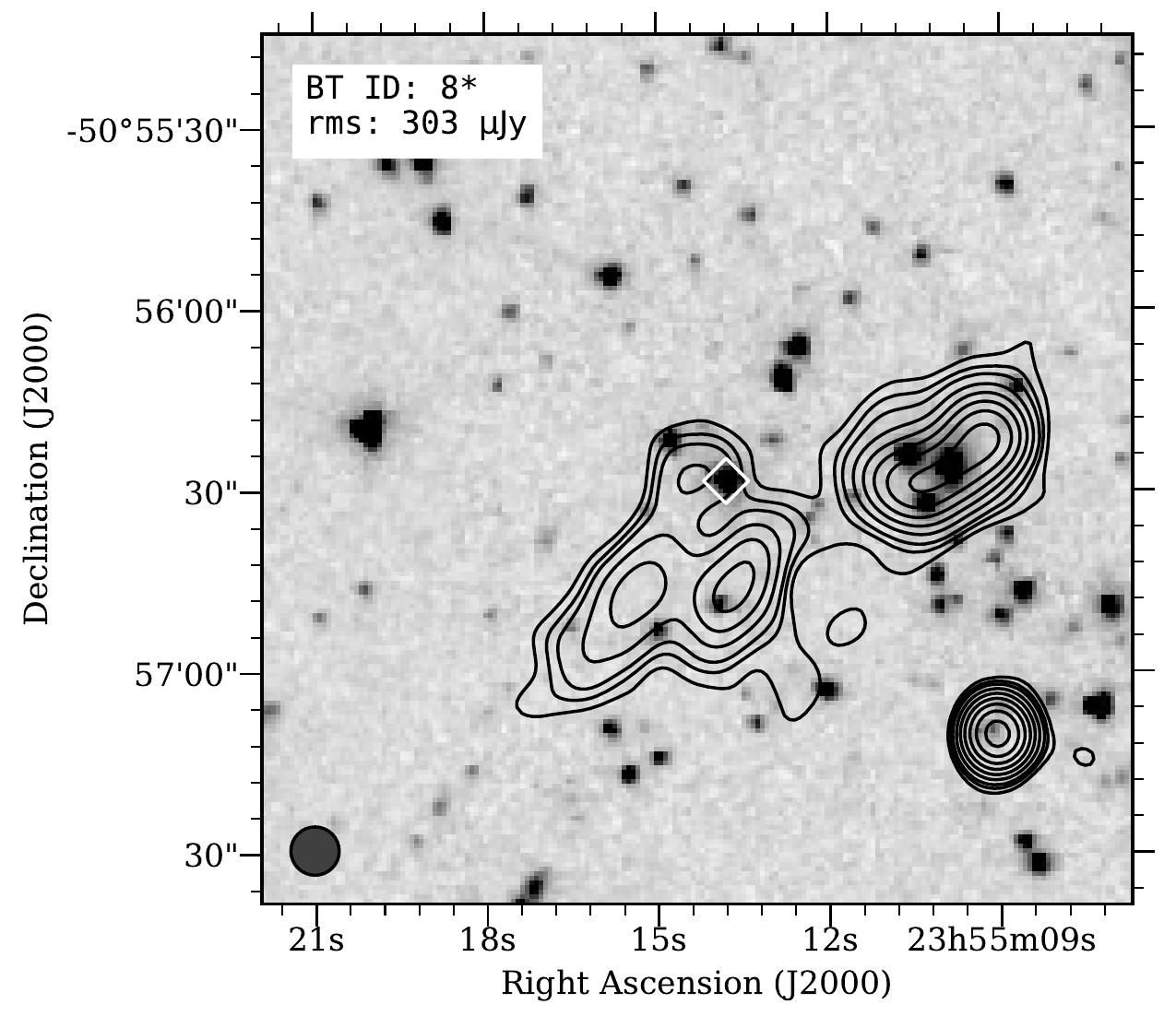}}
\subfloat{\includegraphics[width=0.3\textwidth]{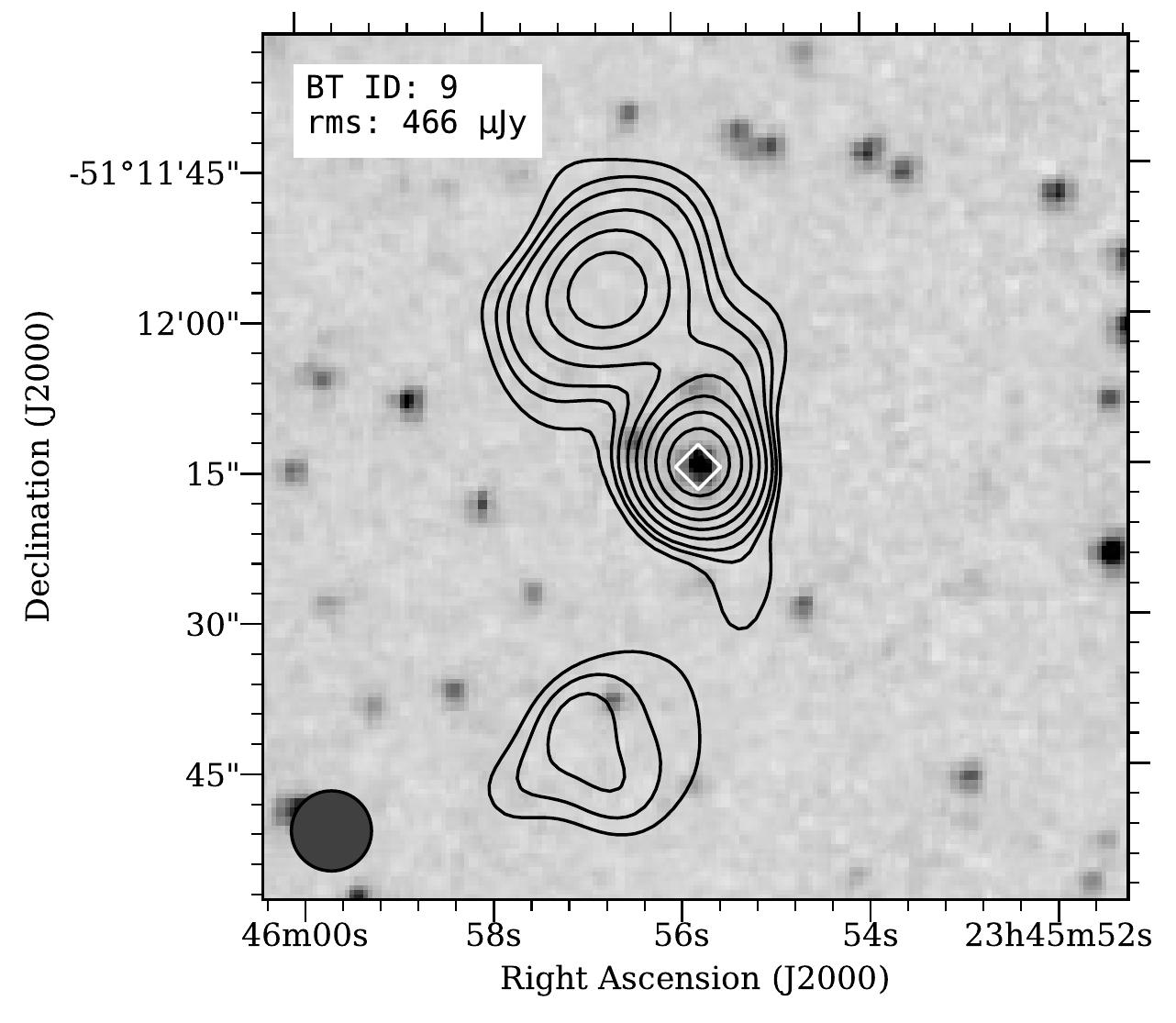}}\\
\subfloat{\includegraphics[width=0.3\textwidth]{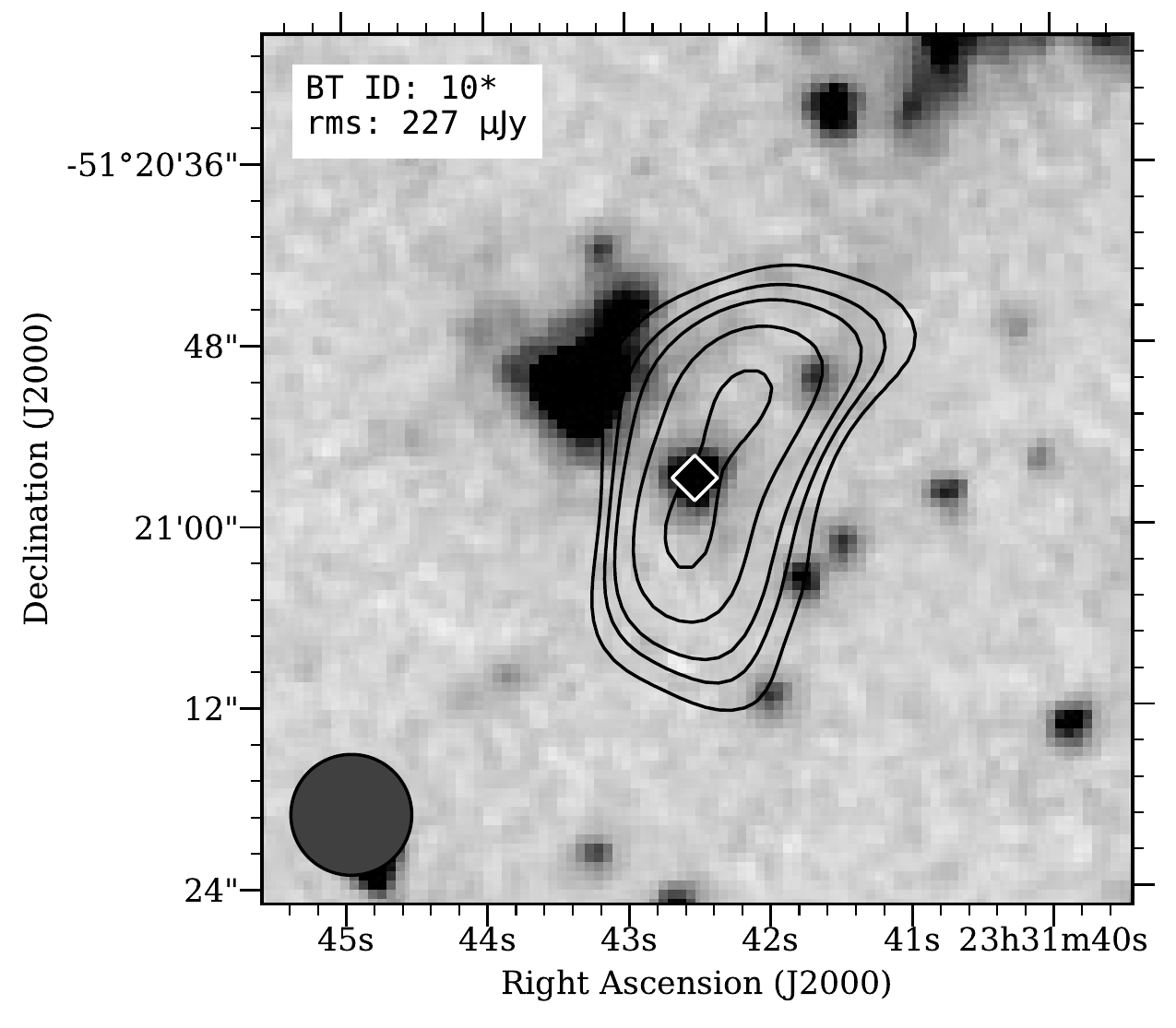}}
\subfloat{\includegraphics[width=0.3\textwidth]{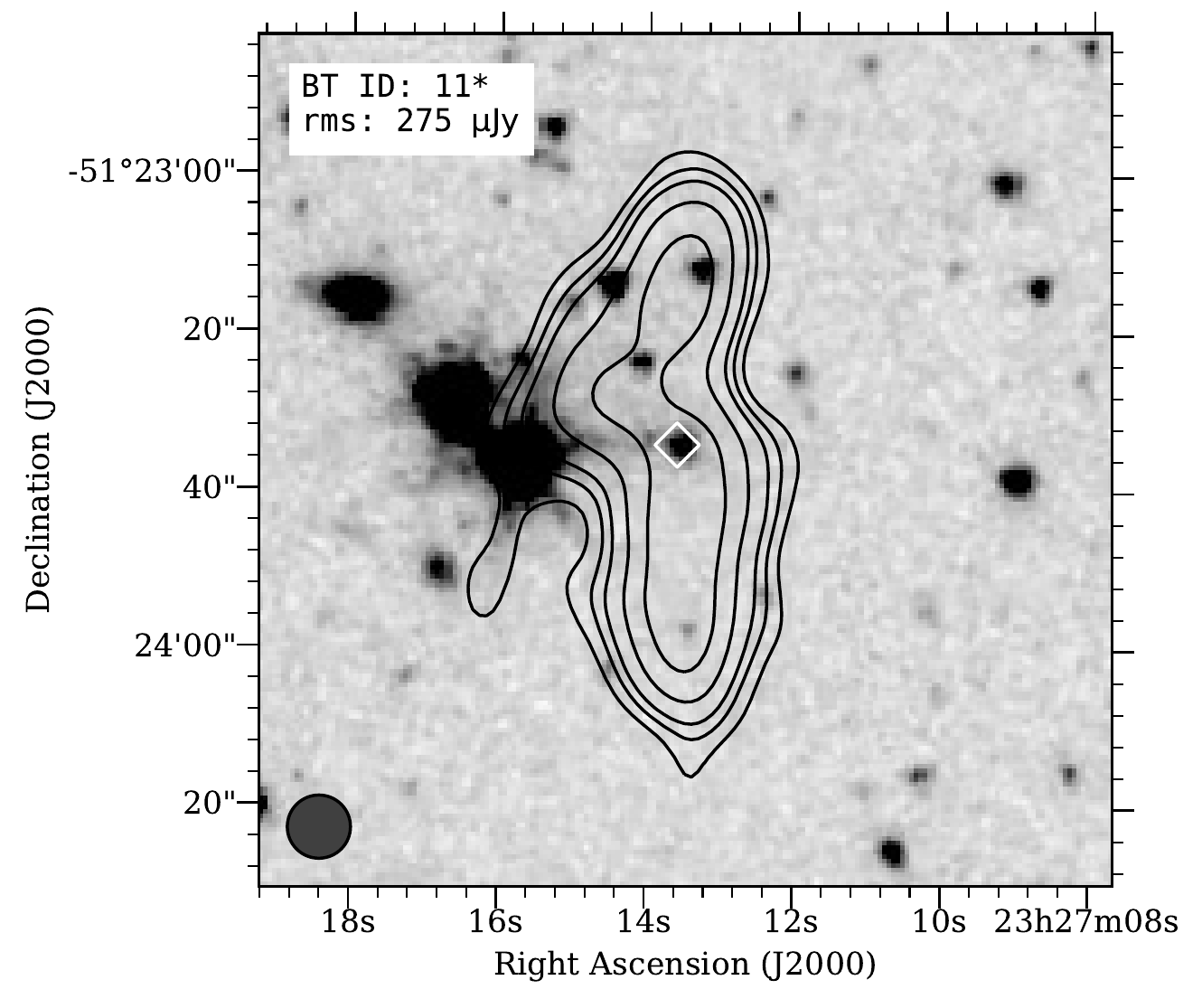}}
\subfloat{\includegraphics[width=0.3\textwidth]{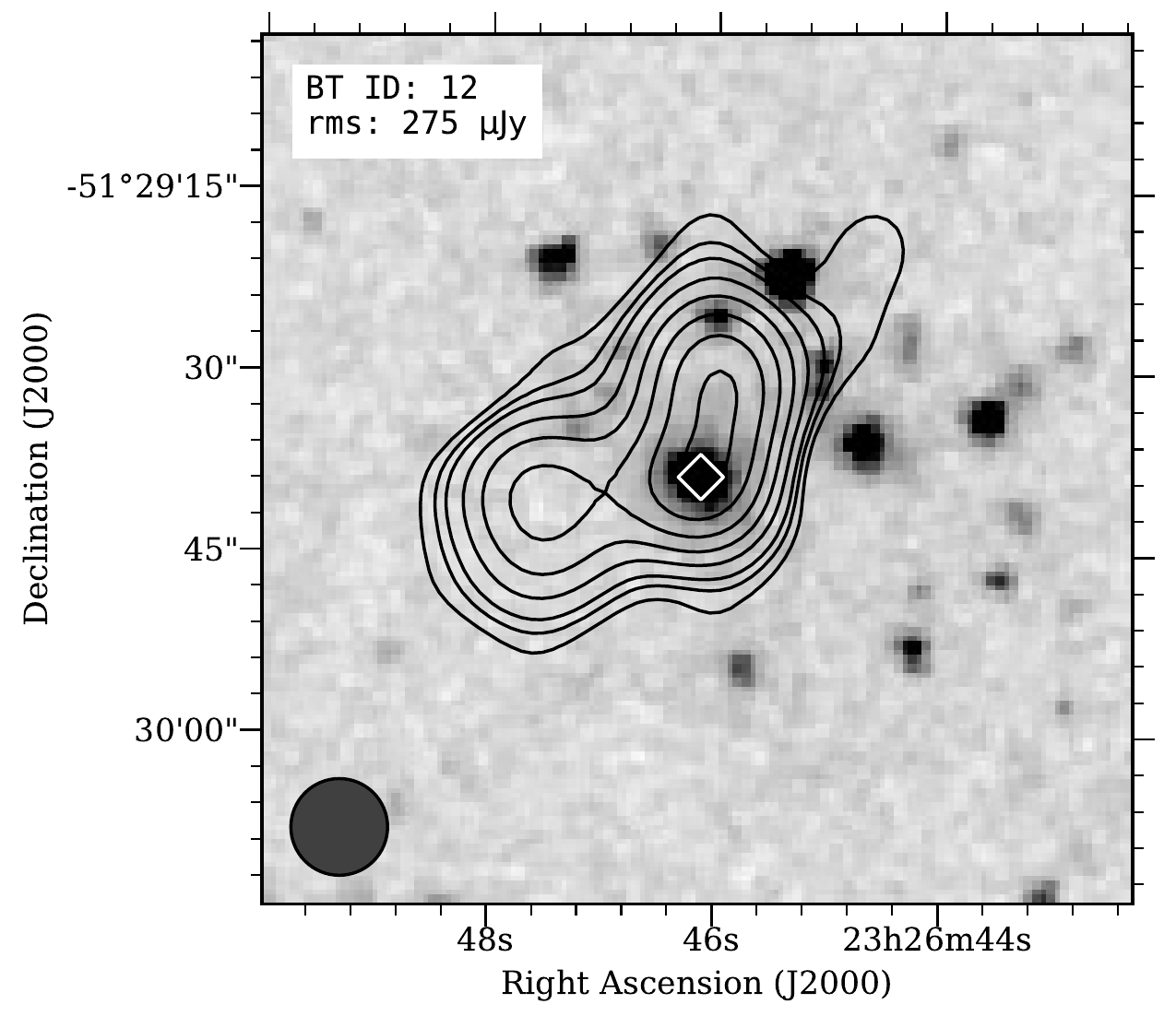}}
\caption{Bent-tail candidates in the ATLAS-SPT survey. Radio structure from this study is shown in black contours overlaid on 3.6\micron{} images from the SSDF. Contours are drawn at $3, 4, 5\sigma$ and then increase by factors of $\sqrt{2}^{n}$ where $n \in \mathbb{N}$. Additional radio contours of higher resolution from the ATCA-XXL survey \citep{butler17} are overlaid as dashed contours when available. SSDF sources identified as the likely host galaxy are marked with a diamond. Optical cross-identifications are indicated with an X.}
\end{figure*}
\begin{figure*}
\subfloat{\includegraphics[width=0.3\textwidth]{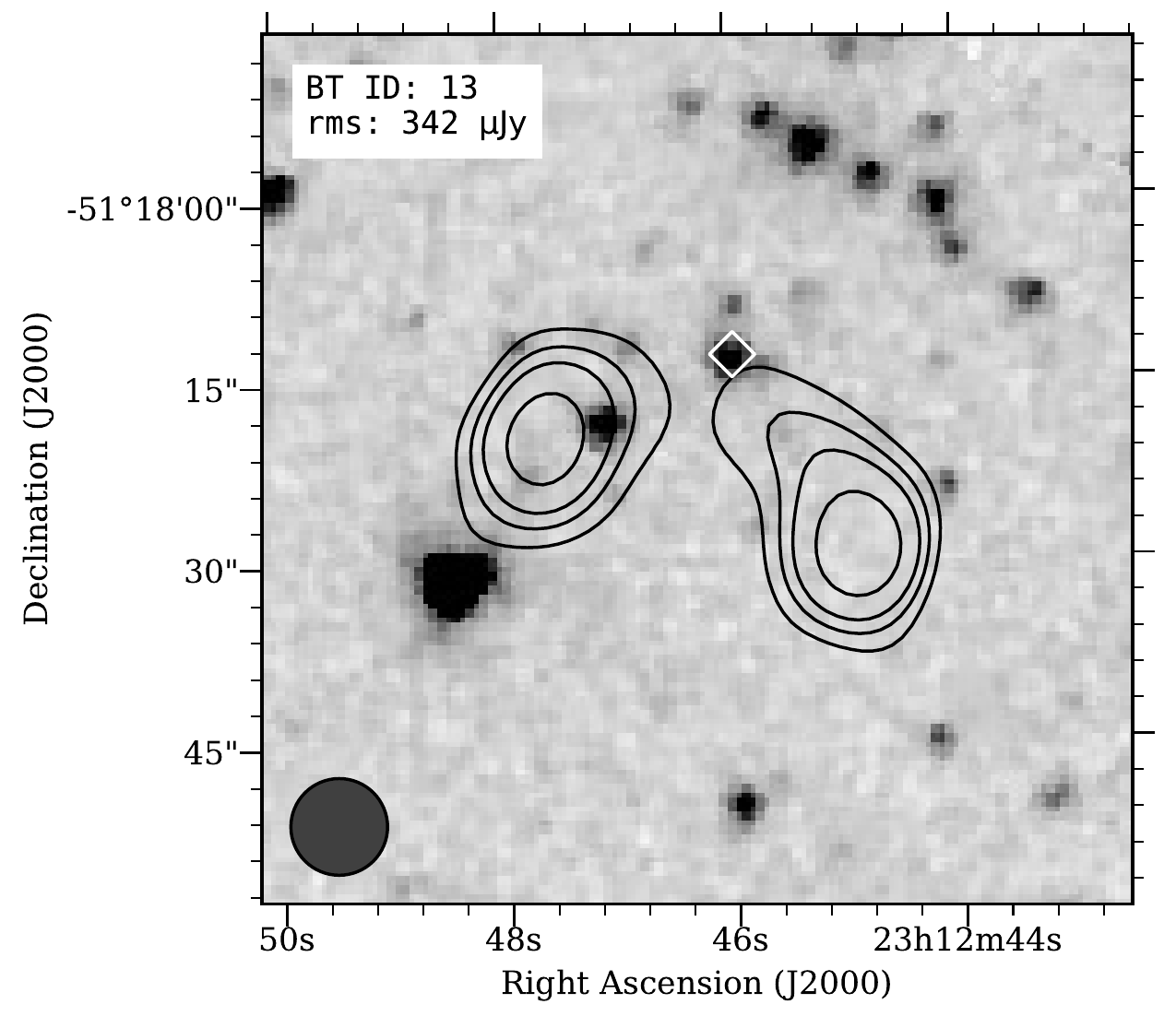}}
\subfloat{\includegraphics[width=0.3\textwidth]{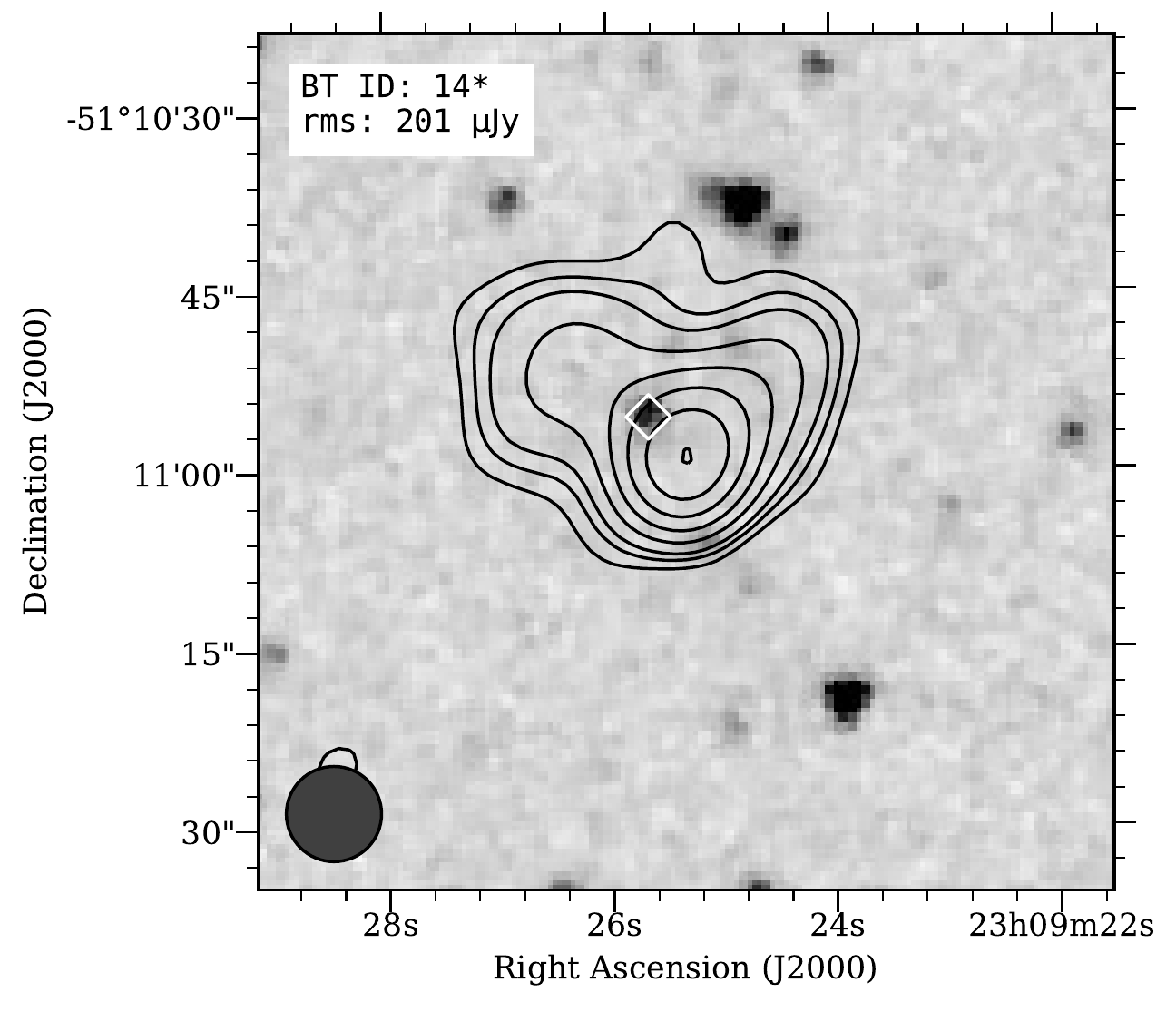}}
\subfloat{\includegraphics[width=0.3\textwidth]{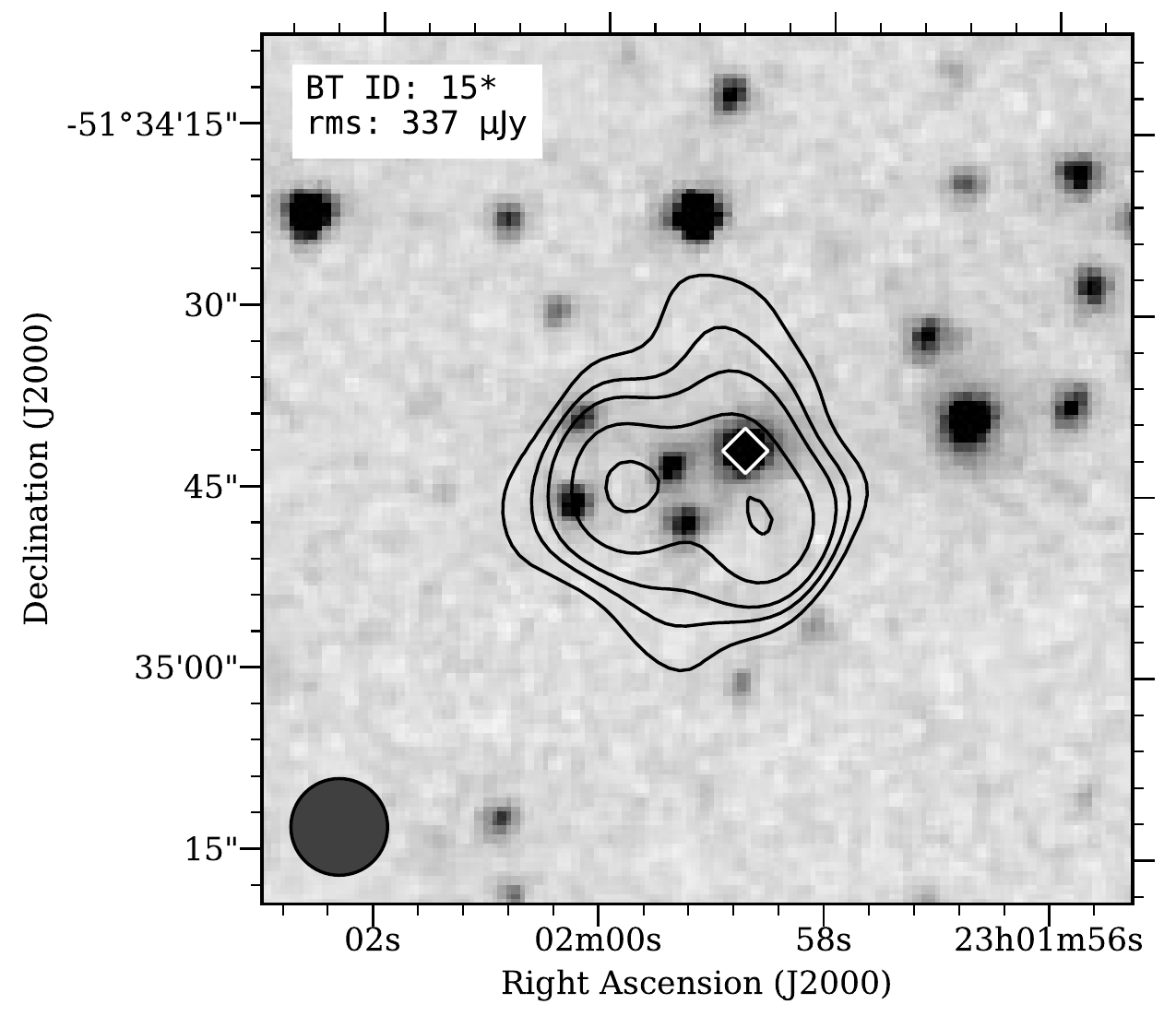}}\\
\subfloat{\includegraphics[width=0.3\textwidth]{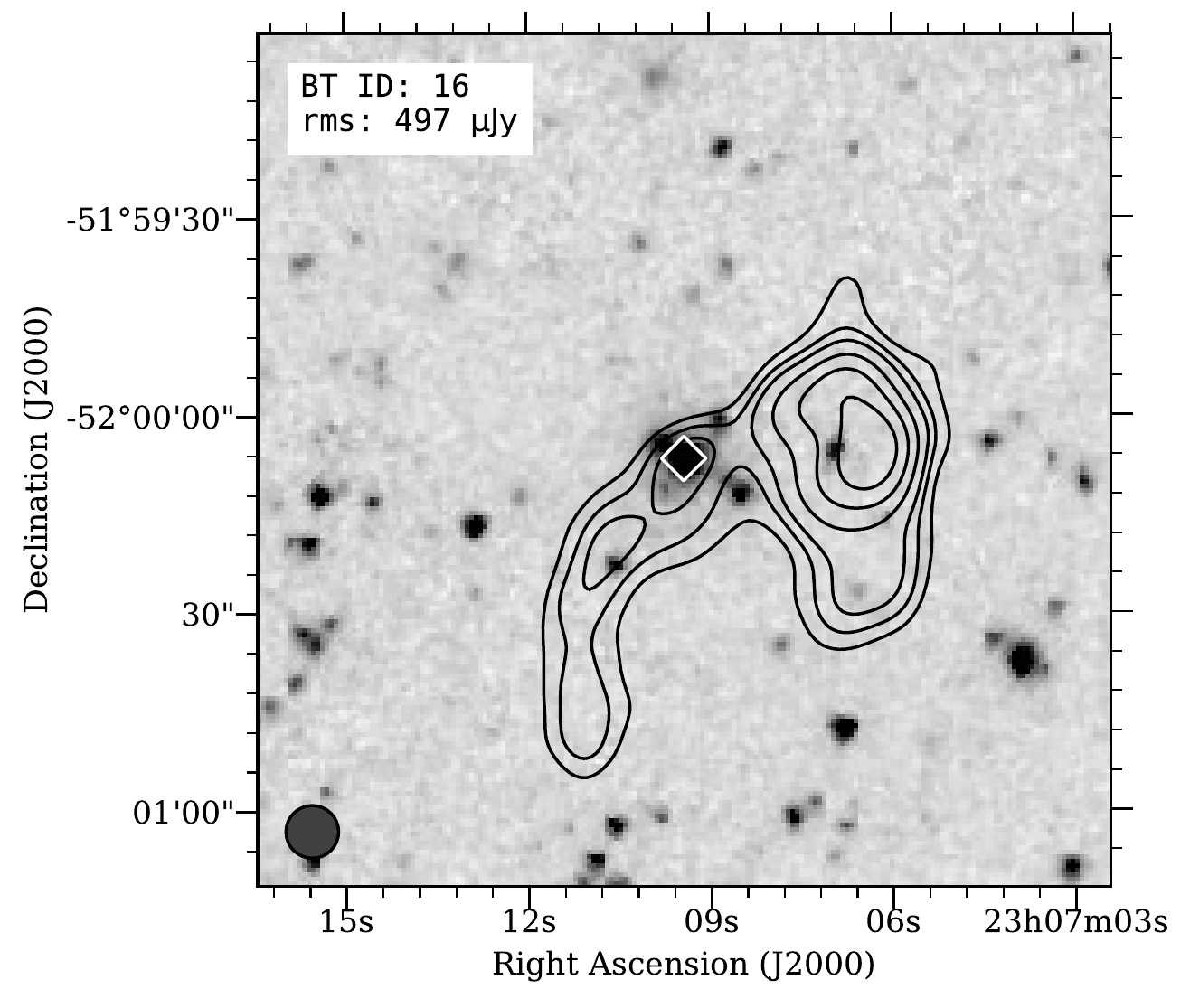}}
\subfloat{\includegraphics[width=0.3\textwidth]{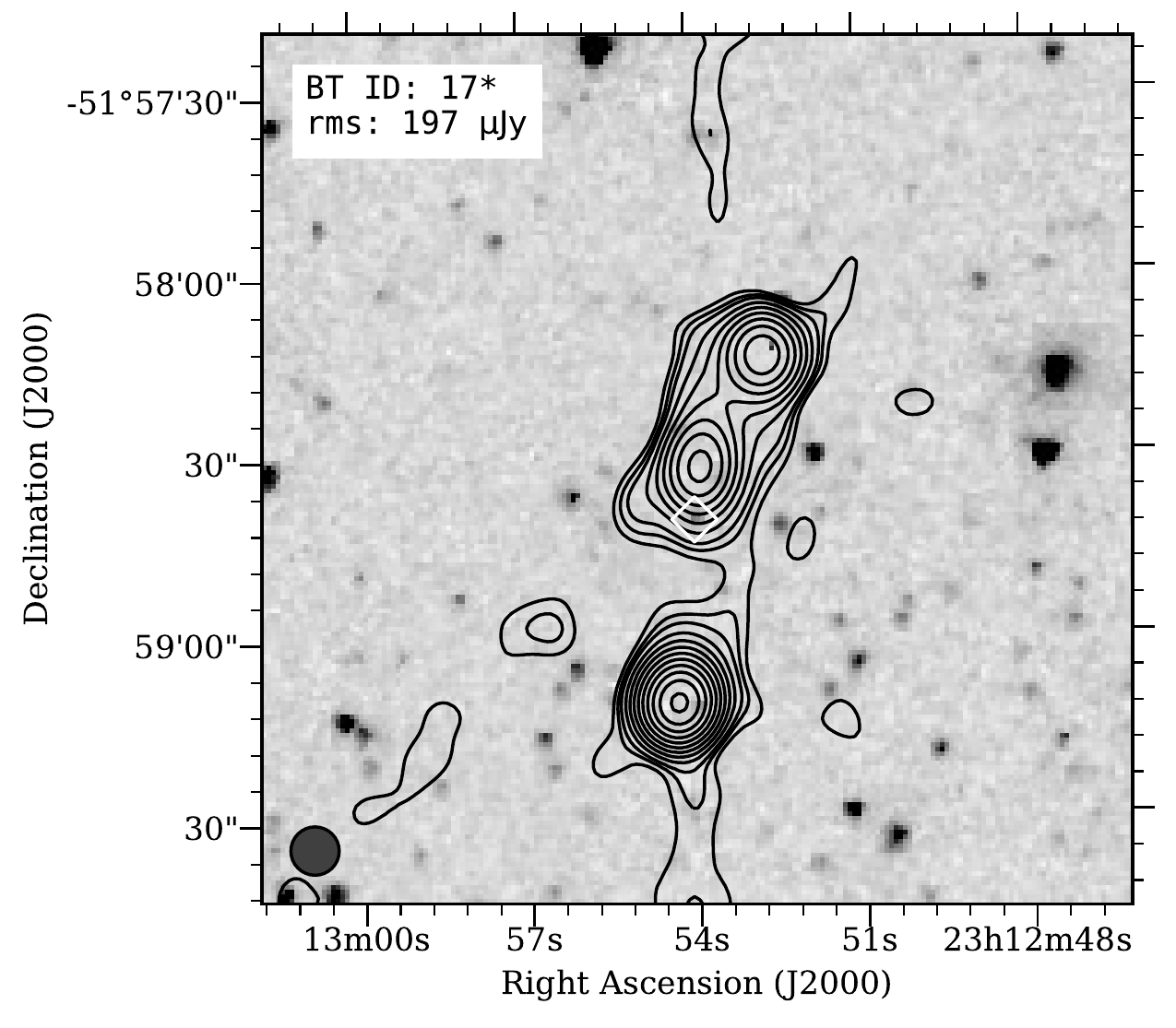}}
\subfloat{\includegraphics[width=0.3\textwidth]{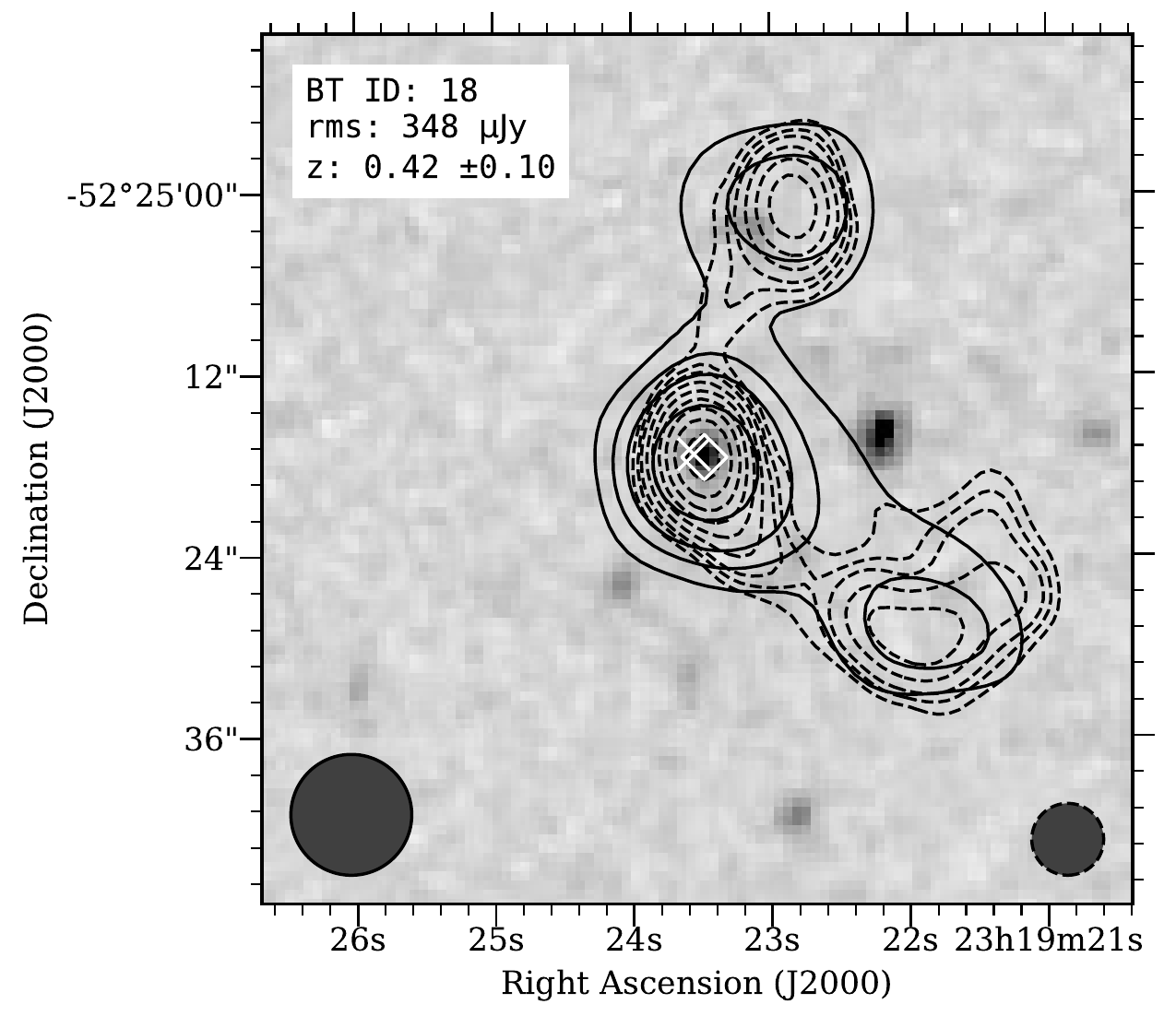}}\\
\subfloat{\includegraphics[width=0.3\textwidth]{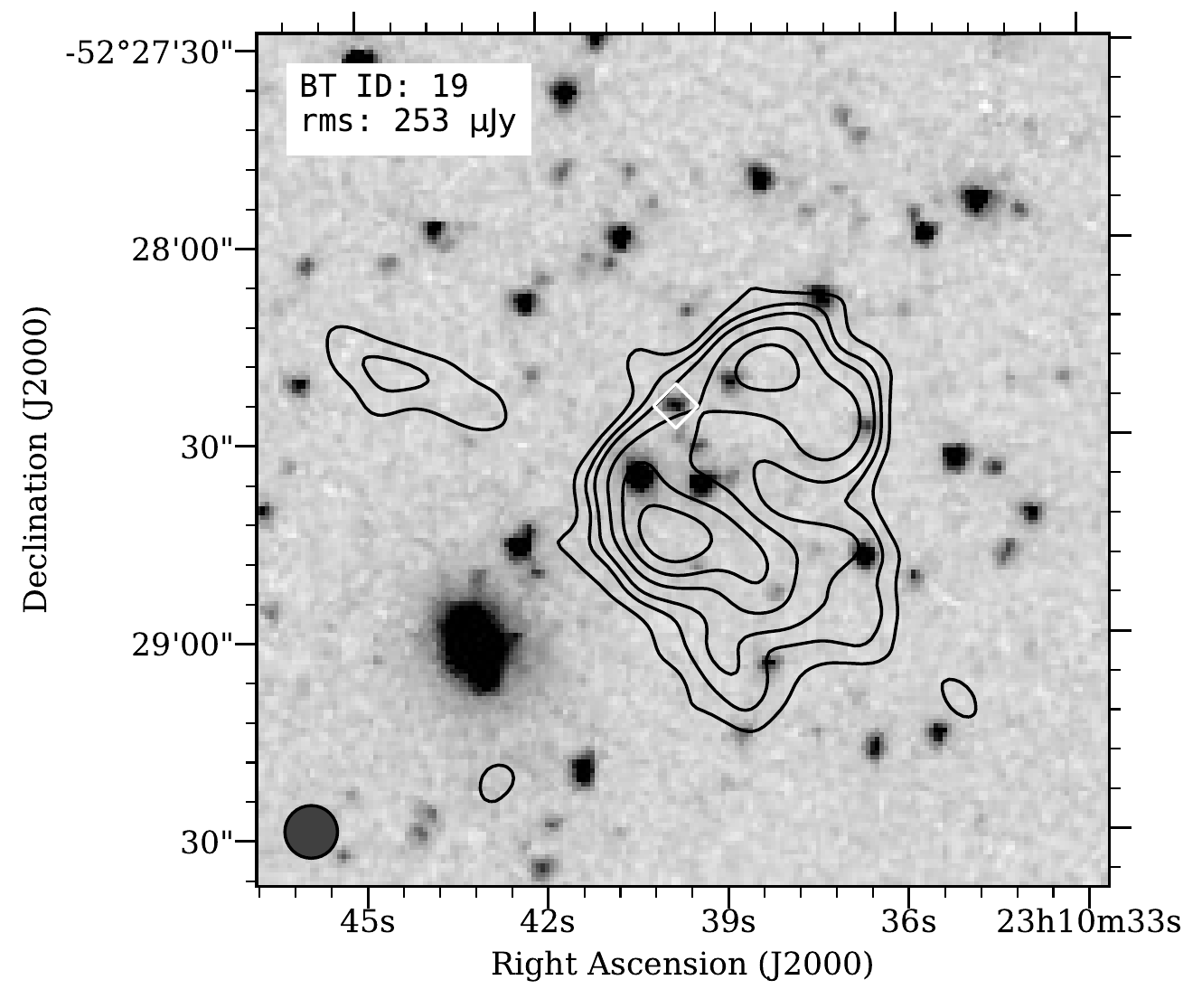}}
\subfloat{\includegraphics[width=0.3\textwidth]{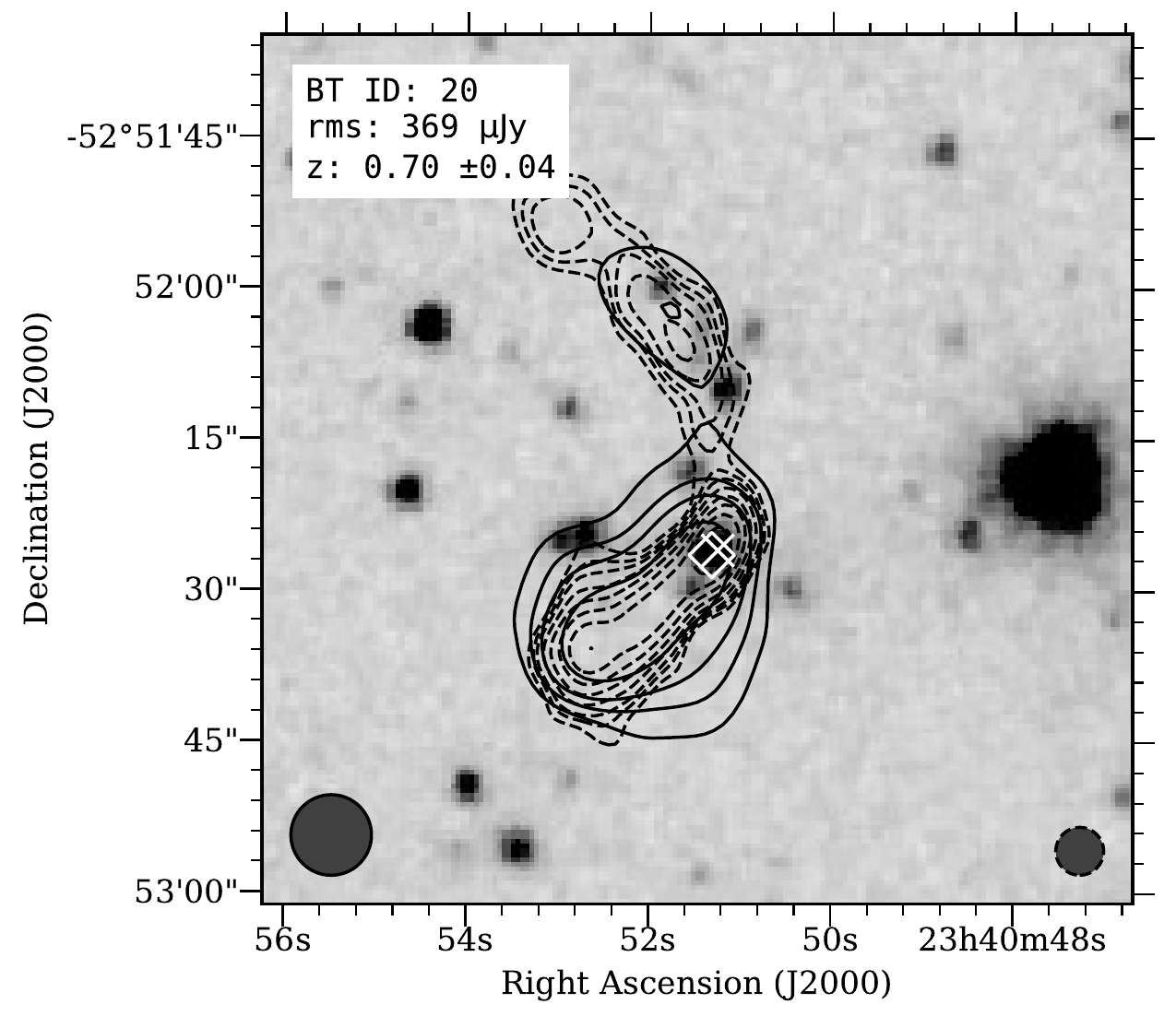}}
\subfloat{\includegraphics[width=0.3\textwidth]{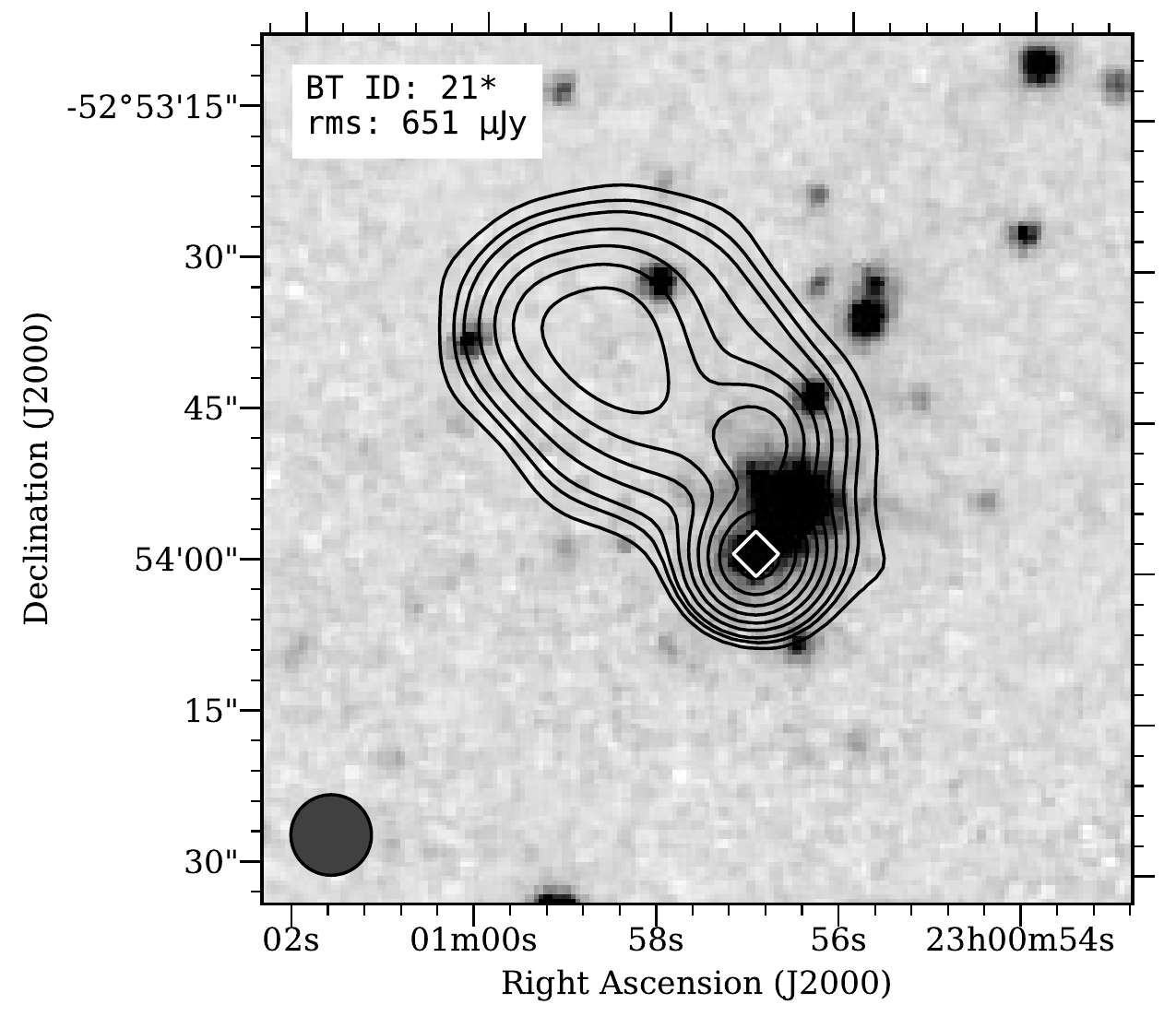}}\\
\subfloat{\includegraphics[width=0.3\textwidth]{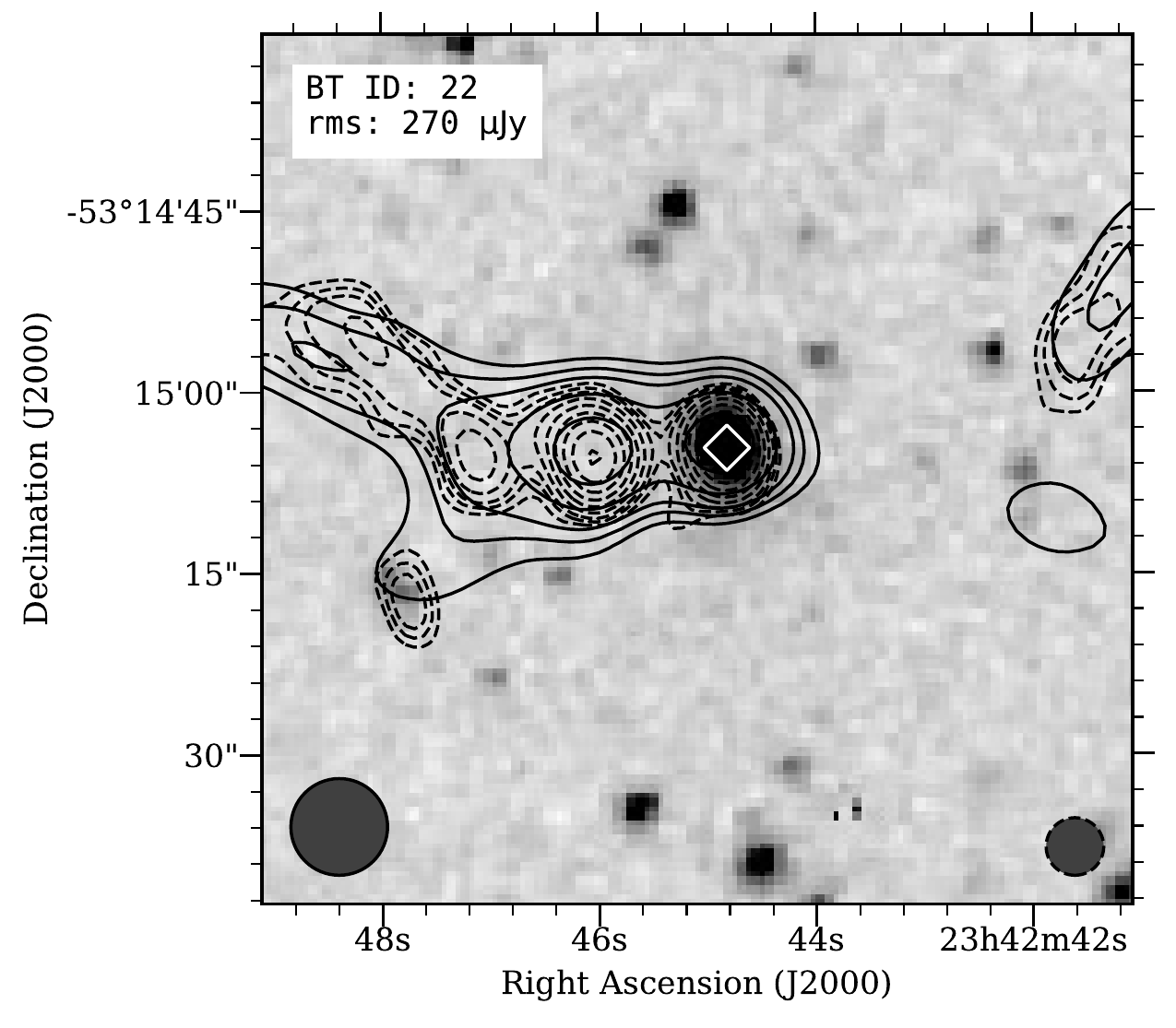}}
\subfloat{\includegraphics[width=0.3\textwidth]{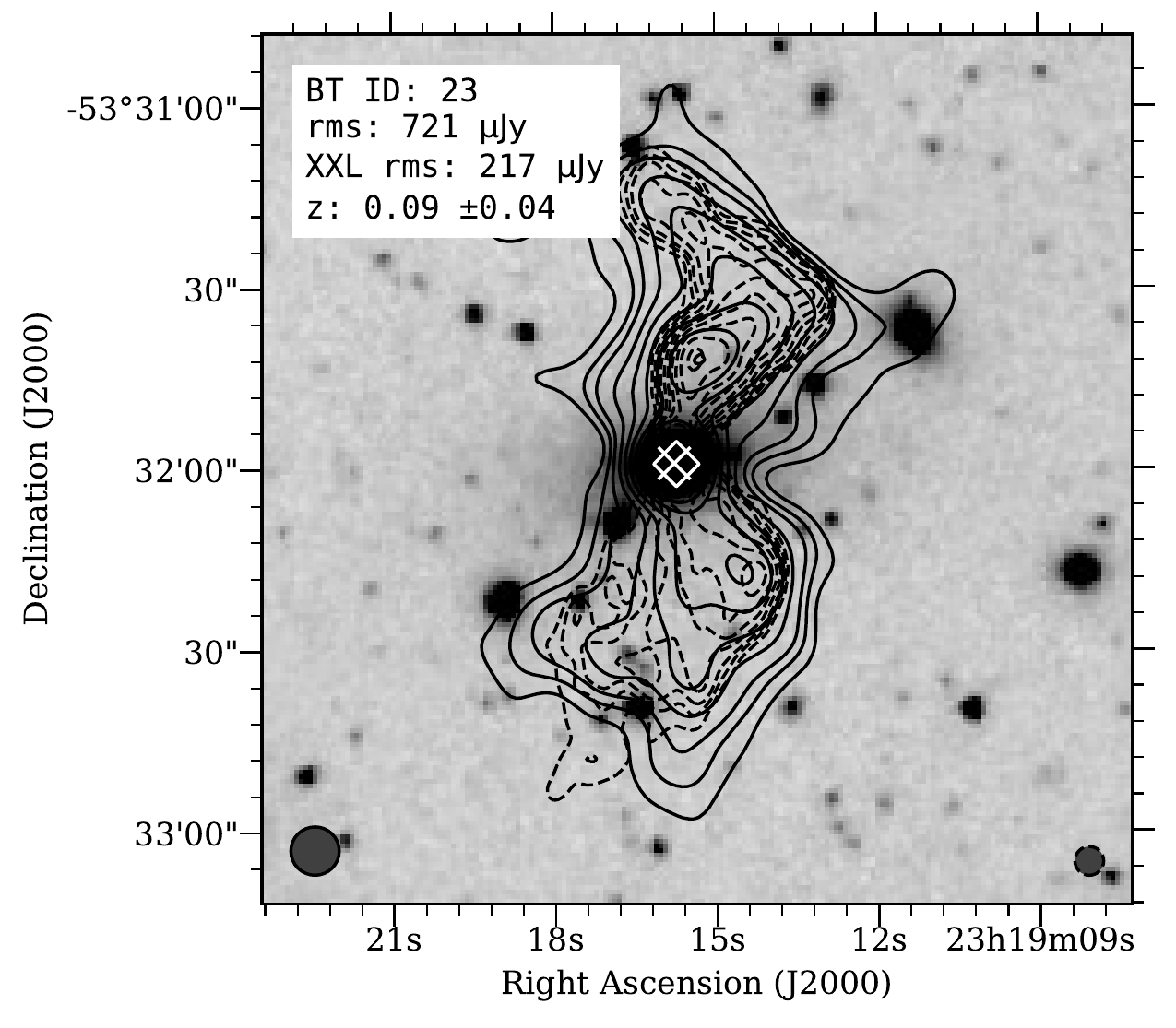}}
\subfloat{\includegraphics[width=0.3\textwidth]{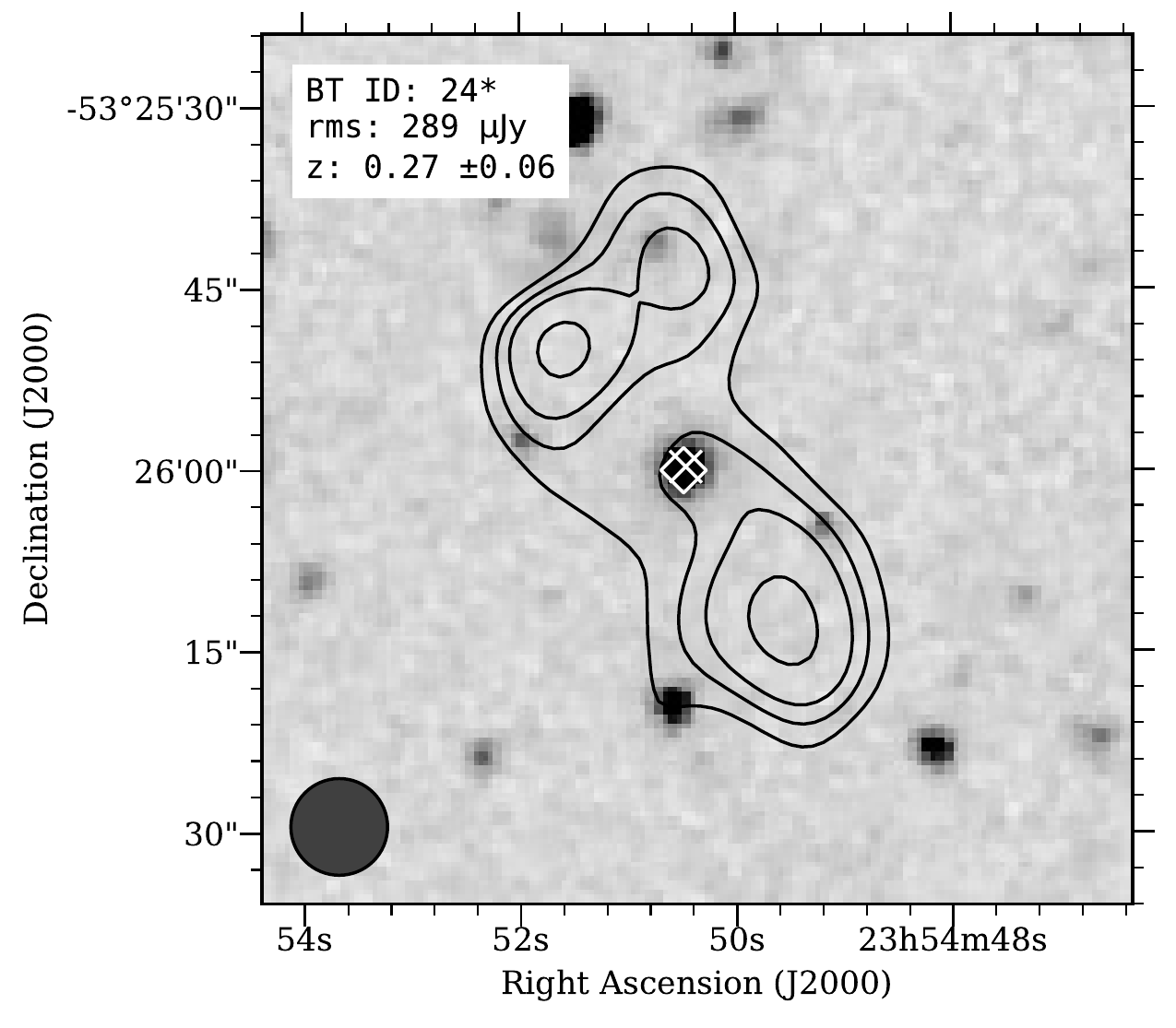}}\\
\subfloat{\includegraphics[width=0.3\textwidth]{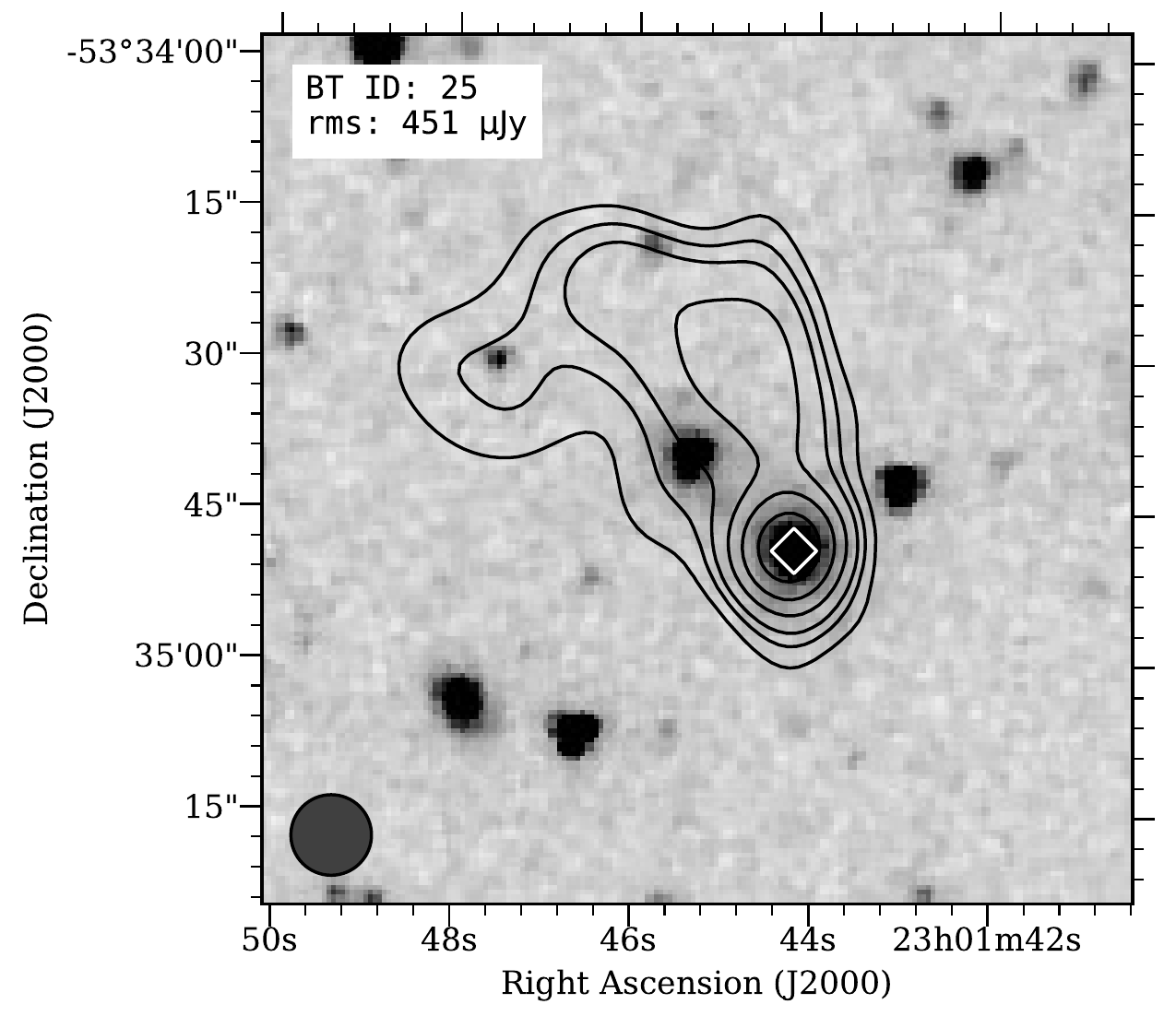}}
\subfloat{\includegraphics[width=0.3\textwidth]{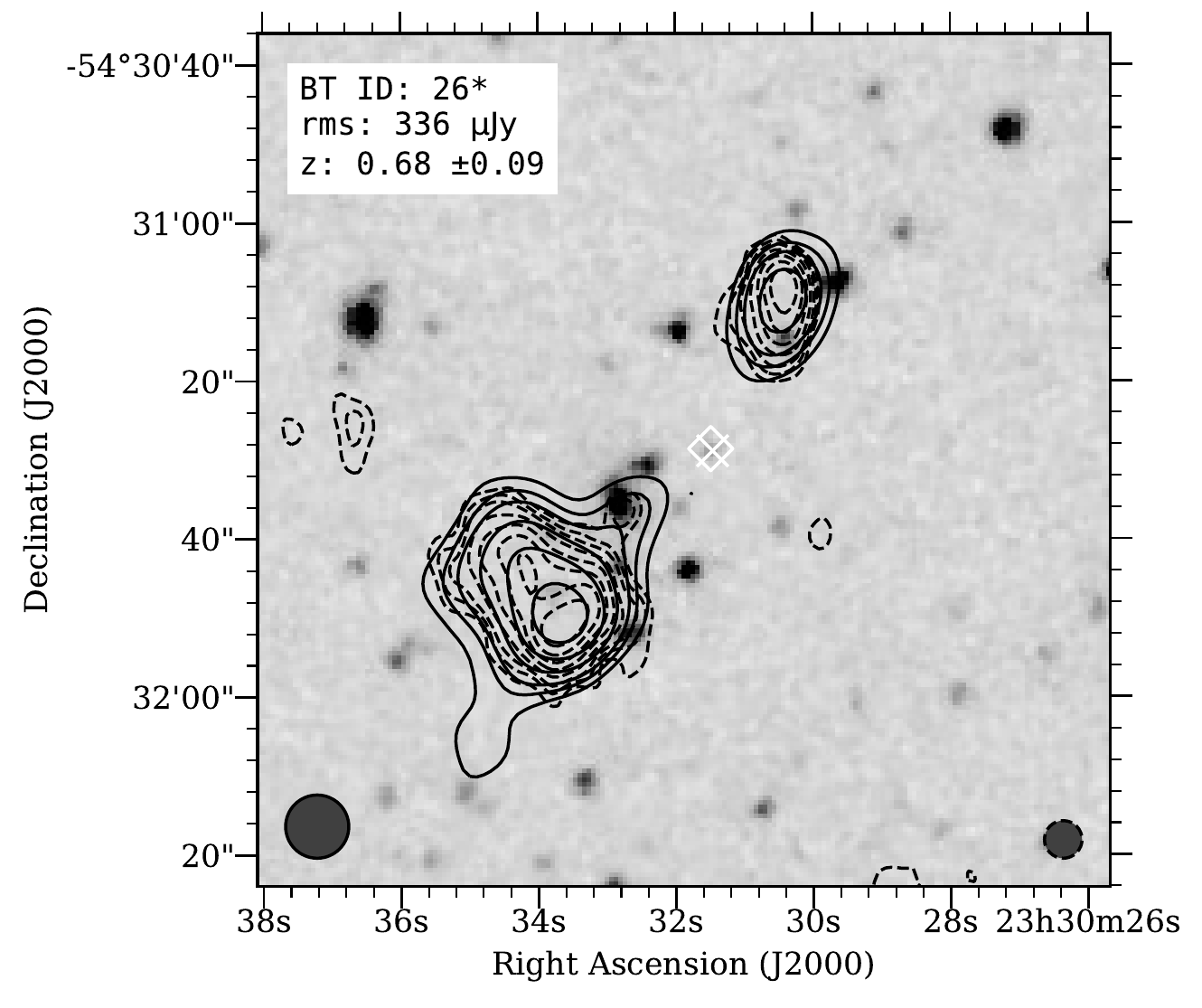}}
\subfloat{\includegraphics[width=0.3\textwidth]{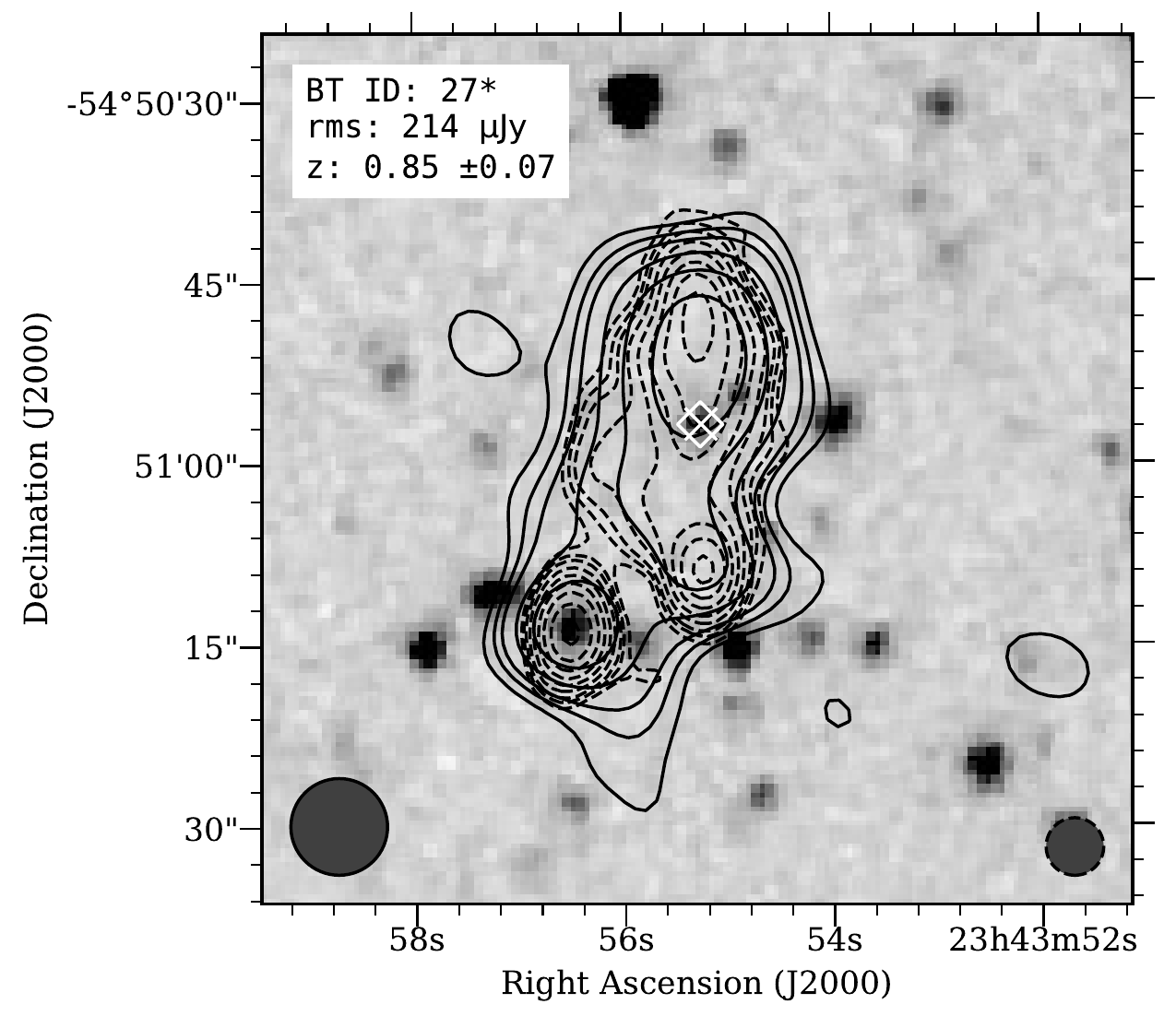}}\\
\contcaption{}
\end{figure*}
\begin{figure*}
\subfloat{\includegraphics[width=0.3\textwidth]{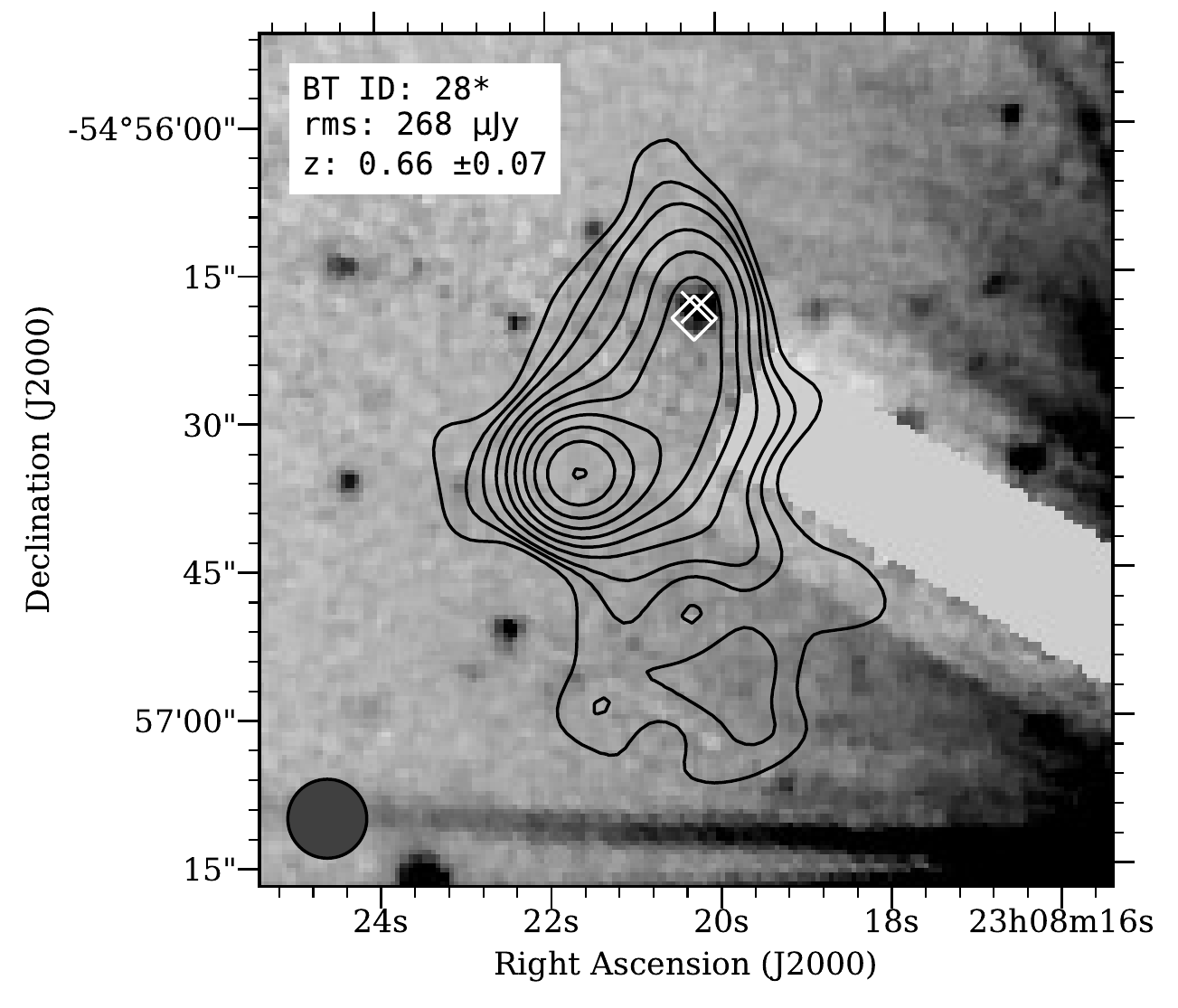}}
\subfloat{\includegraphics[width=0.3\textwidth]{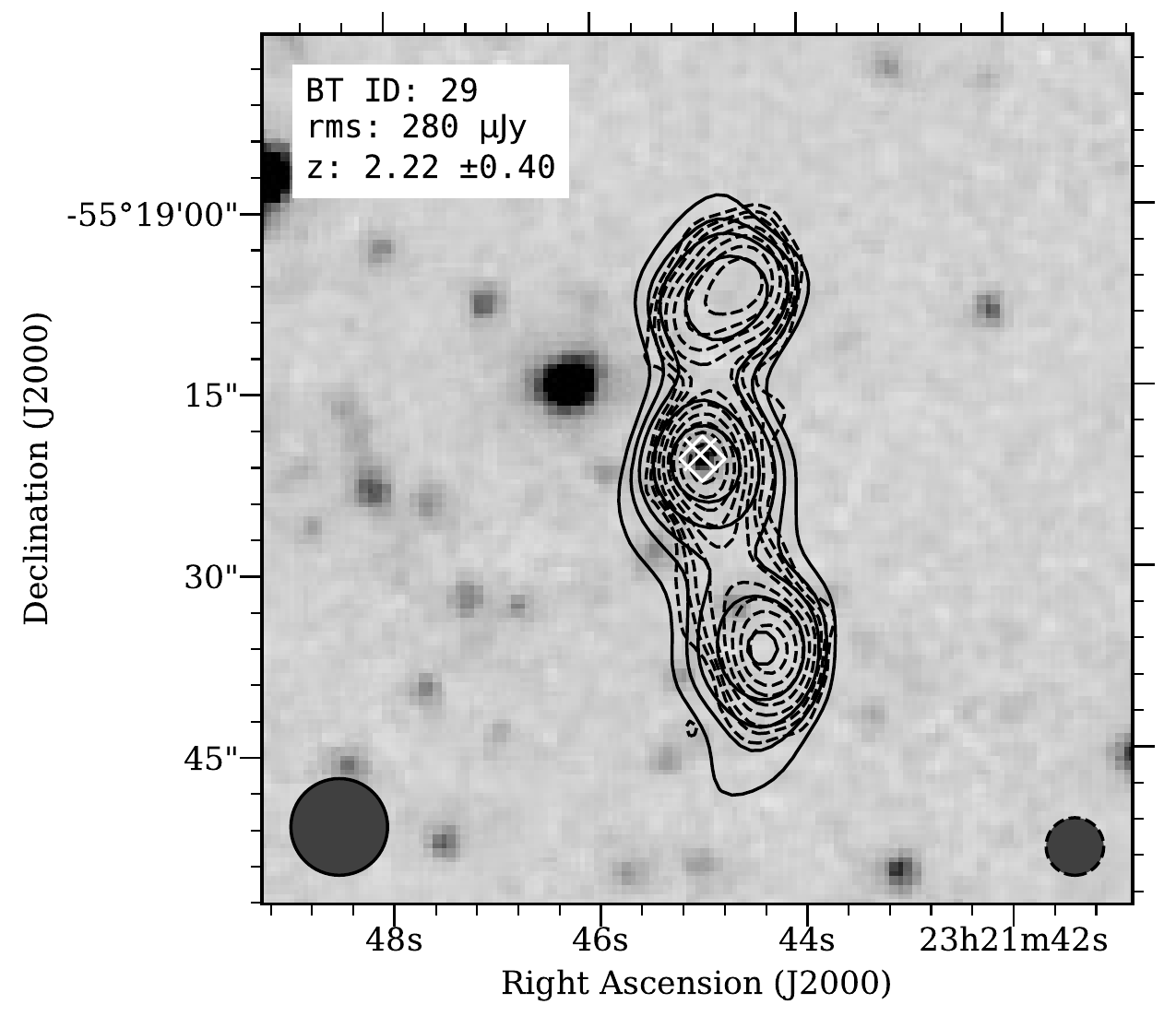}}
\subfloat{\includegraphics[width=0.3\textwidth]{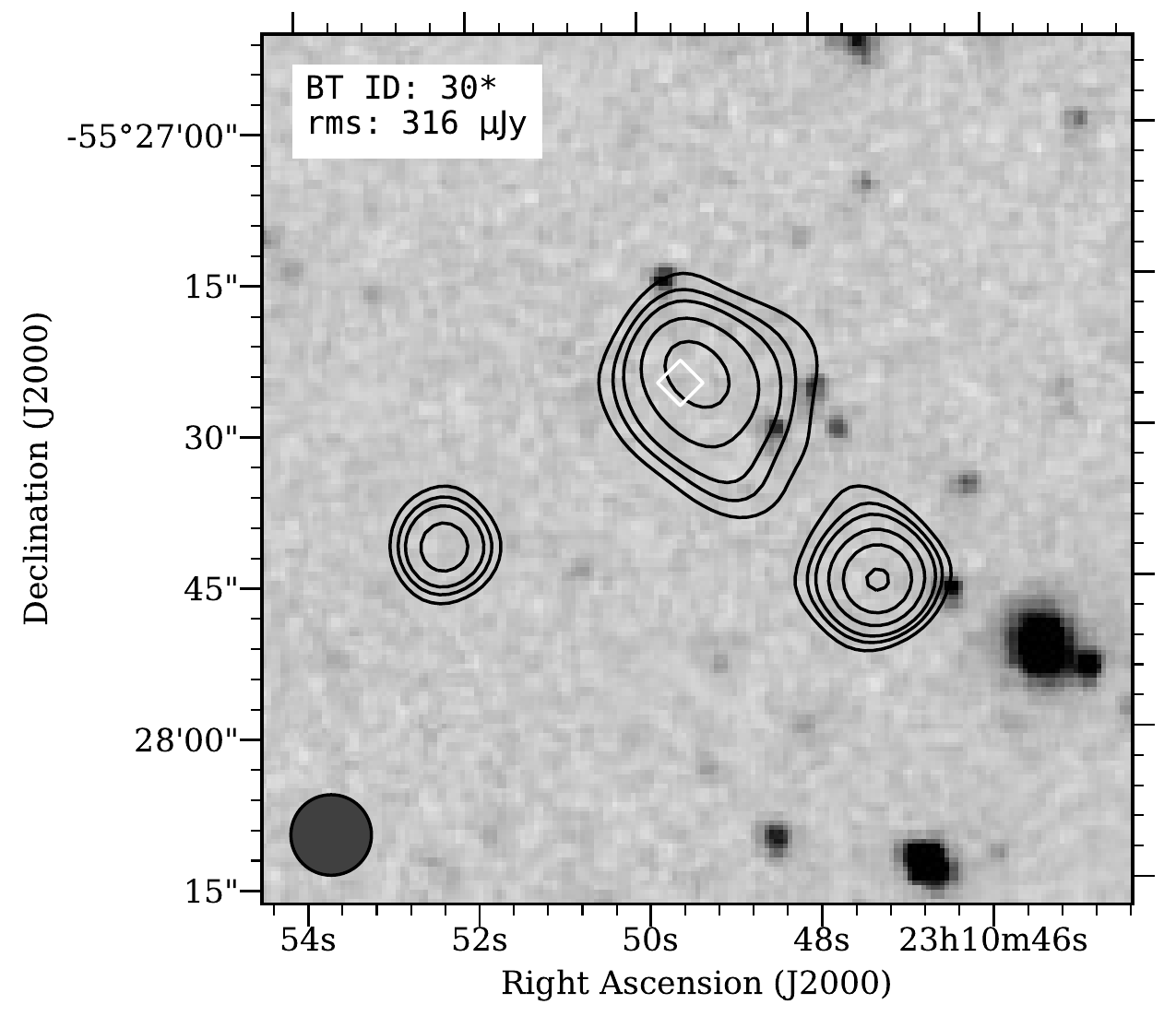}}\\
\subfloat{\includegraphics[width=0.3\textwidth]{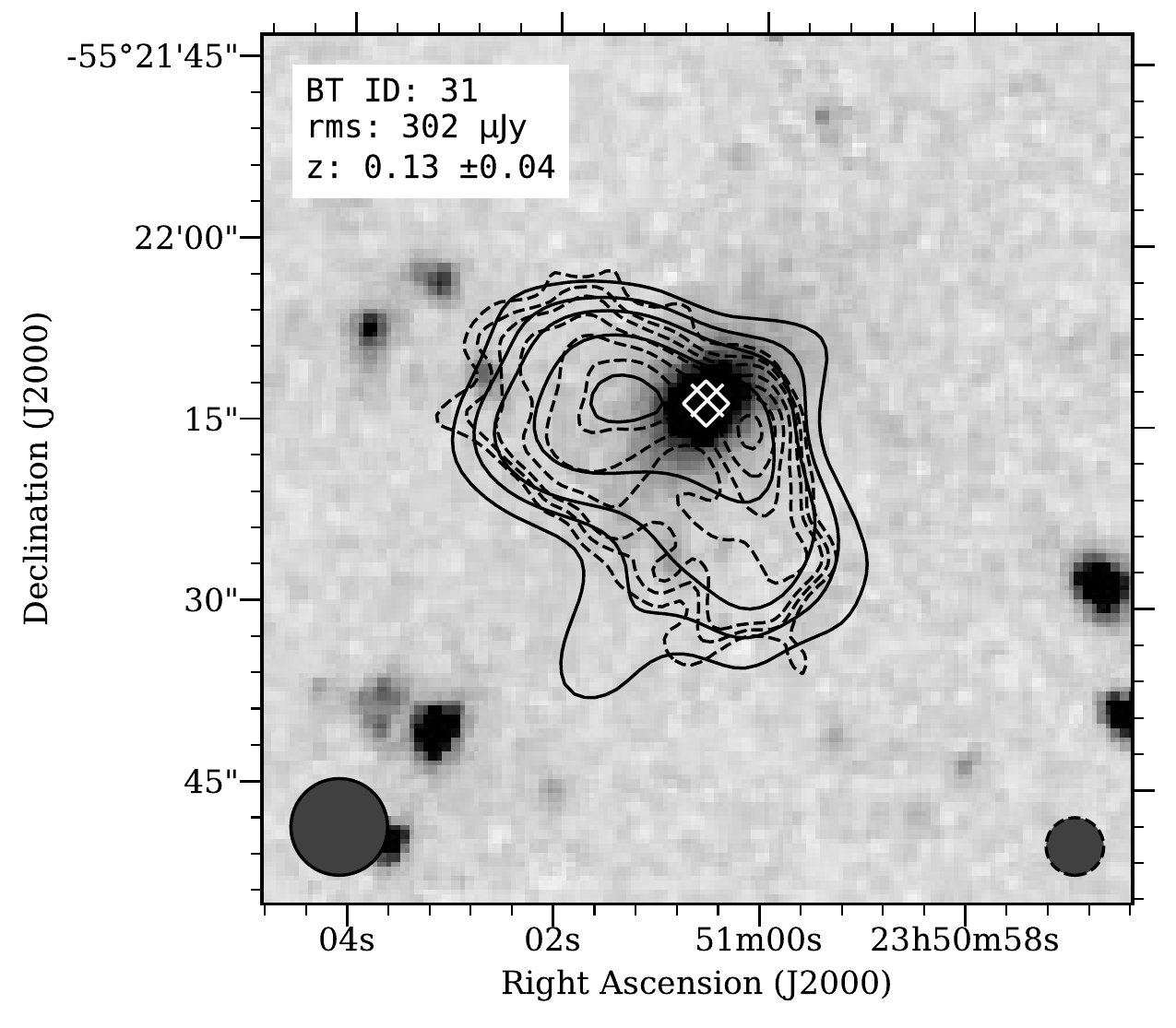}}
\subfloat{\includegraphics[width=0.3\textwidth]{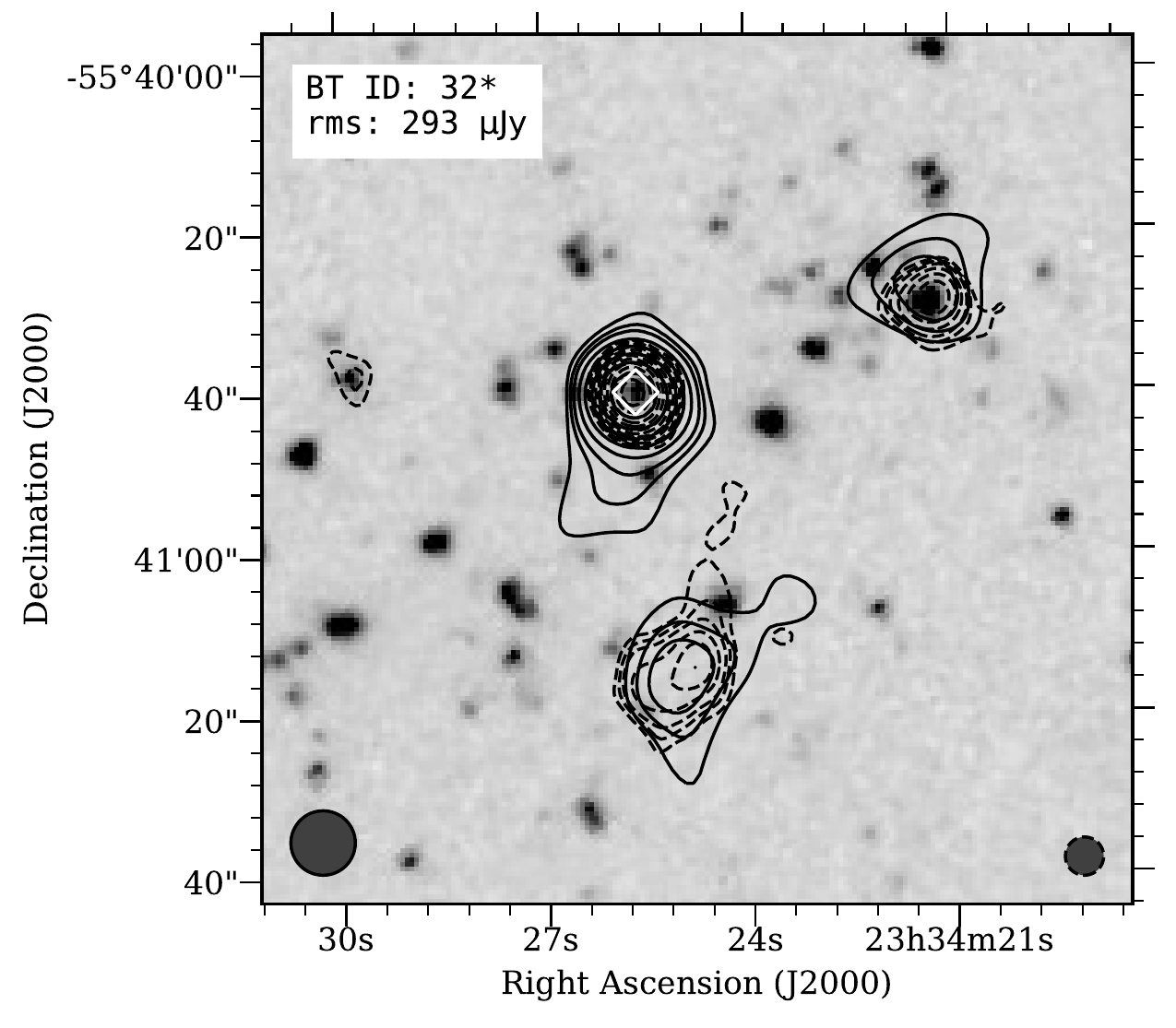}}
\subfloat{\includegraphics[width=0.3\textwidth]{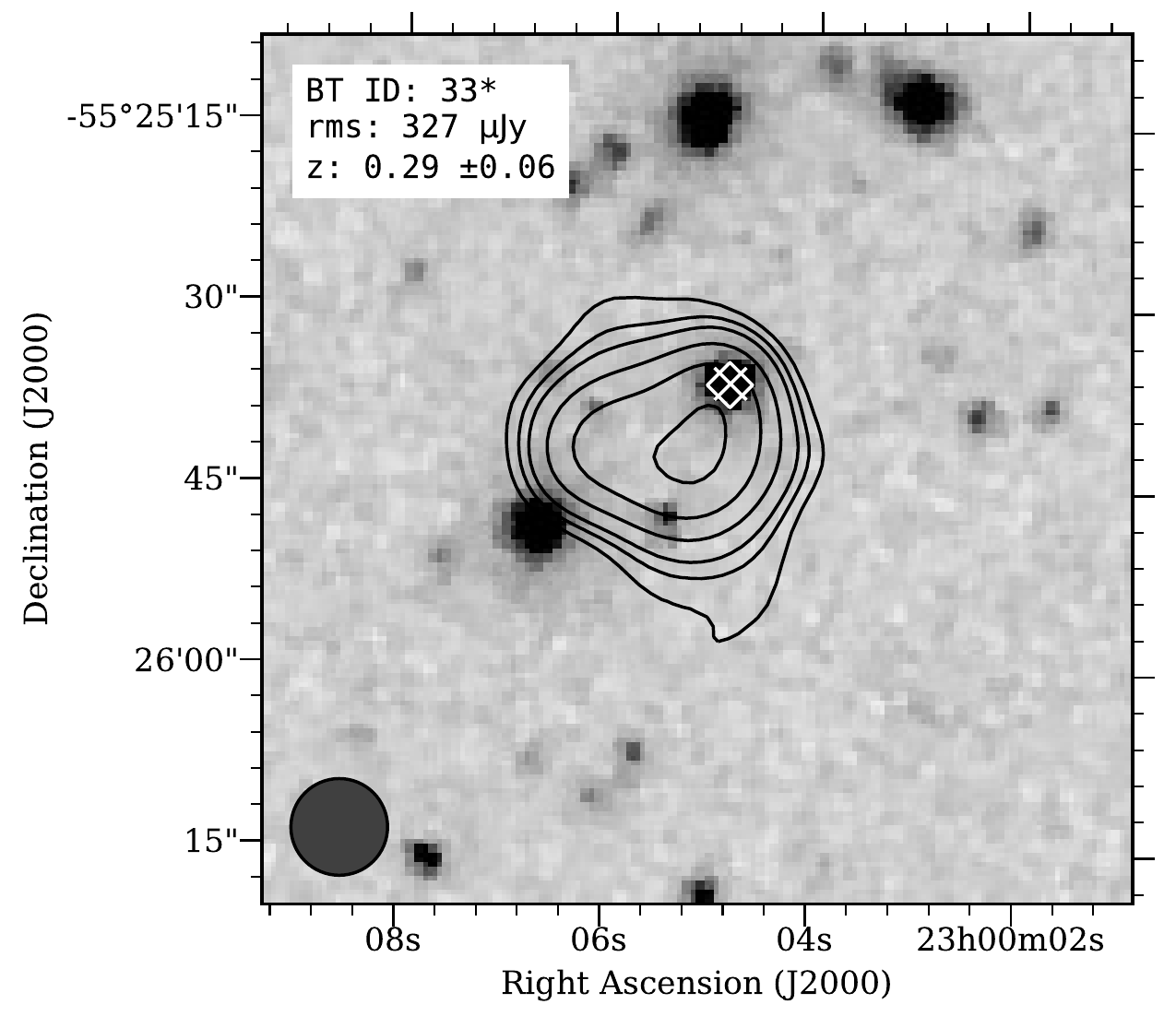}}\\
\subfloat{\includegraphics[width=0.3\textwidth]{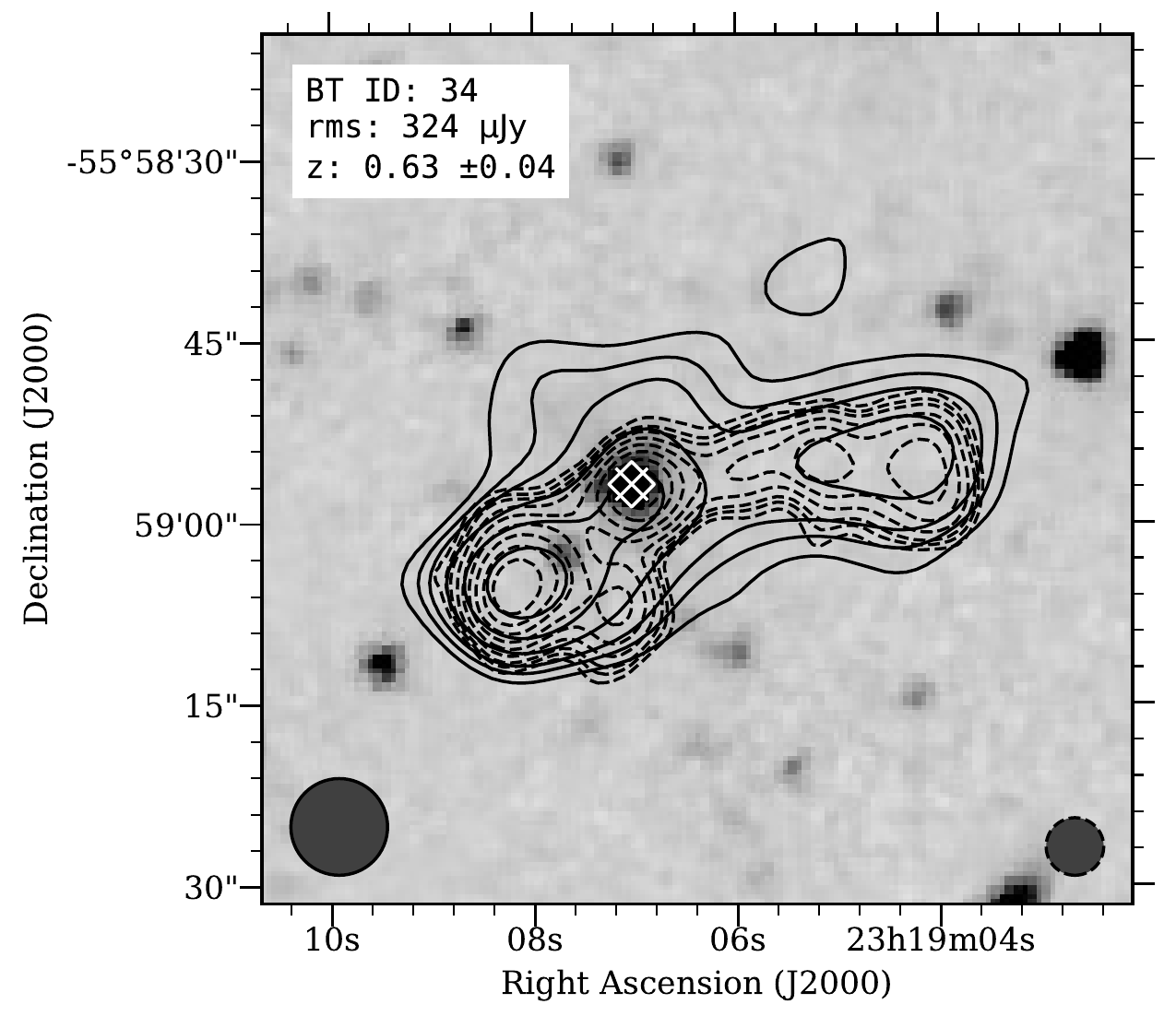}}
\subfloat{\includegraphics[width=0.3\textwidth]{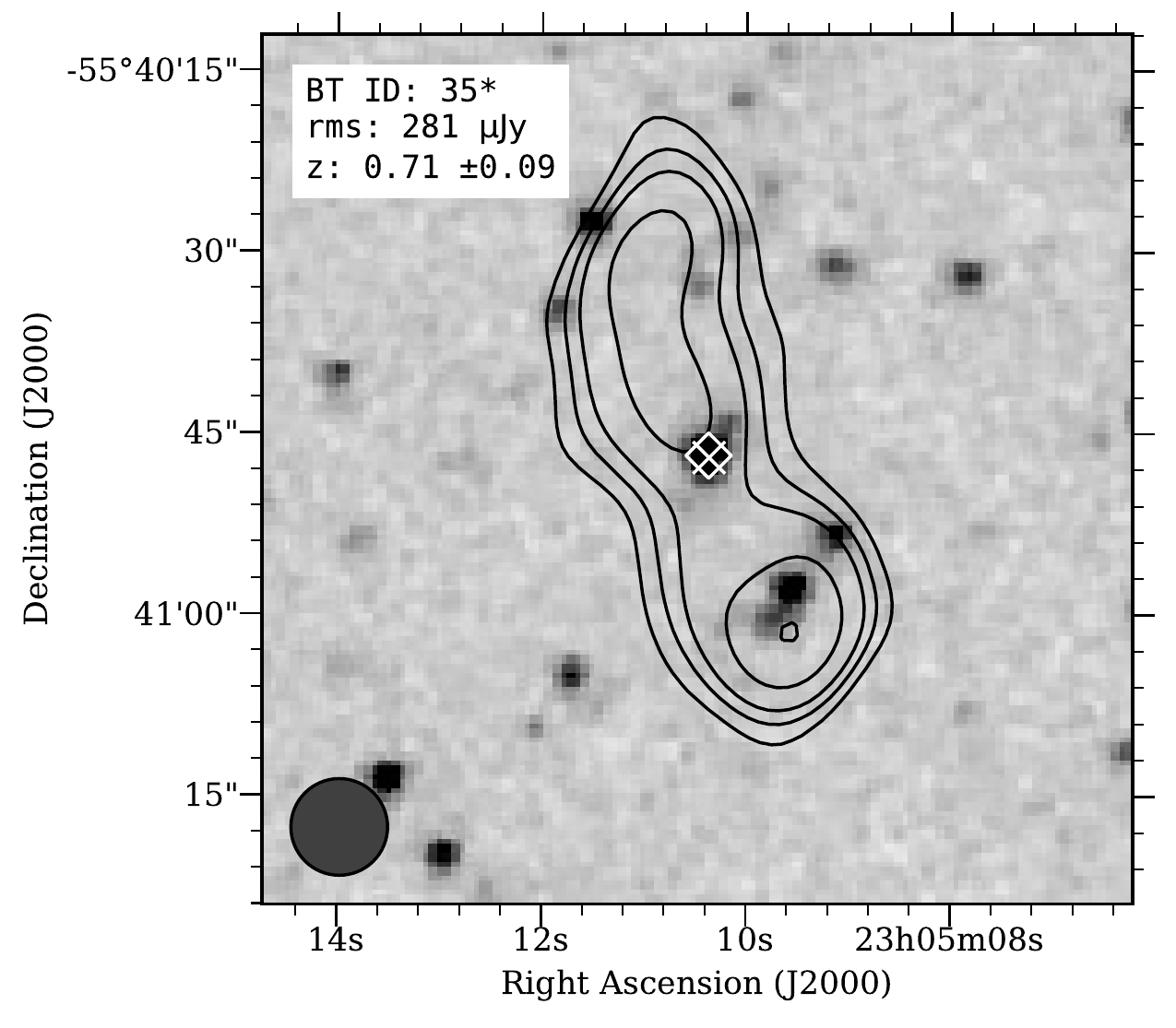}}
\subfloat{\includegraphics[width=0.3\textwidth]{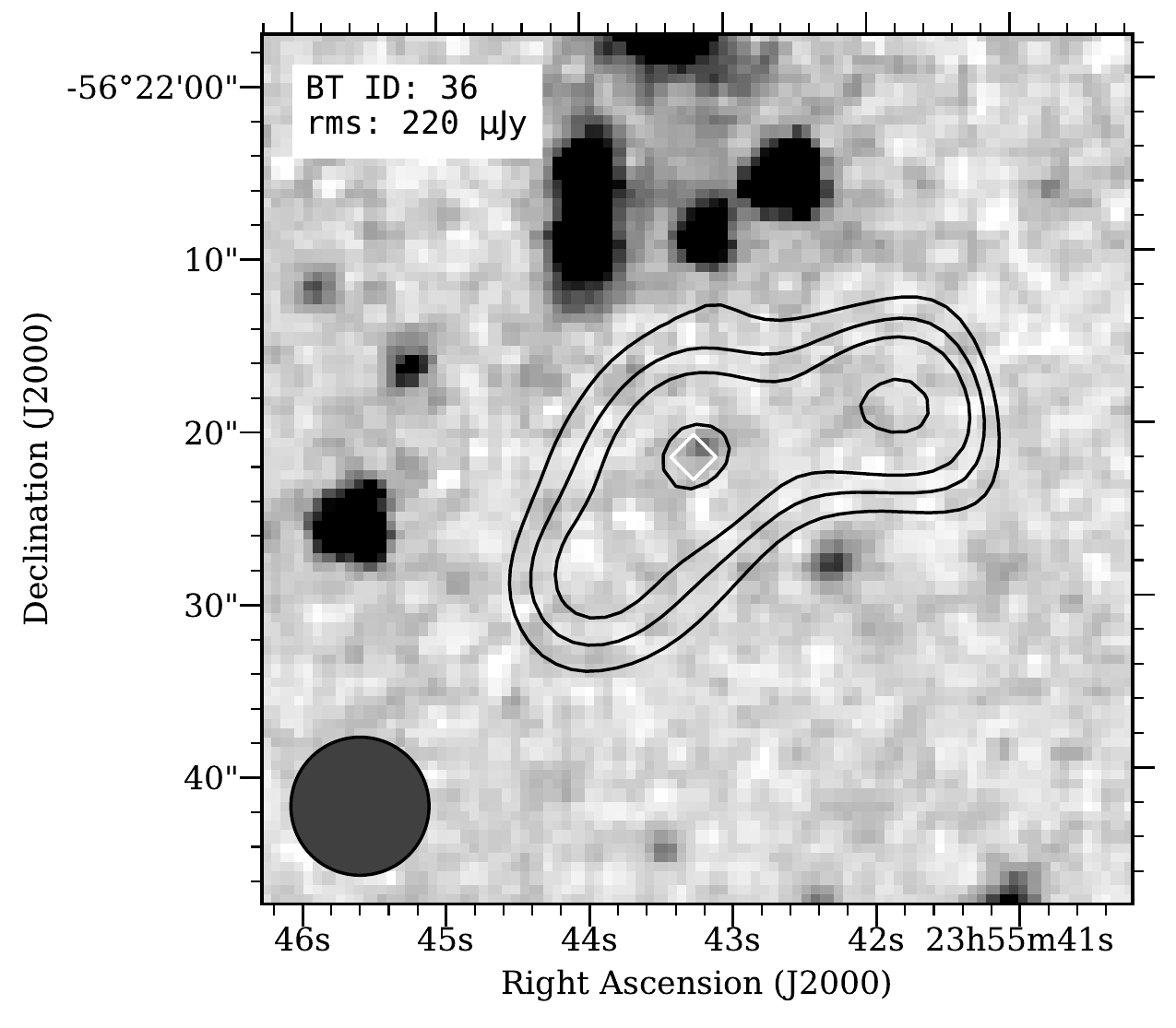}}\\
\subfloat{\includegraphics[width=0.3\textwidth]{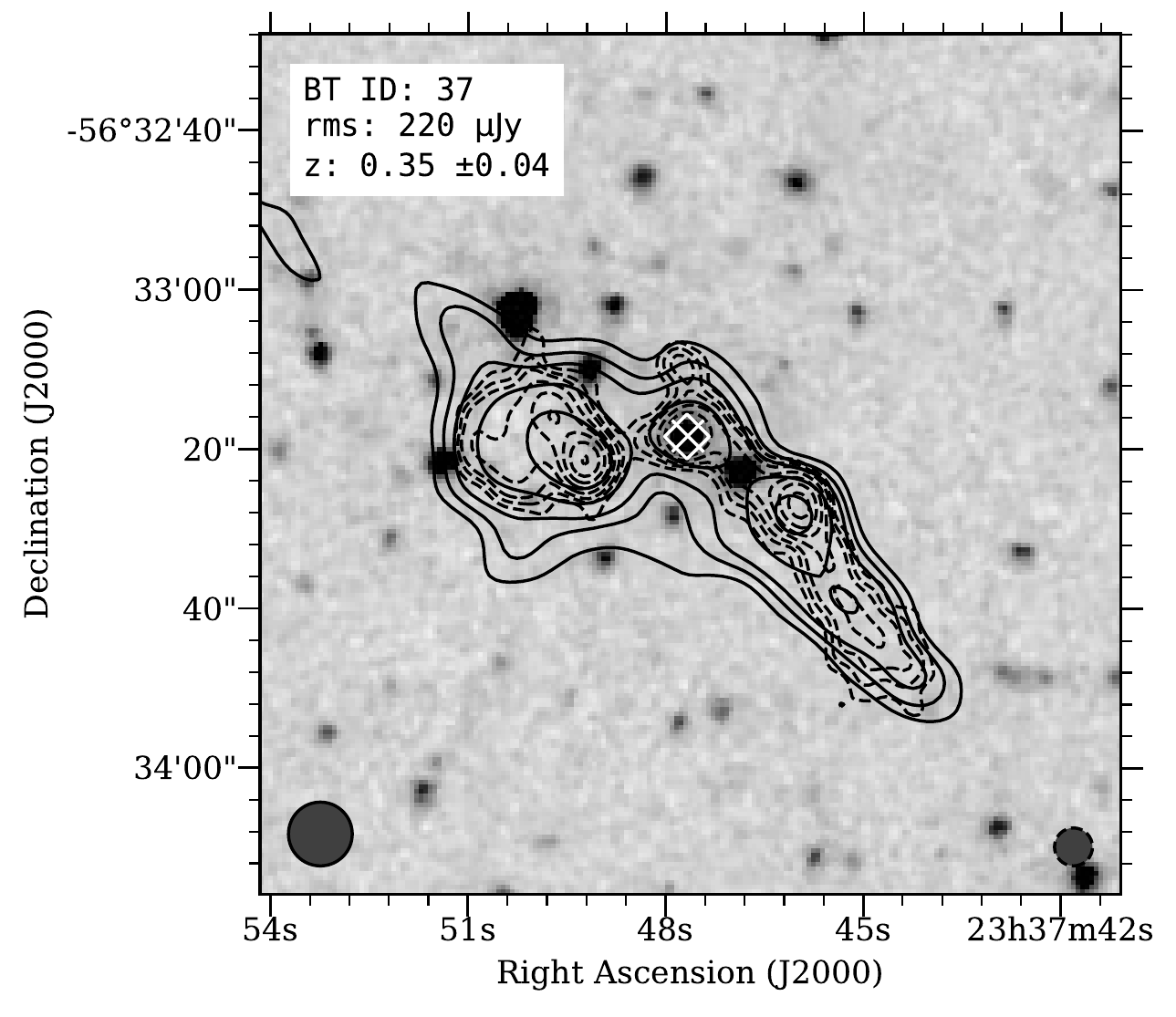}}
\subfloat{\includegraphics[width=0.3\textwidth]{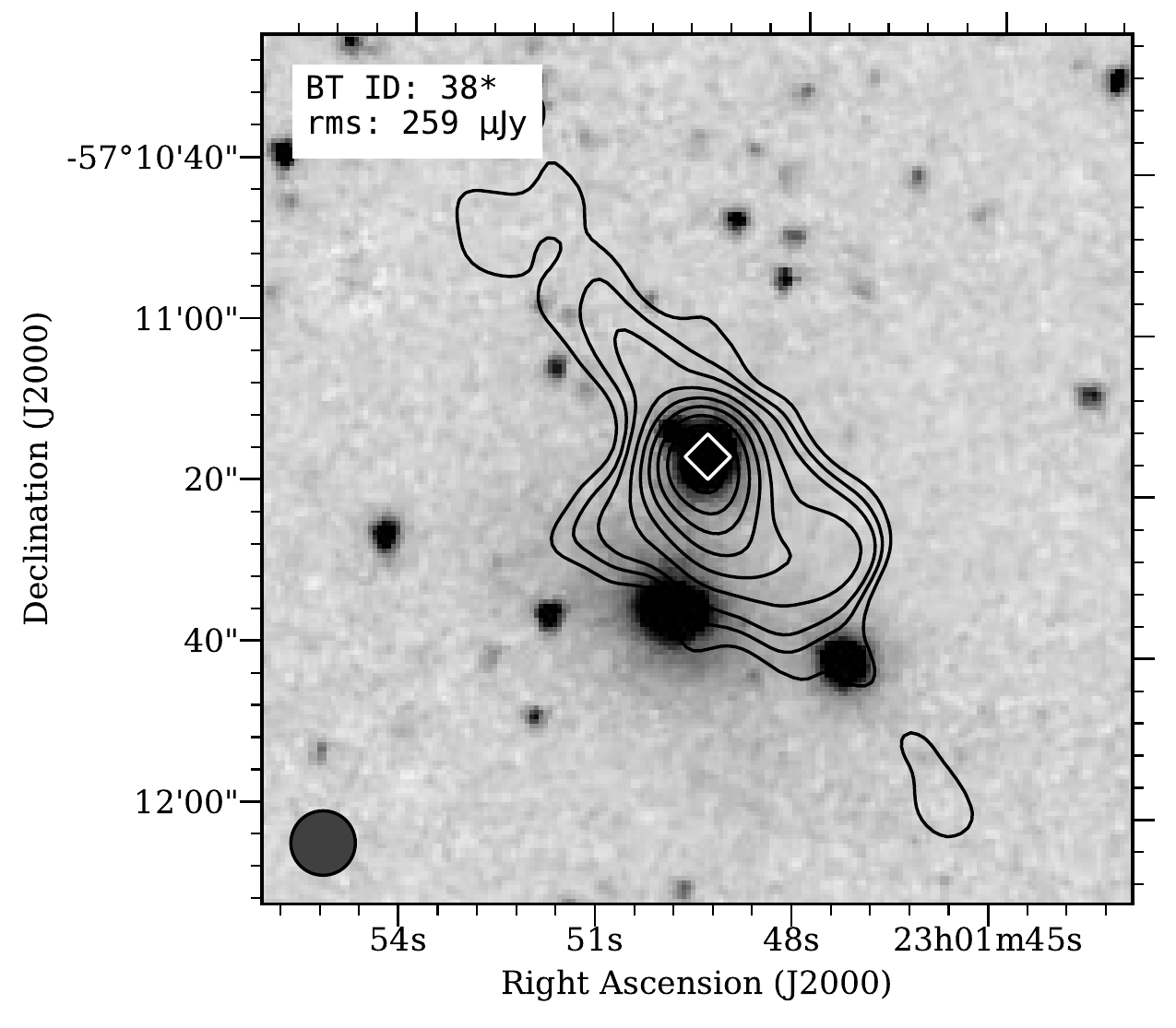}}
\subfloat{\includegraphics[width=0.3\textwidth]{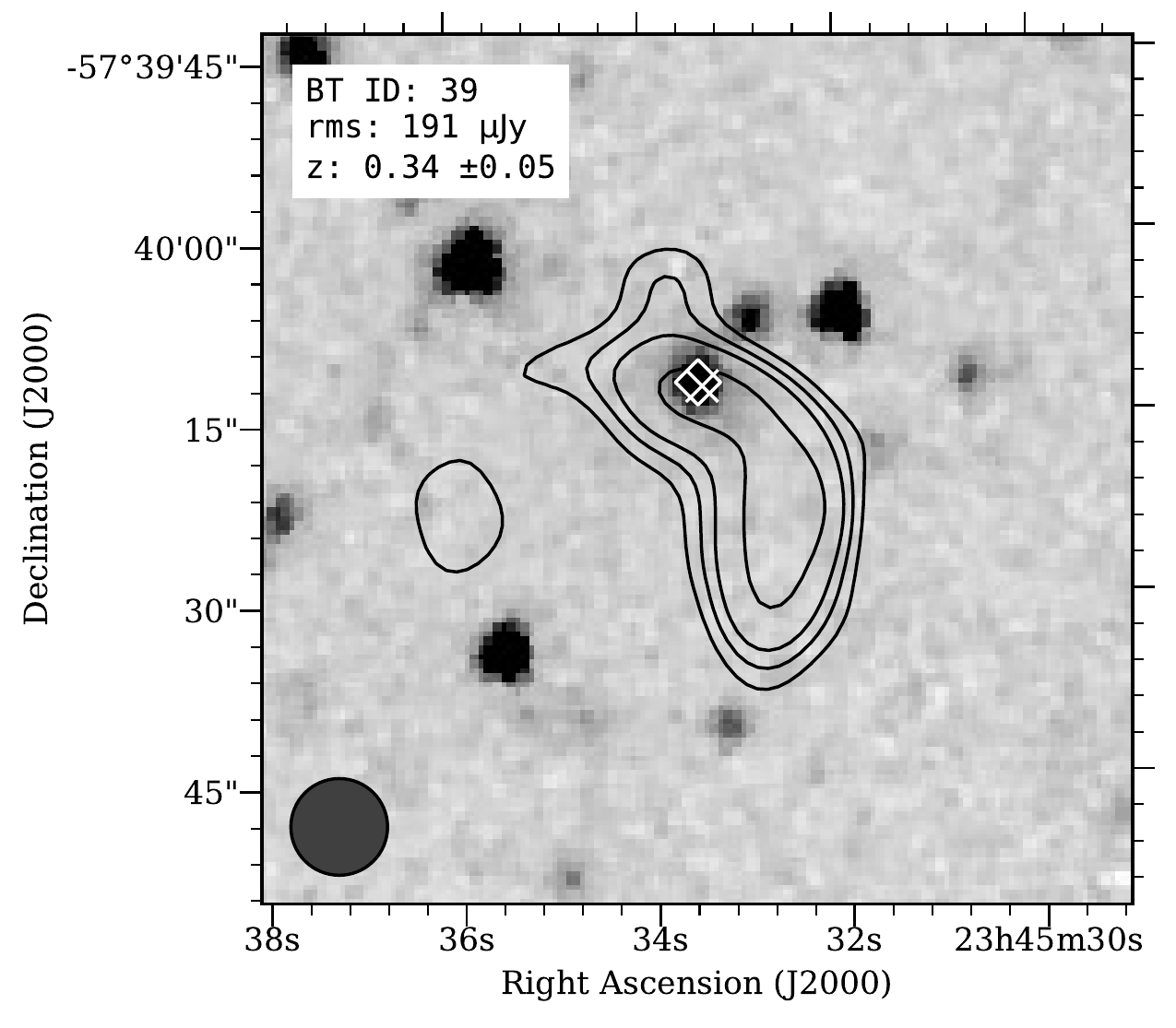}}\\
\subfloat{\includegraphics[width=0.3\textwidth]{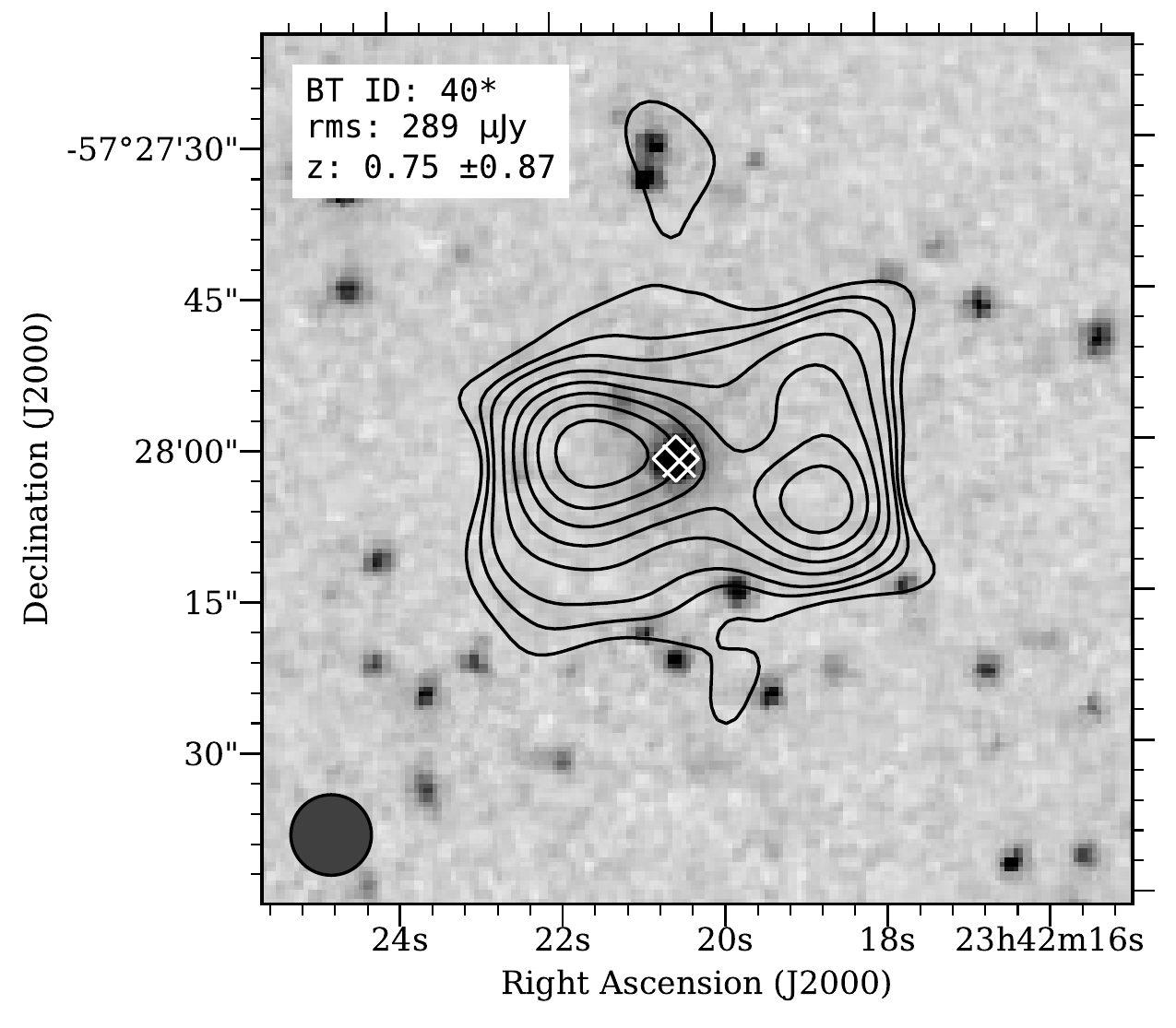}}
\subfloat{\includegraphics[width=0.3\textwidth]{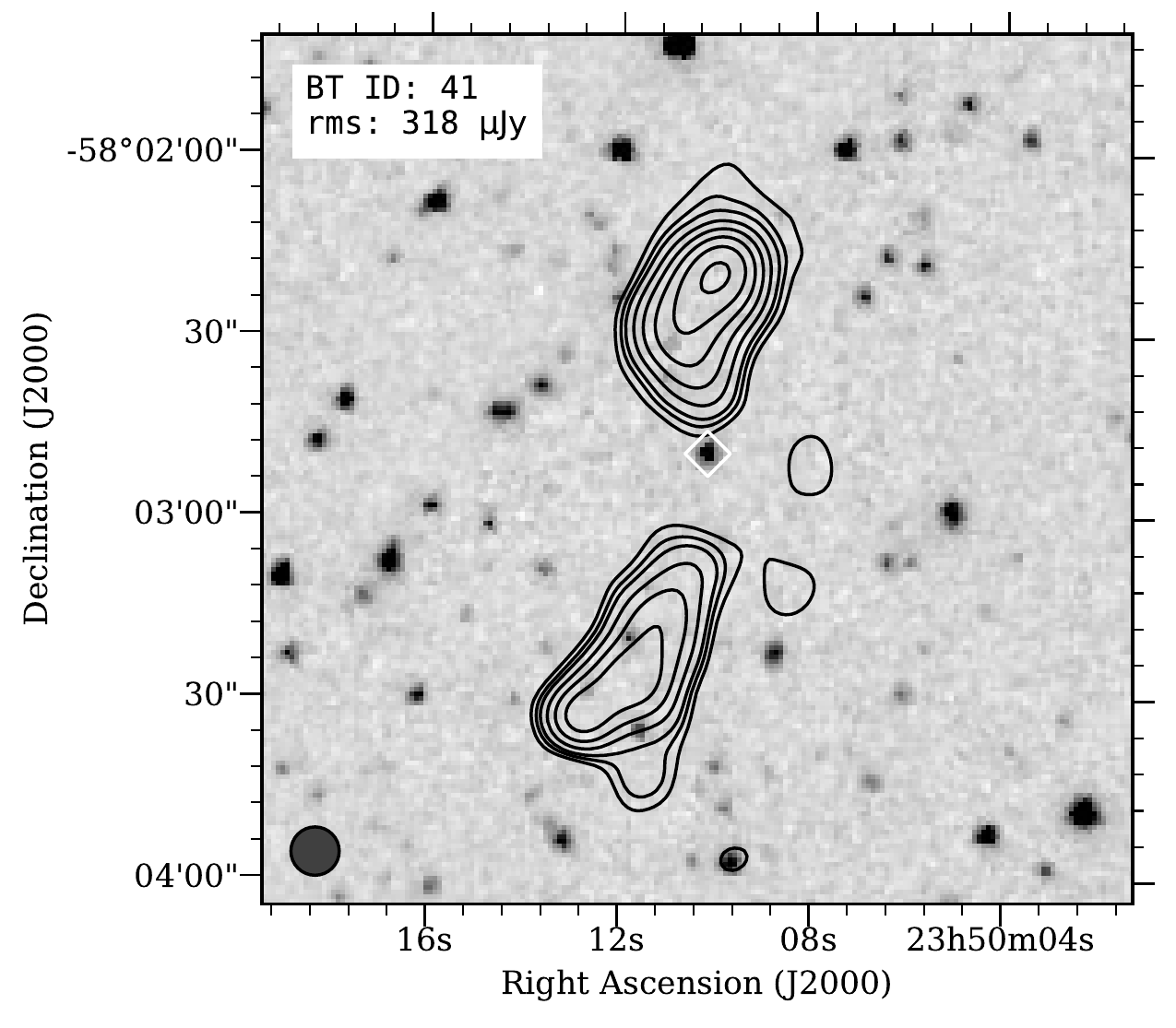}}
\subfloat{\includegraphics[width=0.3\textwidth]{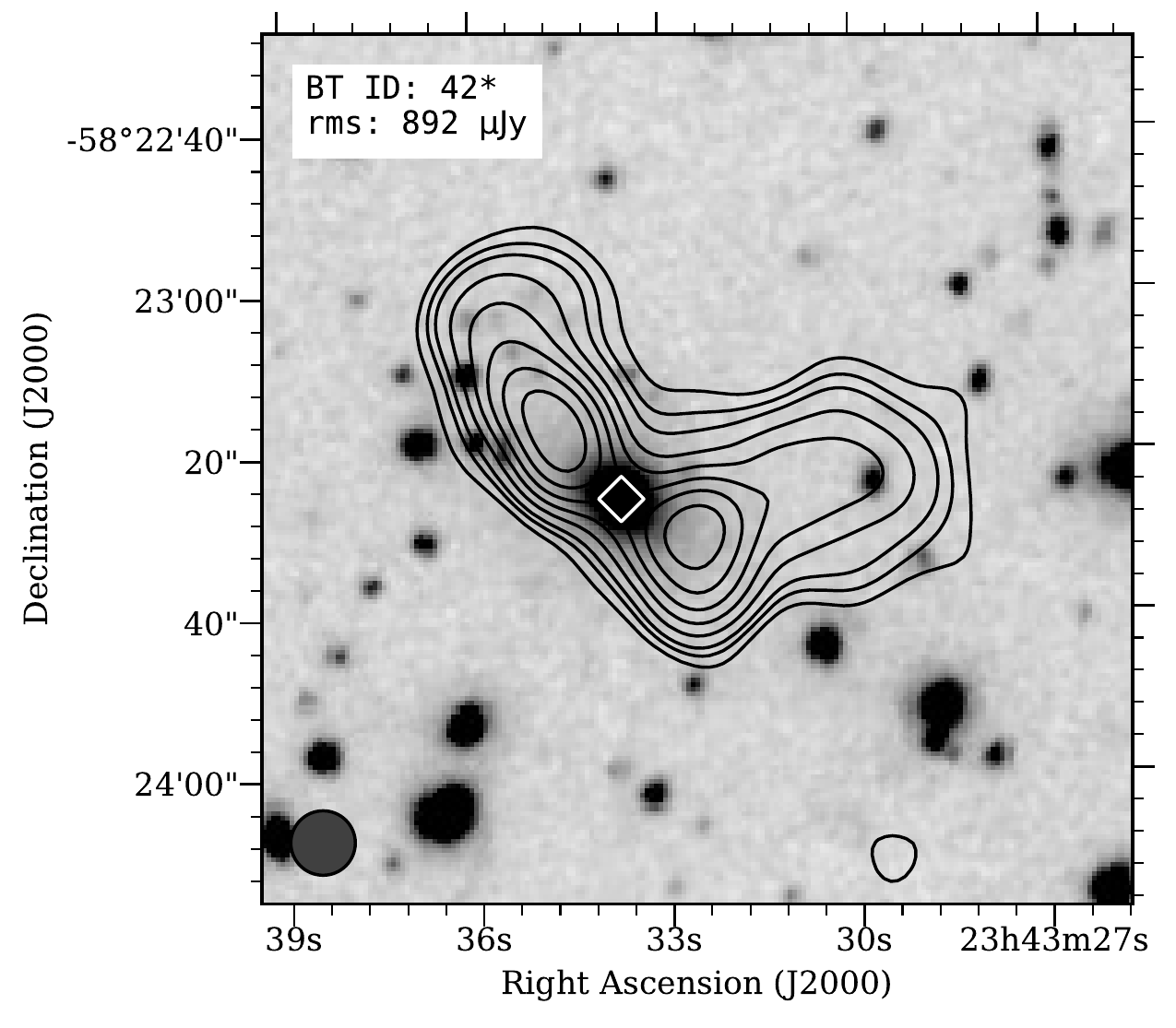}}\\
\contcaption{}
\end{figure*}
\begin{figure*}
\subfloat{\includegraphics[width=0.3\textwidth]{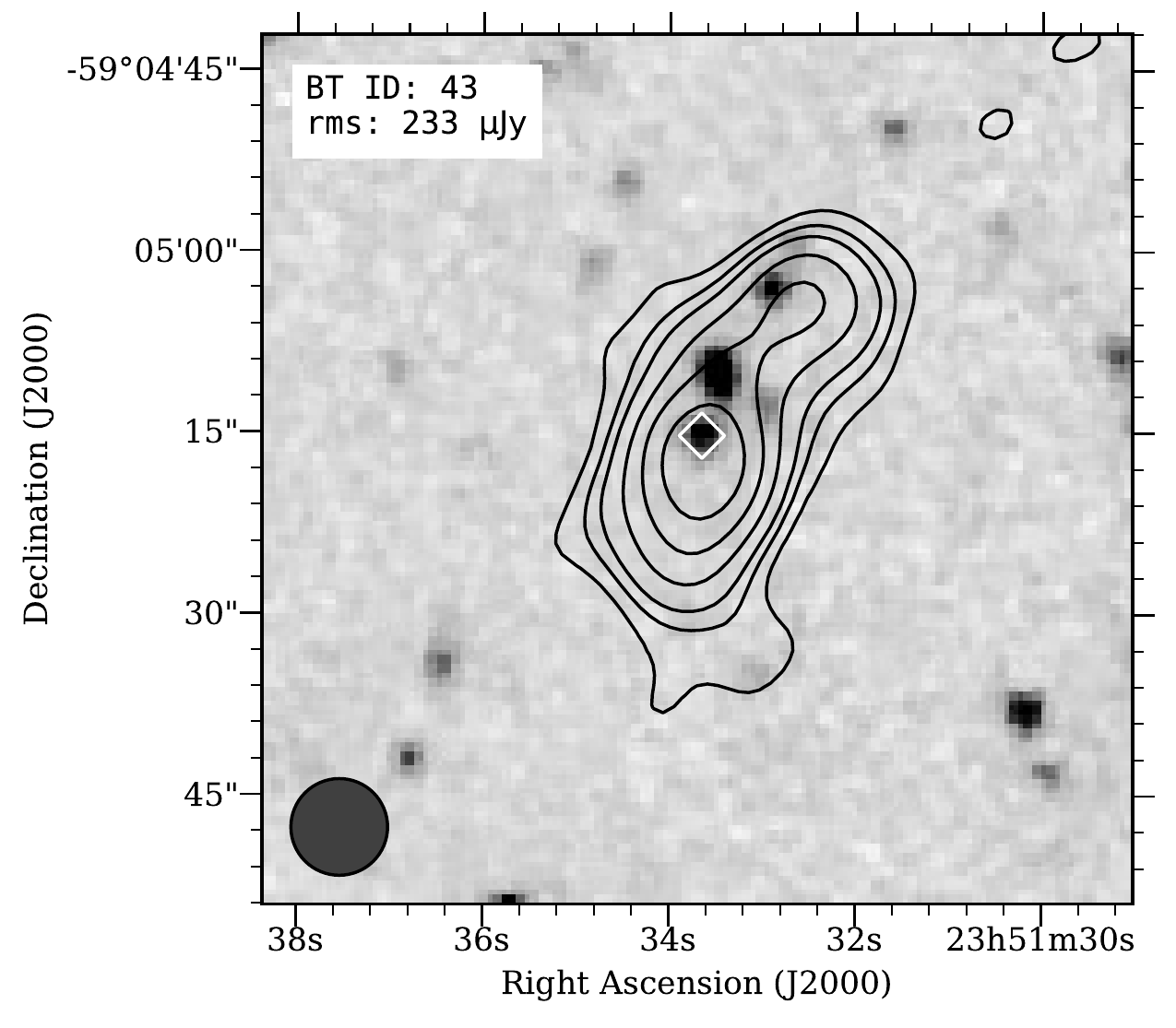}}
\subfloat{\includegraphics[width=0.3\textwidth]{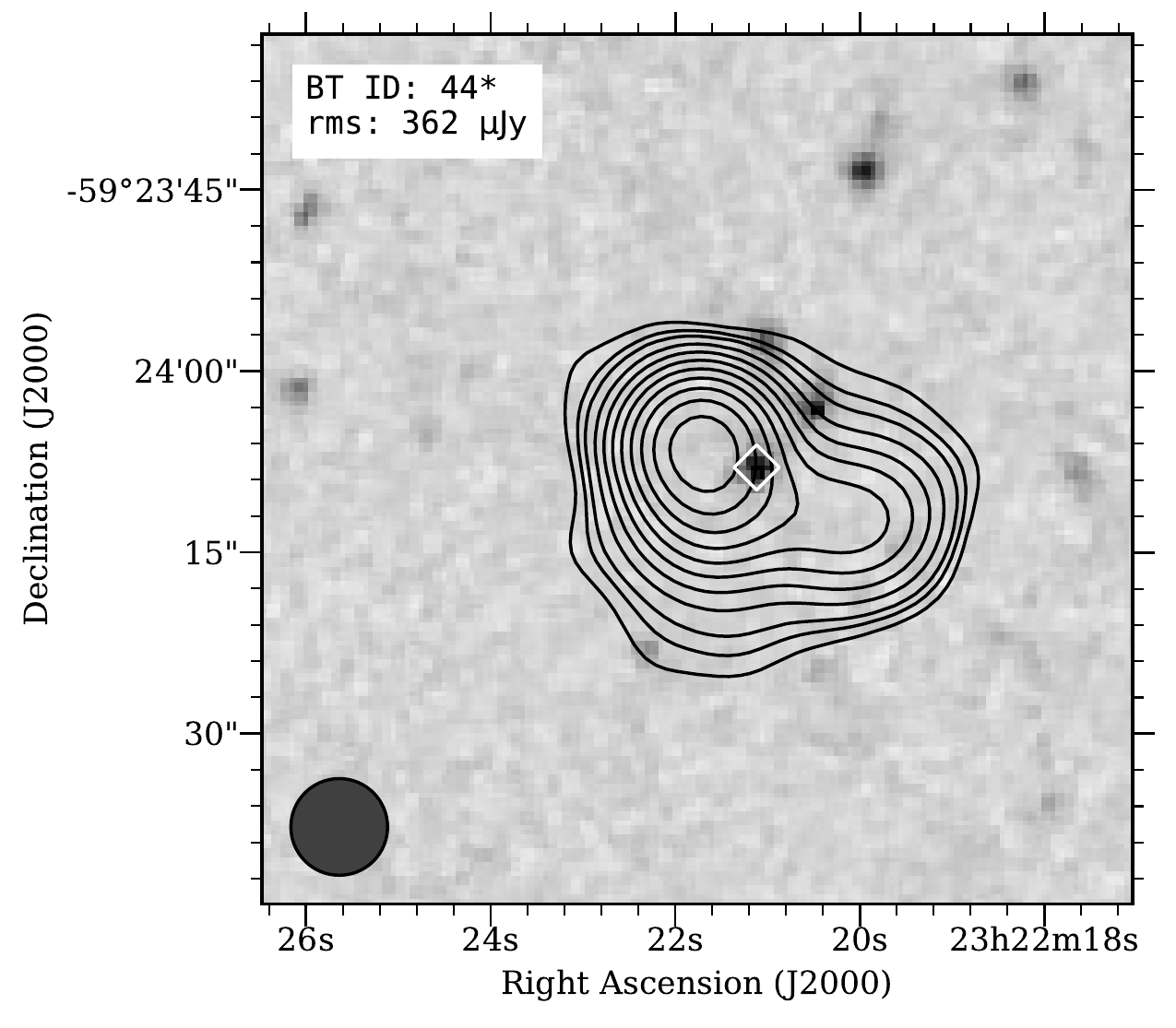}}
\subfloat{\includegraphics[width=0.3\textwidth]{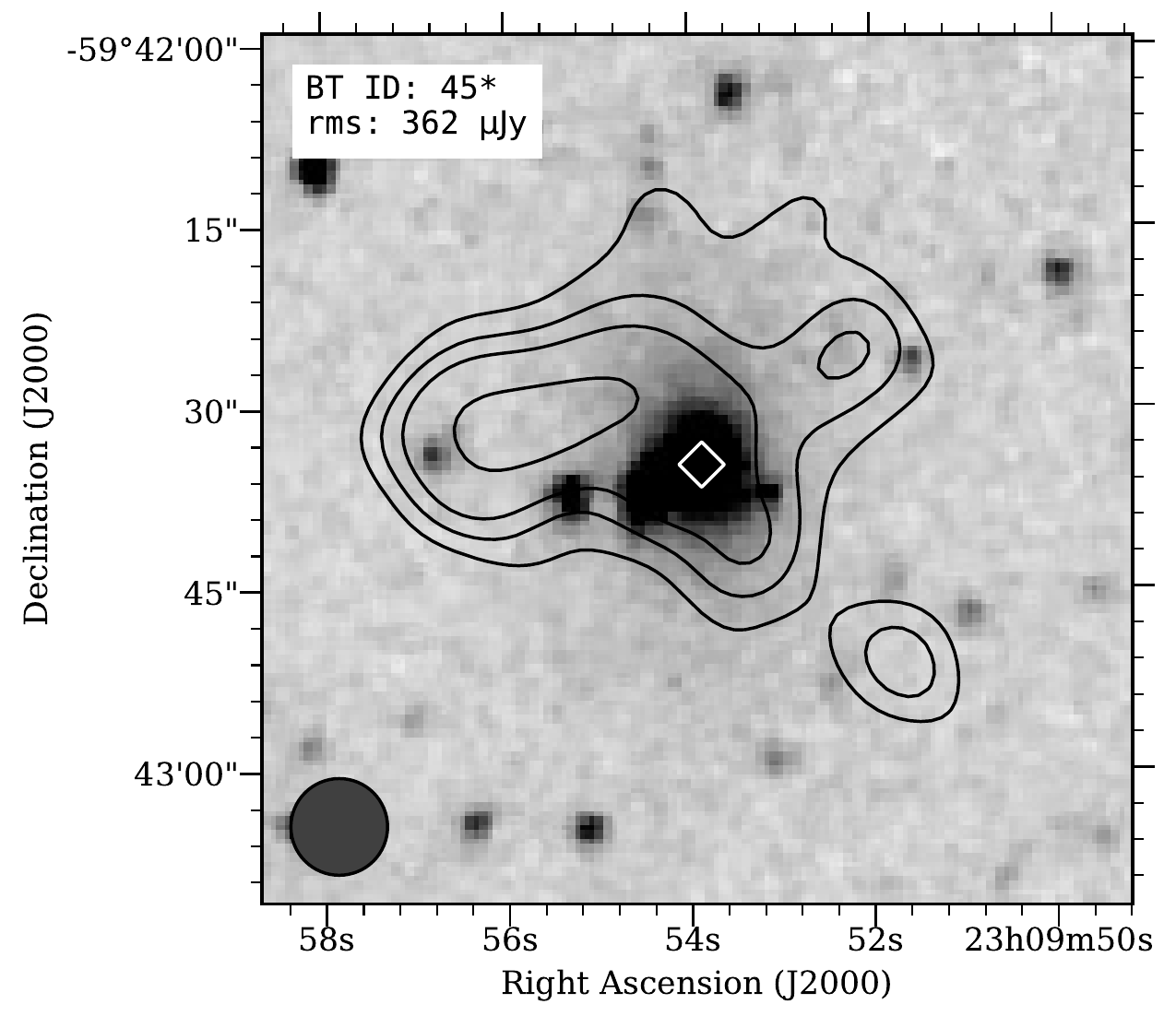}}\\
\subfloat{\includegraphics[width=0.3\textwidth]{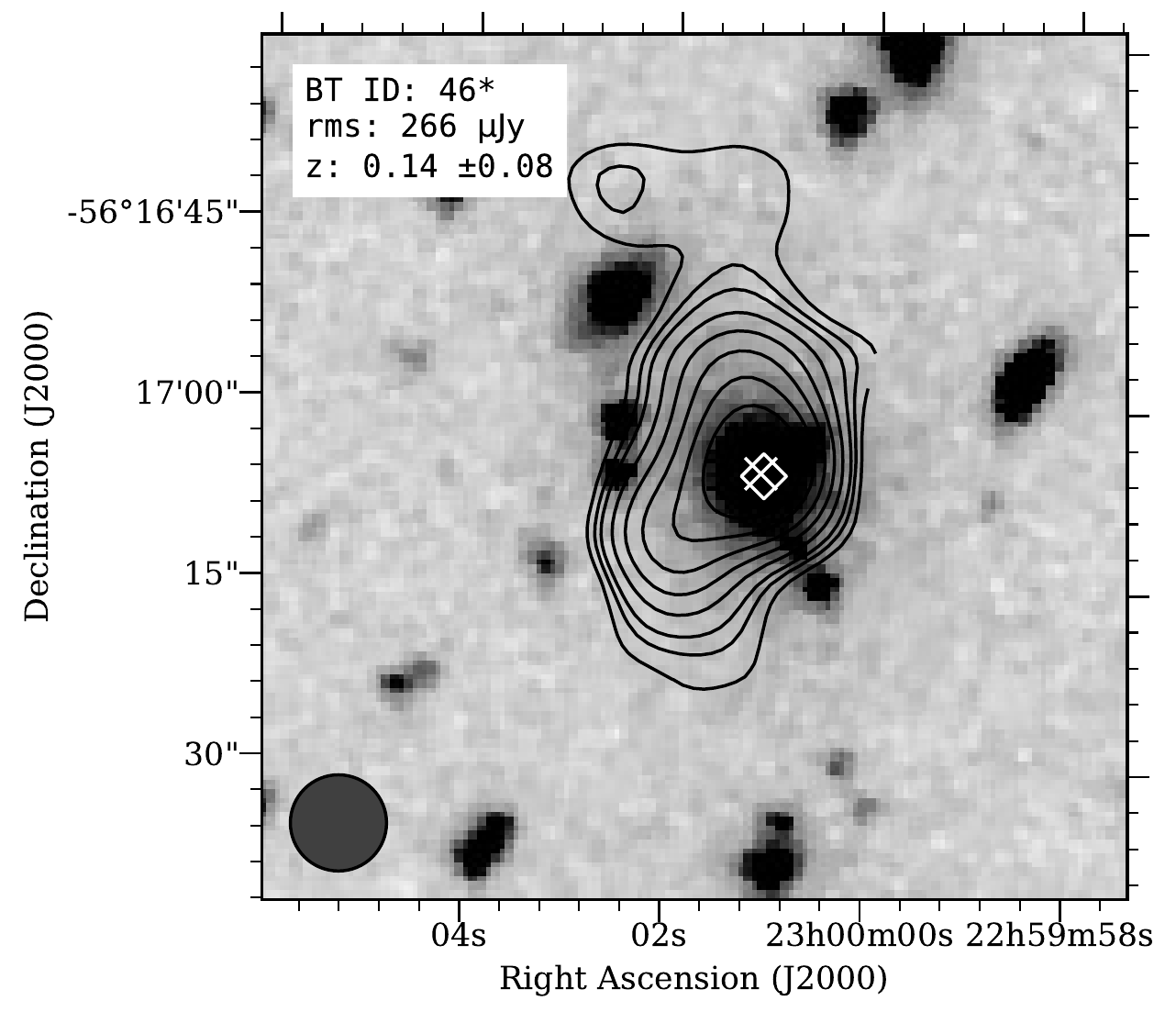}}\\
\contcaption{}
\end{figure*}


\bsp	
\label{lastpage}
\end{document}